\documentclass[11pt]{article}

\usepackage{setspace,graphicx,epstopdf,amsmath,amsfonts,amssymb,amsthm,versionPO}
\usepackage{marginnote,datetime,enumitem,rotating,fancyvrb,booktabs}
\usepackage{jfe}     
\usepackage{natbib}

\usepackage{ amssymb }
\usepackage{bbm} 
\usepackage[toc,page,header]{appendix}
\usepackage{comment}
\usdate
\usepackage[position=top]{subfig}
\usepackage[utf8]{inputenc}
\usepackage{array} 
\usepackage{tabu} 
\usepackage{multirow}
\usepackage{longtable}
\usepackage{hhline}
\usepackage{MnSymbol}

\usepackage{setspace}
\usepackage{caption}
\excludeversion{notes}		
\includeversion{links}          

\iflinks{}{\hypersetup{draft=true}}

\ifnotes{%
	\usepackage[margin=1in,paperwidth=10in,right=2.5in]{geometry}%
	\usepackage[textwidth=1.4in,shadow,colorinlistoftodos]{todonotes}%
}{%
	\usepackage[margin=1in]{geometry}%
	\usepackage[disable]{todonotes}%
}

\makeatletter\let\chapter\@undefined\makeatother 

\usepackage{indentfirst} 
\usepackage{endnotes}    
\usepackage{jfe}          
\usepackage[labelfont=bf,labelsep=period]{caption}   

\theoremstyle{definition}
\theoremstyle{corollary}
\theoremstyle{proposition}
\theoremstyle{lemma}

\usepackage{color,hyperref}
\definecolor{RoyalBlue}{rgb}{0.0,0.08,0.85}
\hypersetup{colorlinks,breaklinks,
	linkcolor=RoyalBlue,urlcolor=RoyalBlue,
	anchorcolor=RoyalBlue,citecolor=RoyalBlue}
\usepackage{hyperref,float}
\usepackage{hyperref}
\usepackage{tikz}
\usepackage{pgfplots, pgfplotstable}
\pgfplotsset{compat=newest} 
\pgfplotsset{plot coordinates/math parser=false}
\pgfplotsset{scaled z ticks=false}
\pgfplotsset{scaled y ticks=false}
\usepackage{subfloat}

\makeatletter
\def\Url@twoslashes{\mathchar`\/\@ifnextchar/{\kern-.2em}{}}
\g@addto@macro\UrlSpecials{\do\/{\Url@twoslashes}}
\let\Date\@date
\makeatletter

\usepackage{fnpct}
\usepackage{pdflscape}
\usepackage{accents}
\newlength{\dhatheight}

\usepackage{tabularx}
\usepackage{array}
\newcolumntype{H}{>{\setbox0=\hbox\bgroup}c<{\egroup}@{}}
\newcolumntype{Y}{>{\centering\arraybackslash}X}

\begin{document}

	\setlist{noitemsep}  
\onehalfspacing      

\renewcommand{\thempfootnote}{\fnsymbol{mpfootnote}}


\begin{center}
	{\huge \bf \hspace{-0.1in}Behavioral Economics of AI: \\\vspace{0.15in} LLM Biases and Corrections$^\smallstar$}
\end{center}

\vspace{0.15in}

\begin{centering}{\Large Pietro Bini$^{a}$\quad Lin William Cong$^{b,c}$\quad Xing Huang$^{b}$\quad Lawrence J. Jin$^{b,c}$ }

\vspace{0.15in}

\noindent $^{a}$\textit{Questrom School of Business, Boston University}

\noindent $^{b}$\textit{SC Johnson College of Business, Cornell University}

\noindent $^{c}$\textit{National Bureau of Economic Research}

\end{centering}
\thispagestyle{empty} 
\vspace{0.15in}

\begin{center}
First draft: January 2025; this draft: January 2026 
\end{center}

\vspace*{0.10in}

\begin{center}
{\bf ABSTRACT}

\end{center}

\begin{quote}

Do generative AI models, particularly large language models (LLMs), exhibit systematic behavioral biases in economic and financial decisions? If so, how can these biases be mitigated? Drawing on the cognitive psychology and experimental economics literatures, we conduct the most comprehensive set of experiments to date---originally designed to document human biases---on prominent LLM families across model versions and  scales. We document systematic patterns in LLM behavior. In preference-based tasks, responses become more human-like as models become more advanced or larger, while in belief-based tasks, advanced large-scale models frequently generate rational responses. Prompting LLMs to make rational decisions reduces biases.\\

\end{quote}

\vspace{\fill}

\noindent {\footnotesize \renewcommand{\baselinestretch}{1.0} $^{\smallstar}$We are grateful to Nicholas Barberis, James Choi, Clifton Green, William Goetzmann, David Hirshleifer, Gerard Hoberg, Camelia Kuhnen, Devin Shanthikumar, Kelly Shue, Siew Hong Teoh, Luyao Zhang, and seminar participants at Cornell University, the Federal Reserve Board, the University of California Los Angeles, the 2024 Conference for Financial Economics and Accounting, the 2025 Conference on Emerging Technologies in Accounting and Financial Economics, the 2025 Chicago Booth Conference on Behavioral Approaches to Financial Decision Making, and the 2026 AFA Annual Meeting for helpful discussions and comments.~Jordan Velte and Shuhuai Zhang provided excellent research assistance. Bini and Cong acknowledge financial support from Ripple's University Blockchain Research Initiative. Please send correspondence to Jin at lawrence.jin@cornell.edu or Cong at will.cong@cornell.edu.}

\newpage

\doublespacing

\setcounter{page}{1}

\doublespacing
	
\section{Introduction} \label{sec:introduction}

Artificial intelligence (AI), especially generative large language models (LLMs), is becoming increasingly essential in daily work and general economic activities. For example, banks and FinTech firms are integrating generative AI (GenAI) technologies into operations management, customer service, financial advice, and risk assessment (\citealp{vidal2023};~\citealp*{tomlinsonetal2024}). Researchers are exploring the potential for LLMs to enhance experimentation that studies human behavior (\citealp*{charness2023generation};~\citealp{korinek2023generative};~\citealp{bail2024can}). However, much remains unknown about how AI algorithms and agents behave systematically, particularly in economic and financial decisions, or whether their behavior resembles human behavior. Understanding the ``behavioral economics'' of AI---potentially a novel class of agents \citep{tegmark2018life}---starting with LLMs is urgent for assessing and improving the technology's utility, safety, and appropriateness.

Recent studies have begun to examine the reliability of LLMs in expectation formation and decision making, with a focus on the behavior of ChatGPT.\footnote{For example, ChatGPT's behavior has been examined in both individual settings (\citealp{chen_etal_2023_emergence};~\citealp*{ma_etal_2023};~\citealp{chen_etal_2024_manager};~\citealp{bybee_2025};~\citealp{hansen_et_al_2025}) and game-theoretic settings (\citealp{bauer_etal_2023_decoding};~\citealp{mei_etal_2024_turing};~\citealp{fan_etal_2024_can};~\citealp{brookins_2024_playing};~\citealp{manning_horton_2025}).} Our paper not only adds to this work but also conceptually introduces a new field---the behavioral economics of AI---by establishing benchmark results: we conduct the most comprehensive set of experiments to date, originally designed to document human biases but now applied to investigate the biases of multiple prominent LLM families; we systematically compare LLM responses with both rational and human responses; and we explore methods for correcting these biases. An important goal of the paper is to develop a public database of experimental questions for ongoing evaluations of behavioral biases across various LLMs.

We start by exploring two broad approaches for conducting experiments that allow us to document the behavioral biases of LLMs. The first approach draws on the cognitive psychology literature, originating  with~\cite{ellsberg1961} and~\citet*{kahneman1973prediction,kahneman1979prospect}, which uses carefully designed experimental questions to assess psychological biases in humans. From this literature, we select a comprehensive set of experiments covering both questions that study the psychology of preferences and questions that study the psychology of beliefs. Our selection ensures the inclusion of experiments that document the psychological biases that are first-order important in financial markets.\footnote{\cite{Barberis:2018} argues that prospect theory preferences, overextrapolation, and overconfidence are the three main psychological biases that drive investor behavior, firm behavior, and asset prices in financial markets. } For each question, we design a prompt suitable for LLMs, hence allowing us to elicit responses from these models and analyze their behavior. The second approach draws on recent experimental economics studies, which, compared to the cognitive psychology literature, feature experimental tasks that are more closely tied to economic and financial settings. We adapt these tasks for LLMs to investigate the behavioral biases they exhibit in financial decision making.

With the experimental questions in hand, we collect responses through an application programming interface (API) from four prominent families of LLMs: OpenAI's ChatGPT, Anthropic Claude, Google Gemini, and Meta Llama.\footnote{These were the extant LLMs when we begun our study in 2023.} For each family, we consider two variations. First, we examine an advanced version of the model alongside an older version; this allows us to study time-series variation in the model's degree of behavioral biases. Second, for the advanced model, we compare a version with a large parameter scale to one with a smaller scale; this allows us to study cross-sectional variation in the model's degree of behavioral biases. 

Analyzing the responses reveals five observations. First, for questions from the cognitive psychology literature that document biases in preferences, the LLMs' answers exhibit a clear pattern: as models become more advanced or larger, responses become increasingly human-like and they are irrational according to the Expected Utility framework. For example, Claude 3 Opus, an advanced large-scale LLM, answers four out of six preference-based questions in a way that is consistent with human responses. In comparison, Claude 3 Haiku, another advanced model with a smaller scale, gives human-like answers to three out of the six questions, while Claude 2, an older version, gives human-like answers only to one out of the six questions. 

Second, for questions from the cognitive psychology literature that document biases in beliefs, the LLMs' answers exhibit the \textit{opposite} pattern: more advanced or larger models produce increasingly rational responses. For example, Gemini 1.5 Pro, a highly advanced large-scale LLM, answers all ten belief-based questions correctly. By contrast, Gemini 1.5 Flash, another advanced but smaller model, answers five questions correctly, while Gemini 1.0 Pro, an older version, answers only two questions correctly. Overall, three of the four advanced large-scale LLMs we examine---GPT-4, Claude 3 Opus, and Gemini 1.5 Pro---produce predominantly rational answers to belief-based questions.

Third, substantial heterogeneity emerges when we compare responses across LLM families. For preference-based questions, Gemini's responses are less rational and more human-like compared to those from ChatGPT, while the responses from Claude or Llama are generally similar to those from ChatGPT. For belief-based questions, Meta Llama's responses are less rational and more human-like compared to those from GPT, while the responses from Anthropic Claude or Google Gemini are similar to those from GPT. 
 
Fourth, we examine LLMs' responses to questions from two recent experimental economics studies.~\cite{Afrouzi2023} ask human participants to observe a sequence of past realizations of a random variable and then forecast its future values; the random variable follows an autoregressive process. We elicit LLMs' responses in this setting and find that advanced small-scale LLMs---GPT-4o, Claude 3 Haiku, and Gemini 1.5 Flash---produce forecasts that are human-like and irrational: they perceive an autoregressive process that is more persistent than the true process. By contrast, their larger-scale counterparts generate more rational forecasts, with perceived persistence similar to the true persistence.

\cite{bose2022} present human participants with stock price trajectories and ask them how much to invest in each stock. Replicating this setting for LLMs, we find that large-scale models---GPT-4, Claude 3 Opus, and Gemini 1.5 Pro---make investment decisions that are more human-like than their smaller-scale counterparts: investment depends more strongly on the visual salience of a stock's price trajectory, a preference-based variable identified by~\cite{bose2022} as driving human investment behavior.~Taken together, these results suggest that the patterns documented with cognitive psychology questions also hold in experimental economics studies: for preference-based questions, larger models produce increasingly human-like responses, while for belief-based questions, larger models produce increasingly rational responses.\footnote{For the~\cite{Afrouzi2023} and~\cite{bose2022} tasks, LLMs must process graphical inputs. Six of the twelve LLMs we examine do not support graphical inputs, so the tasks are not run on these models. See Section~\ref{subsec:prompt_design} for a detailed discussion.} 

Finally, we explore methods for correcting the observed behavioral biases. Among the methods we test, one seems effective while the others are not. The effective method involves a brief role-priming instruction that asks an LLM to think of itself as a rational investor who makes decisions using the Expected Utility framework; this instruction is provided before the LLM answers any question.~We find that, for both preference-based and belief-based questions, this role-priming instruction makes LLM responses more rational and less human-like, although the magnitude of improvement is economically modest. We also find that role priming affects LLM responses through the confidence levels LLMs assign to their choices as well as their self-reported reasoning type, where type A corresponds to intuitive thinking and type B corresponds to analytical thinking and calculations. The other methods, which combine the role-priming instruction with additional genuinely useful information, are found to be ineffective in reducing biases. Overall, these results indicate that debiasing LLMs remains a challenging task.

The five observations stated above are descriptive, yet informative. Although a full analysis of the mechanisms driving these results is beyond the scope of the paper, we put forth two conjectures. First, why do more advanced or larger-scale models become more human-like when responding to preference-based questions? We conjecture this is partly because advanced, large-scale LLMs increasingly rely on Reinforcement Learning from Human Feedback (RLHF), a training process that aligns the underlying model with human preferences as reflected in human feedback \citep{stiennon2020learning}. Second, why do more advanced or larger-scale models become more rational when responding to belief-based questions? We conjecture this is partly because the larger training data and greater computational capacity of advanced, large-scale LLMs enable them to better identify statistical ground truths based on which they respond to belief-based questions. Investigating these conjectures could provide guidance for the design of future LLMs.

\paragraph{Literature.} Over the past five decades, the cognitive psychology literature (\citealp{ellsberg1961};~\citealp{kahneman1973prediction,kahneman1979prospect};~\citealp{tversky1981framing};~\citealp{rapoport1992generation,rapoport1997randomization};~\citealp*{frederick2002};~\citealp{barberis2003survey}) and the experimental economics literature (\citealp*{lian2018};~\citealp{bose2022};~\citealp{Afrouzi2023}) have systematically documented behavioral biases exhibited by \textit{human} participants. A related strand of research focuses on developing methods to mitigate these human biases  (\citealp{choi2004};~\citealp{thaler:2008};~\citealp{dellavigna2022}). In this paper, we extend these research fields by studying  the behavioral economics of AI---doing so is important for addressing two fundamental questions.

The first question concerns the use of LLMs as research tools for studying human behavior. As GenAI advances, LLMs are increasingly used for social science research: recent studies highlight LLMs' potential to enhance research design, experimentation, data analysis, and agent-based modeling of complex activities (\citealp*{charness2023generation};~\citealp{korinek2023generative};~\citealp{bail2024can}). These studies generally treat LLMs as neutral, unbiased tools. We challenge this assumption by systematically investigating LLM behavior using insights from cognitive psychology and experimental economics. We document LLM biases and their heterogeneity across preference-based and belief-based questions, model versions, and scales. Understanding these biases is critical for evaluating the reliability of LLM-based experiments and simulations. 

The second question concerns the behavior of AI agents in tasks traditionally performed by humans.~These agents are increasingly deployed in economic and financial settings, yet the reliability of their performance remains unclear. Recent work shows mixed patterns:~\cite{chen_etal_2023_emergence} find that GPT-3.5 Turbo exhibits higher economic rationality and lower choice heterogeneity than humans in multiple domains of individual decision making;~\cite{mei_etal_2024_turing} show that GPT-4 exhibits human-like traits in games;~\citet{chen2024does} document biased beliefs in LLM forecasts of stock returns that are commonly observed among humans;~\cite{bowen2024measuring} show that LLMs exhibit strong racial biases in mortgage underwriting, which can be mitigated with prompts that require unbiased decisions;~\cite{cook_et_al_2025} find that, in allocation games, LLM preferences exhibit inequality aversion and are malleable under interventions; and~\cite*{ouyang2024ethical} study how LLM risk preferences in financial settings can be aligned with human ethical standards. Together, these studies suggest that LLM behavior sometimes mirrors human behavior but is sensitive to prompt framing, training data, and model architecture. 

Compared with the studies mentioned above and the broader literature on LLM performance and algorithmic biases, our work is among the first to advocate  a systematic exploration of the behavioral economics of AI, treating GenAI agents as a novel class of economic agents. We advance the literature in several ways. First, we analyze multiple prominent LLM families and examine both cross-sectional and time-series variations within each family.\footnote{Our work is contemporaneous with~\cite{chen_etal_2023_emergence},~\cite{mei_etal_2024_turing},~\cite{ouyang2024ethical},~\cite{chen2024does}, and~\cite{bowen2024measuring}.~Nonetheless, a longer data sample is needed to study time-series variation in LLM responses.} Second, we document biases across both preference-based and belief-based questions, drawing on the cognitive psychology and experimental economics literatures, and covering behavioral biases that are first-order important in financial markets.\footnote{Our approach is consistent with~\cite{binz_schulz_2023} and~\cite{shiffrin2023psychology}: treating an LLM as a participant in a psychology experiment and studying its responses can help understand its mechanisms of reasoning and decision making.} Third, we investigate methods to mitigate these biases, comparing different approaches and propose new ones. Overall, our work lays the foundation for systematically documenting LLM biases and exploring debiasing methods, contributing to the emerging literature on LLM evaluations.\footnote{Recent work by~\cite*{vafa2024large} finds that many LLMs, in particular the highly capable models such as GPT-4, perform poorly on tasks that humans expect them to perform well; this discrepancy highlights the necessity of systematic LLM evaluations. See~\cite{chang2024survey} for a review of LLM evaluations across multiple domains.} 

The rest of the paper proceeds as follows. Section~\ref{sec:exp_design} discusses the experimental design. Section~\ref{sec:results} presents our results on LLM responses to preference-based and belief-based questions. Section~\ref{sec:corrections} explores methods for correcting the observed behavioral biases of LLMs, and Section~\ref{sec:conclusion} concludes. 

\section{Experimental Design}\label{sec:exp_design}

This section describes the experimental design. We first discuss the selection of questions that study either the psychology of preferences or the psychology of beliefs. We then describe the selection of LLMs. Finally, we outline the design of API prompts that allow us to systematically elicit LLM responses to the experimental questions.

\subsection{Selection of Experimental Questions}
\label{subsec:experimental_questions}

Traditional theories in economics and finance posit that economic agents make rational decisions. Here, rationality has two components. The first is rational preferences, namely that agents make decisions according to the Expected Utility framework proposed by~\cite{vonneumann:1944}. The second is rational beliefs, namely that agents incorporate new information into their beliefs according to Bayes' law.

While traditional theories serve as a rational benchmark for economic studies, decades of research from cognitive psychology cast doubt on such theories. Specifically, through carefully designed experimental questions, the psychology literature has documented~\textit{actual} behaviors of human participants that systematically deviate from rational decision making. To illustrate, consider the following question posed to human participants by~\cite{kahneman1979prospect}:

\begin{quote}
	``In addition to whatever you own, you have been given 1,000. You are now asked to choose between A: (1,000, .50), and B: (500)." 
\end{quote}

\noindent{Here, (1,000, .50) means winning \$1,000 with 50\% probability and winning zero with 50\% probability, while (500) means winning \$500 with certainty. For this question, the majority of participants would choose option B. The same set of participants are then asked a separate question:} 

\begin{quote}
	``In addition to whatever you own, you have been given 2,000. You are now asked to choose between C: (--1,000, .50), and D: (--500)." 
\end{quote}

\noindent{Here, (--1,000, .50) means losing \$1,000 with 50\% probability and losing zero with 50\% probability, while (--500) means losing \$500 with certainty. For this question, the majority of participants would choose option C.} 

It is easy to verify that, in terms of monetary payoffs, option A from the first question is equivalent to option C from the second question, and option B from the first question is equivalent to option D from the second question. As a result, the same participant choosing option B from the first question and option C from the second question is a clear violation of the Expected Utility framework. 

Through experimental questions such as the one described above, cognitive psychologists have carefully examined human psychology of preferences---including both risk and time preferences---and human psychology of beliefs, and they have documented a comprehensive set of behavioral biases. In this paper, we ask LLMs to answer the same experimental questions and collect their responses through the prompt design described in Section~\ref{subsec:prompt_design}; in other words, we replace human participants by LLMs. This approach allows us to systematically document the behavioral biases of LLMs and compare their behavior with human behavior. Table~\ref{tab:psych_questions} provides a summary of all the cognitive psychology questions studied in this paper. 

\begin{center}
    [Place Table~\ref{tab:psych_questions} about here]
\end{center}

Two observations are worth noting. First, for each question in Table~\ref{tab:psych_questions}, an LLM response can be classified into one of three categories: a rational response that is derived from rational preferences and rational beliefs; a human-like (irrational) response that corresponds to the response from the majority of human participants; or a non-human-like response, which is neither rational nor human-like. Second, Table~\ref{tab:psych_questions} includes experimental questions that are designed to document prospect theory preferences (Questions 1 to 3), overextrapolation (Questions 7 to 10), and overconfidence (Questions 15 and 16). These three psychological biases, according to~\cite{Barberis:2018}, are the ``big three" biases that are of first-order importance for understanding investor behavior, firm behavior, and asset prices in financial markets. 

Compared to the cognitive psychology literature, the experimental economics literature examines human behavior by designing and conducting experimental tasks that are more closely tied to real-world economic and financial settings. To broaden the scope of our analysis, we also collect LLM responses to a set of tasks from this literature, focusing on two recent works: those of~\cite{Afrouzi2023}, which study investor beliefs, and those of~\cite{bose2022}, which study investor preferences.

We begin with the experiments of~\cite{Afrouzi2023}, in which human participants first observe a sequence of past realizations of a random variable $x_t$ and then forecast its future values; the time-series evolution of this random variable is governed by the following autoregressive process:
\begin{equation}\label{eq:autoregressive_process}
    x_t = \mu + \rho x_{t-1} + \epsilon_{t},
\end{equation}
where $\rho$ measures the persistence of the process and $\epsilon_t$ is an i.i.d. Gaussian shock.

We elicit LLM responses in this setting. As in~\cite{Afrouzi2023}, we consider three experiments. In the baseline experiment, an LLM is endowed with the knowledge that the evolution of $x_t$ follows a ``stable random process." The LLM first observes 40 past realizations of $x_t$, ranging from $x_1$ to $x_{40}$, and is then asked, at time 40, to forecast the next two outcomes, $x_{41}$ and $x_{42}$. Subsequently, it observes the realization of $x_{41}$ and is asked, at time 41, to forecast $x_{42}$ and $x_{43}$. This procedure continues until the LLM observes 44 past realizations of $x_t$ and is asked, at time 44, to forecast $x_{45}$ and $x_{46}$. 

The second and third experiments are variants of the baseline.~In the second experiment, at each time $t$, the LLM is asked to forecast $x_{t+1}$ and $x_{t+5}$; for example, at time 40, it observes $x_1$ to $x_{40}$ and then forecasts $x_{41}$ and $x_{45}$. In the third experiment, the LLM is endowed with more detailed knowledge that the evolution of $x_t$ follows ``a fixed and stationary AR(1) process: $x_{t} = \mu + \rho x_{t-1} + \epsilon_{t}$, with a given $\mu$, a given $\rho$ in the range [0,1], and an $\epsilon_{t}$ that is an i.i.d. random shock." 

For each experiment and a wide range of $\rho$ values, we compare the true $\rho$ with $\hat{\rho}$, the ``perceived" autoregressive coefficient implied by LLM forecasts. This comparison allows us to document biases in LLM beliefs through experiments that mimic real-world forecasting tasks. 

We next turn to the experiment of~\cite{bose2022}, in which human participants first observe a stock price trajectory and then makes an investment decision: assuming an endowment of 1,000 monetary units, they indicate how much to invest in the stock over a 12-month period. We replicate this setup with LLMs by eliciting their responses in the same setting. We then assess whether LLM responses are human-like by examining whether the same preference-based factors that drive investment decisions of human participants also explain the LLMs' decisions. This analysis allows us to understand LLM preferences through experiments that mimic real-world investment tasks.

\subsection{Selection of LLMs}

We select twelve LLMs from four prominent families of Generative Pre-trained Transformers (GPT): ChatGPT, Anthropic Claude, Google Gemini, and Meta Llama. For each family, we select three models: a benchmark model defined as the most recent and best-performing one available at the time of writing, its smaller-scale version, and its predecessor. For ChatGPT, we use GPT-4 as the benchmark, GPT-4o as the smaller-scale version, and GPT-3.5 Turbo as the predecessor. For Anthropic Claude, we use Claude 3 Opus as the benchmark, Claude 3 Haiku as the smaller-scale version, and Claude 2 as the predecessor. For Google Gemini, we use Gemini 1.5 Pro as the benchmark, Gemini 1.5 Flash as the smaller-scale version, and Gemini 1.0 Pro as the predecessor. Finally, for Meta Llama, we use Llama 3 70B as the benchmark, Llama 3 8B as the smaller-scale version, and Llama 2 70B as the predecessor.~Table~\ref{tab:LLM_description} summarizes all the LLMs we examine.

\begin{center}
    [Place Table~\ref{tab:LLM_description} about here]
\end{center}

These twelve models differ both across and within families, particularly along three dimensions: the size of the training data, the design of the model architecture, and the reinforcement learning algorithm. In terms of training data, newer models are generally trained on larger datasets. For example, Meta Llama reports that the training dataset of Llama 3 consists of over 15 trillion tokens, while Llama 2 consists of 1.8 trillion tokens only.\footnote{For more details about the characteristics of Meta Llama 3, see: \href{https://ai.meta.com/blog/meta-llama-3/}{https://ai.meta.com/blog/meta-llama-3/}. For the other three LLM families, the exact training data are not publicly disclosed; nonetheless, newer models in general tend to be trained on more data.~\cite{Brown_gpt3.5} report that OpenAI used around 500 billion tokens to train GPT-3.5. The exact number of tokens used to train GPT-4 is unknown, although unofficial sources suggest around 13 trillion tokens; see \href{https://semianalysis.com/2023/07/10/gpt-4-architecture-infrastructure/}{https://semianalysis.com/2023/07/10/gpt-4-architecture-infrastructure/}.} 

In terms of model architecture, we note three differences across models. First, model specifics such as the context window---the maximum number of words that a model can take as input---and the number of parameters vary widely from one model to another. For example, among older models, Claude 2 has an estimate of 200 billion parameters and a context window of approximately 100,000 tokens, whereas Llama 2 has 70 billion parameters and a context window of approximately 4,000 tokens. Second, within each family, model architectures have evolved significantly across the two generations that we consider. In particular, for ChatGPT, Anthropic Claude, and Google Gemini, the most significant evolution is the transition from a single-transformer architecture to a multi-transformer mixture-of-experts architecture.\footnote{Mixture-of-experts architectures use a ``router," or gating network, to activate specific experts for each input token (\citealp{shazeer2017}). The sparsity that arises from activating only a fraction of parameters for each input enables better scaling. For example, GPT-3.5 Turbo uses a single-expert architecture with 175 billion parameters, while  unofficial sources suggest that GPT-4 uses a mixture-of-experts architecture that consists of multiple transformers with approximately 110 billion parameters each for a total of over 1 trillion parameters.} Third, within each family and each generation, model architecture may differ between the benchmark model and its smaller-scale version. The smaller versions are often obtained by applying compression techniques to the benchmark model; for example, Gemini 1.5 Flash is a distilled version of Gemini 1.5 Pro.\footnote{Common compression techniques include quantization, which reduces parameter precision, and pruning, which removes less important connections from the neural network.}\footnote{In contrast, Llama 3 8B and Llama 3 70B share similar architectures and training data but differ in scale (8 billion versus 70 billion parameters).}  

In terms of reinforcement learning algorithms, each model relies on a different implementation of the Reinforcement Learning from Human Feedback (RLHF) algorithm to align outputs with human preferences. For example, Anthropic Claude combines RLHF with Constitutional AI, a method designed to align model behavior with human principles of helpfulness, harmlessness, and honesty (\citealp{bai2022}). 


\subsection{Prompt Design}
\label{subsec:prompt_design}

We collect LLM responses to each experimental question through an application programming interface (API). The API takes as input a ``prompt," which is a text file submitted to an LLM to elicit a response. Below, we describe the prompt design that allows for elicitation of desired responses from LLMs.

A proper prompt needs to satisfy two requirements. First, it should instruct the LLM to provide standardized responses suitable for subsequent analysis. Second, it should phrase questions in a manner that is comparable to the original experimental questions used to study human behavior. Given these requirements, Fig.~\ref{prompt: ds_pt} provides an example: the prompt we use to elicit LLM responses to a question that~\cite{kahneman1979prospect} designed to document diminishing sensitivity as a key element of prospect theory.

\begin{center}
    [Place Fig.~\ref{prompt: ds_pt} about here]
\end{center}

Fig.~\ref{prompt: ds_pt} shows that the prompt is structured in three parts; this structure applies to all experimental questions listed in Table~\ref{tab:psych_questions}. The first part contains general instructions that ask the LLM to consider a specific set of experimental scenarios; in Fig.~\ref{prompt: ds_pt}, this part begins with ``Instructions" and ends with ``completely separate from the other." two lines below. The second part contains a code block that instructs the LLM to format its responses in a standardized JSON format; in Fig.~\ref{prompt: ds_pt}, this part begins with ``The output should be" and ends with ``\} \texttt{```}".\footnote{This part of the prompt requires formatting LLM responses as a snippet that contains a JSON object within a code block. JSON is a widely used format that stores data as key-value pairs. Encapsulating the JSON object within a code block helps ensure that LLM responses adhere to the pre-specified format.} The third part is the main element of the prompt. It contains the precise experimental questions originally designed by psychologists to study human behavior; in Fig.~\ref{prompt: ds_pt}, this part begins with ``Scenario A:" and ends with ``calculations)." from Scenario B. At the end of each scenario, additional instructions are given to the LLM, to ensure that it provides the specific set of responses we elicit.

Two observations are worth noting. First, under our prompt design, an LLM response typically contains, for each experimental scenario, four components: choice, confidence, explanation, and reasoning. Here, ``choice" refers to the explicit decision made by the LLM---for example, whether it accepts or turns down a risky gamble.\footnote{Instead of eliciting a ``choice" among multiple options, Question 10 (regarding ``base rate neglect") and Question 12 (regarding ``gambler's fallacy") ask for an estimate of a probability, while Question 14 (regarding ``anchoring") asks for an estimate of a percentage number.} ``Confidence" refers to the  confidence level the LLM assigns to its choice, measured on a score between 0 and 1. ``Explanation" refers to a brief explanation that the LLM provides to justify its choice. And ``reasoning" requires the LLM to select between two reasoning types: type ``A," which corresponds to reasoning that is based more on intuitive thinking, and type ``B," which corresponds to reasoning that is based more on analytical thinking and calculations.


Second, many of the experimental questions we examine document behavioral biases by eliciting responses from the \textit{same} participant across different scenarios. For example, as discussed in Section~\ref{subsec:experimental_questions}, \cite{kahneman1979prospect} document diminishing sensitivity by having the same human participant answer two different questions---one that frames lottery payoffs as gains and another that frames them as losses. Such experimental questions require a within-subject design, which allows us to treat an LLM as a participant and elicits its responses across different scenarios. To implement this design, we combine multiple questions into a single API call; we treat each API call as an individual participant; and we include in the prompt an instruction that asks the LLM to ``treat each scenario as completely separate from the other."\footnote{LLM responses can be stochastic---identical questions posed multiple times to the same model can yield different responses. We therefore view each API call as an individual participant.} The Internet Appendix provides the prompt designs for all sixteen questions listed in Table~\ref{tab:psych_questions}.

The above discussion is concerned with the prompt design that implements experimental questions from the cognitive psychology literature. We conclude this section by making three observations about two separate prompts, each corresponding to one of the experiments: the~\cite{Afrouzi2023} experiments and the~\cite{bose2022} experiment described in Section~\ref{subsec:experimental_questions}. First, these experiments require not only textual inputs but also \textit{graphical} inputs: participants are presented with textual instructions and figures that plot either past realizations of a random variable or stock price trajectories. To satisfy this requirement, an LLM must support graphical inputs; this leads to the exclusion of six LLM platforms.\footnote{All three Meta Llama models---Llama 3 70B, Llama 3 8B, and Llama 2 70B---as well as GPT-3.5 Turbo, Claude 2, and Gemini 1.0 Pro do not support graphical inputs.} For the remaining LLMs, we follow platform-specific guidelines when uploading figures.\footnote{Google Gemini directly processes figures uploaded as .jpg files.~ChatGPT and Anthropic Claude, however, require encoding a binary image into bytes, which are then converted into a regular UTF-8 string format.} Second, LLMs do not always provide precise forecasts or investment decisions when presented with figures; sometimes, they refuse to respond. To address this issue, for both the~\cite{Afrouzi2023} and~\cite{bose2022} experiments, we include the following sentences in the instruction: ``For the following question, please provide an estimate to the best of your knowledge. Please ensure that you always provide a concrete numerical answer when prompted to do so." Third, the~\cite{Afrouzi2023} experiments require that the same individual makes multiple rounds of forecasts, with each round depending on the textual and graphical inputs presented up to that point in time. To enforce this sequential dependence, we implement a sequence of API calls. In particular, for each call, we feed the entire conversation history---including all previous prompts and responses---into the LLM, thereby preserving the structure of the original experiments.

\section{Behavioral Biases of LLMs} 
\label{sec:results}

We now document patterns in LLM responses to the experimental questions drawn from the cognitive psychology literature and the experimental economics studies. We begin with a baseline analysis of the four highly advanced large-scale LLMs, which we treat as our benchmark models.  We analyze how they respond to the cognitive psychology questions, with a focus on whether these models are more likely to produce rational or human-like responses. A central feature of this analysis is to draw distinction between LLM responses to preference-based questions and their responses to belief-based questions. We then explore heterogeneity in LLM responses across LLM families, model generations, and model scales. Finally, we examine LLM responses to questions from the experimental economics tasks, which are more closely tied to real-world economic and financial decision making.

\subsection{Baseline Results}
\label{subsec:baseline}

This section presents our baseline results. We first describe the procedure for data collection. We then analyze the responses of the four benchmark models---GPT-4, Claude 3 Opus, Gemini 1.5 Pro, and Llama 3 70B---to the sixteen experimental questions drawn from the psychology literature, as listed in Table~\ref{tab:psych_questions}.

For each question and each model, we collect 100 responses; that is, for each LLM, we iterate over each question 100 times. Each iteration consists of an API call submitted to the model, with the prompt for the specific question provided as input along with a key temperature parameter. This parameter controls the randomness of the model. For our baseline analysis, we set the temperature parameter to 0.5, the recommended value for most LLM families.\footnote{The range of the temperature parameter varies across platforms: for ChatGPT and Anthropic Claude, it is [0, 1]; for Meta Llama, it is [0, 5]; for Gemini 1.0 Pro, it is [0, 1]; and for Gemini 1.5 Pro and Gemini 1.5 Flash, it is [0, 2].} Note that setting the temperature parameter to zero results in deterministic outputs, while higher values increase the randomness of LLM responses.\footnote{Specifically, during the iterative process of generating each word (token) in a response, the LLM first forms a probability distribution over all possible tokens in its dictionary and then draws the next token from this distribution. The temperature parameter reshapes the distribution: higher values make it more uniform, therefore increasing the randomness of the output token. Two other parameters, $k$ and $p$, also affect the selection of the output token: top-$k$ sampling restricts the selection to the top-$k$ most probable tokens, while top-$p$ sampling retains a subset of the top-$k$ tokens whose cumulative probability, when normalized by the total probability of the top-$k$ tokens, exceeds the threshold $p$. In our analysis, we set $k$ to its default value of 50 and $p$ to its default value of 0.9.}

We collect and analyze each LLM response, categorizing it into one of three groups: rational, human-like, or other. A response is categorized as rational if the LLM’s choice or estimate aligns with that of an agent who has rational preferences and rational beliefs; it is categorized as human-like if it is irrational but aligns with the most common behavior observed in human participants from prior psychology research; and it falls into the ``other" category if it is neither rational nor human-like. Take the diminishing sensitivity question from Fig.~\ref{prompt: ds_pt} as an example. A rational response, according to the Expected Utility framework, is to choose option B, the option that indicates risk aversion, in both Scenario A and Scenario B. \cite{kahneman1979prospect} show that the majority of human participants choose option B in Scenario A and option A in Scenario B; if an LLM makes the same choices, the response is categorized as human-like. If, however, the LLM selects option A in Scenario A and option B in Scenario B, or selects option A in both scenarios, the response is categorized as ``other."

\begin{center}
    [Place Fig.~\ref{fig_large_advanced} and Table~\ref{tab:summary_advanced} about here]
\end{center}

Fig.~\ref{fig_large_advanced} summarizes responses from the four benchmark models of GPT-4, Claude 3 Opus, Gemini 1.5 Pro, and Llama 3 70B. For each model, the sixteen experimental questions from cognitive psychology are divided into two groups: preference-based questions (left panel) and belief-based questions (right panel). The results are presented using bar charts that show the proportion of responses categorized as rational (blue), human-like (red), or other (gray).\footnote{Fig.~\ref{fig_small_advanced} and Fig.~\ref{fig_older} in the Internet Appendix present the proportion of rational, human-like, or other responses for advanced small-scale models and older models.} Table~\ref{tab:summary_advanced} presents the same results in tabular form and includes a binomial test for each question, where the null hypothesis states that the proportion of rational (or human-like) responses is less than or equal to 50\%.

Two observations are worth emphasizing. First, the majority of LLM responses fall into either the rational category or the human-like category, with the responses classified as ``other" in just a few cases. For GPT-4, such ``other” responses arise only for Question 3, which pertains to the probability weighting element of prospect theory. For Claude 3 Opus, ``other” responses appear in two preference-based questions---Question 3 on probability weighting and Question 4 on narrow framing---and two belief-based questions---Question 10 on base rate neglect and Question 15 on overprecision. For Gemini 1.5 Pro, ``other" responses occur only in Question 3. Finally, for Llama 3 70B, ``other” responses are observed in one preference-based question---Question 3 on probability weighting---and one belief-based question---Question 7 on sample size neglect.

Second, a comparison between the left and right panels of Fig.~\ref{fig_large_advanced} reveals a clear pattern: LLM responses to preference-based questions tend to be more human-like, whereas their responses to belief-based questions tend to be more rational. Table~\ref{tab:summary_advanced} confirms this result. For a large fraction of preference-based questions, a binomial test confirms, with a confidence level greater than 99\%, that LLMs produce human-like responses more than 50\% of the time. Specifically, Gemini 1.5 Pro has the majority of its responses categorized as human-like in five out of six questions; Claude 3 Opus has the majority of its responses categorized as human-like in four out of six questions; and GPT-4 and Llama 3 70B have the majority of their responses categorized as human-like in three out of six questions. By contrast, for most belief-based questions, LLMs produce rational responses more than 50\% of the time. Specifically, Gemini 1.5 Pro has the majority of its responses categorized as rational in all ten questions; both GPT-4 and Claude 3 Opus have the majority of their responses categorized as rational in eight out of ten questions; and Llama 3 70B has the majority of its responses categorized as rational in five out of ten questions.

\subsection{Heterogeneity in LLM Responses}
\label{subsec:model_heterogeneity}

While Section~\ref{subsec:baseline} documents systematic patterns in LLM responses for the four benchmark models, we now broaden the analysis to include all twelve models. Within each LLM family, we consider three types of models: a highly advanced, large-scale benchmark model; a highly advanced model with a smaller scale; and a large-scale model from an older generation. We begin by examining variations in responses across the four LLM families. We then, controlling for LLM family fixed effects, analyze how model generation and model scale influence patterns in LLM responses.~As in Section~\ref{subsec:baseline}, we conduct separate analyses for the six preference-based questions and the ten belief-based questions.

\subsubsection{Heterogeneity across LLM families}

We first examine variations in LLM response across the four LLM families. Fig.~\ref{fig_large_advanced} provides preliminary graphical evidence of differences among the four benchmark models. For preference-based questions, Gemini 1.5 Pro, relative to GPT-4, produces a lower share of rational responses and a higher share of human-like responses. For belief-based questions, Llama 3 70B, relative to GPT-4, produces a lower share of rational responses and a higher share of human-like responses.

To formally assess heterogeneity in responses across the LLM families, we estimate a series of probit regressions using all twelve LLMs. The regression specification is:
\begin{equation}\label{eq:probit_across_families}
\Pr(Y_{iqk}=1) = \Phi(\alpha + \beta_1 \cdot Claude_i + \beta_2 \cdot Gemini_i + \beta_3 \cdot Llama_i + \epsilon_{iqk})
\end{equation}
for model $i$, question $q$, and iteration $k$, where $\Phi(\cdot)$ denotes the cumulative distribution function of a standard Normal random variable. When studying how variation in LLM families affects the likelihood of observing a rational response, $Y_{iqk}$, the dependent variable in~\eqref{eq:probit_across_families}, is a binary variable that takes the value of one if model $i$'s response to question $q$ in iteration $k$ is classified as rational, and zero otherwise. When studying how variation in LLM families affects the likelihood of observing a human-like response, $Y_{iqk}$ is a binary variable that takes the value of one if model $i$'s response to question $q$ in iteration $k$ is classified as human-like, and zero otherwise. For both cases, the independent variables---$Claude_i$, $Gemini_i$, and $Llama_i$---are indicators for the respective LLM families, with the LLM family of GPT serving as the omitted baseline. 

\begin{center}
    [Place Table~\ref{tab:across_family} about here]
\end{center}

 Table~\ref{tab:across_family} reports marginal effects from these regressions, where each reported coefficient represents the change in the predicted probability of observing $Y_{iqk} = 1$ that is associated with changing the LLM from GPT to Claude, Gemini, or Llama. Consistent with the heterogeneity observed from Fig.~\ref{fig_large_advanced}, for preference-based questions, Gemini models are 22.9\% less likely to produce a rational response than GPT models; this effect is significant at the 1\% level. At the same time, Gemini models are 16.7\% more likely to produce a human-like response than GPT models; this effect is significant at the 5\% level. Moreover, responses from Claude or Llama models to the preference-based questions are statistically similar to those from GPT models.

 For belief-based questions, Llama models are 25.0\% less likely to produce a rational response than GPT models; this effect is significant at the 5\% level. Llama models are 21.0\% more likely to produce a human-like response than GPT models; this effect is also significant at the 5\% level. Finally, responses from Claude or Gemini models to the belief-based questions are statistically similar to those from GPT models. Overall, these findings highlight meaningful LLM family-level differences in responses to experimental questions drawn from cognitive psychology. In subsequent analyses of heterogeneity across model generations and scales, we control for LLM family fixed effects.

\subsubsection{Heterogeneity across model generations and model scales}

We next examine variations in LLM responses across model generations and model scales. Changes in model generation and  scale capture key aspects of LLM development, including improvements in model architectures and advancements of reinforcement learning algorithms. To study the effect of model generation on LLM responses, we compare advanced models with older models of a similar scale. To study the effect of model scale, we compare large-scale models with smaller-scale ones of the same generation. In both analyses, we control for LLM family fixed effects.

\begin{center}
    [Place Fig.~\ref{fig_heterogeneity} about here]
\end{center}

We begin by presenting graphical evidence on differences in LLM responses across model generations and scales. Fig.~\ref{fig_heterogeneity} displays radar charts that summarize the number of preference-based questions and the number of belief-based questions for which each model produces predominantly rational or human-like responses. These visualizations offer a compact view of cross-model variation.~For example, Claude 3 Haiku produces predominantly rational responses for three out of six preference-based questions, while Claude 3 Opus does not produce predominantly rational responses for any preference-based question.

The radar charts reveal a striking contrast between LLM responses to preference-based questions and their responses to belief-based questions. For preference-based questions, the left panel shows that, as LLMs become more advanced or larger, the number of questions with predominantly rational responses tends to decrease, while the number of questions with predominantly human-like responses increases. For belief-based questions, the right panel shows the \textit{opposite} pattern: more advanced and larger-scale models tend to generate predominantly rational responses for a large number of questions.

To formally examine heterogeneity in LLM responses across model generations and scales, we estimate a series of probit regressions. In particular, we conduct two analyses. First, to study the effect of model generation on LLM responses, we restrict the sample to responses from either the four advanced large-scale models or the four older models. The regression specification is: 
\begin{equation}\label{eq:probit_across_generation}
\Pr(Y_{iqk}=1)=\Phi(\alpha + \beta \cdot Advanced_i + \gamma_f + \epsilon_{iqk})
\end{equation}
for model $i$, question $q$, and iteration $k$. When studying the effect of a change in model generation on the likelihood of observing a rational response, $Y_{iqk}$ is a binary variable that takes the value of one if model $i$'s response to question $q$ in iteration $k$ is classified as rational,
and zero otherwise. When studying the effect of a change in model generation on the likelihood of observing a human-like response, $Y_{iqk}$ is a binary variable that takes the value of one if model $i$'s response to question $q$ in iteration $k$ is classified as human-like,
and zero otherwise. For both cases, the key independent variable, $Advanced_i$, is an indicator for the advanced models. Moreover, $\gamma_f$ captures LLM family fixed effects. 

Second, to study the effect of model scale, we restrict the sample to responses from either the four advanced large-scale models or the four advanced smaller-scale models. The regression specification is: 
\begin{equation}\label{eq:probit_across_scale}
\Pr(Y_{iqk}=1)=\Phi(\alpha + \beta \cdot LargeScale_i + \gamma_f + \epsilon_{iqk}).
\end{equation}
When studying the effect of a change in model scale on the likelihood of observing a rational response, $Y_{iqk}$ is a binary variable that takes the value of one if model $i$'s response to question $q$ in iteration $k$ is classified as rational,
and zero otherwise. When studying the effect of a change in model scale on the likelihood of observing a human-like response, $Y_{iqk}$ is a binary variable that takes the value of one if model $i$'s response to question $q$ in iteration $k$ is classified as human-like,
and zero otherwise. The key independent variable, $LargeScale_i$, is an indicator for the large-scale models.

\begin{center}
    [Place Table~\ref{tab:heterogneity} about here]
\end{center}

 Table~\ref{tab:heterogneity} reports marginal effects from these regressions, where the reported coefficients represent the change in the predicted probability of observing $Y_{iqk} = 1$ that is associated with either moving from an older to an advanced model or from a smaller-scale to a large-scale model. The regression results are consistent with the variation in LLM responses observed in Fig.~\ref{fig_heterogeneity} across model generations and scales. For preference-based questions, Columns (1) to (4) in Panel A show that, as models become more advanced, their responses are less likely to be categorized as rational and more likely to be categorized as human-like; Columns (1) to (4) in Panel B shows that, as models become larger in scale, the same patterns occur---the models' responses are less likely to be rational and more likely to be human-like. For both Panels A and B, most coefficients reported in Columns (1) to (4) are statistically significant; the only exception is that, as models become larger, the increase in human-like responses to preference-based questions is insignificant.
 
 For belief-based questions, Columns (5) to (8) in Panel A show that more advanced models generate responses that are more likely to be categorized as rational and less likely to be categorized as human-like; Columns (5) to (8) in Panel B shows the same patterns as models become larger in scale. For both Panels A and B, all coefficients reported in Columns (5) to (8) are statistically significant.
 
In summary, Fig.~\ref{fig_heterogeneity} and  Table~\ref{tab:heterogneity} show systematic heterogeneity in LLM responses across model generations and scales. As LLMs become more advanced or larger, their responses to preference-based questions become increasingly human-like, while their responses to belief-based questions become more rational. These opposing patterns highlight the importance of separately studying preferences and beliefs when evaluating LLM behavior.


\subsection{LLM Responses to Questions from the~\cite{Afrouzi2023} Experiments}
\label{subsec:results_exp_econ}

We now examine LLM responses to questions from the three~\cite{Afrouzi2023} experiments described in Section~\ref{subsec:experimental_questions}; we label these experiments as ``Experiment 1," ``Experiment 2," and ``Experiment 3." For each experiment, we simulate the autoregressive process: 
\begin{equation*}
    x_{t} = \mu + \rho x_{t-1} + \epsilon_{t}
\end{equation*}
specified in~\eqref{eq:autoregressive_process}, by setting $\mu$, the constant term, to 0 and setting $\sigma$, the standard deviation of $\epsilon_{t}$, to 20; these parameter values are adopted from~\cite{Afrouzi2023}. For $\rho$, the persistence parameter, we examine six values: 0, 0.2, 0.4, 0.6, 0.8, and 1. For each value, we generate 100 simulated paths, yielding a total of 600 paths per experiment. 

For a given experiment and a given simulated path, we ask LLMs to make five rounds of forecasts. Take Experiment 1 as an example. In the first round, we present each LLM with a figure that displays the first 40 realizations of $x_{t}$ from this simulated path, ranging from $x_1$ to $x_{40}$. Then, a prompt requests the LLM to provide its forecasts for the next two outcomes, $x_{41}$ and $x_{42}$. The model’s response is recorded to establish the beginning of a conversation history. In the second round, we first update the conversation history by adding the previous figure, prompt, and LLM response to it. We then present a new figure that extends the observed sequence to $x_{41}$ and prompt the LLM to forecast the next two outcomes, $x_{42}$ and $x_{43}$. We record the LLM's response and add it to the conversation history. This iterative process continues until five rounds of forecasts are completed. 

To evaluate the extent to which the LLM's forecasts are biased, we estimate $\hat{\rho}$, the ``perceived" autoregressive coefficient implied by these forecasts. Specifically, for each LLM $i$, each value of $\rho$, and each forecasting horizon $s$ of 1, 2, or 5, we estimate the perceived persistence $\hat{\rho}$ using the following regression: 
\begin{equation}
    F_{it}x_{t+s}=c_{is}+(\hat{\rho}_{is})^s x_t+u_{is,t},
\end{equation}
where $F_{it} x_{t+s}$ represents model $i$’s time-$t$ forecast of $x_{t+s}$, $x_t$ is the time-$t$ realization of $x$, and $u_{is,t}$ is an error term. 
 
Fig.~\ref{fig_afrouzi} presents estimates based on LLM responses to questions from Experiment 1 of~\cite{Afrouzi2023}; here, we focus on short-term forecasts, with the forecasting horizon $s$ set to one. The top panel displays the $\hat{\rho}$ values estimated for the three baseline models: GPT-4, Claude 3 Opus, and Gemini 1.5 Pro. The bottom panel displays the $\hat{\rho}$ values estimated for the three smaller-scale models: GPT-4o, Claude 3 Haiku, and Gemini 1.5 Flash.\footnote{Llama models do not support graphical inputs and are excluded from this analysis.} For each $\hat{\rho}$ estimate, we also plot the $95\%$ confidence interval. The results above are compared with a 45-degree line, which represents the persistence implied by full information rational expectations (FIRE).

\begin{center}
    [Place Fig.~\ref{fig_afrouzi} about here]
\end{center}

Two observations emerge from Fig.~\ref{fig_afrouzi}. First, for the advanced large-scale models (top panel), LLM forecasts are largely rational: for each model and each $\rho$, the estimated $\hat{\rho}$ is close to the true $\rho$. Second, for the smaller-scale models (bottom panel), LLM forecasts are human-like: consistent with the findings of~\cite{Afrouzi2023} for human participants, the persistence $\hat{\rho}$ implied by LLM forecasts is significantly higher than the true persistence $\rho$, and the difference between $\hat{\rho}$ and $\rho$ is larger for lower values of $\rho$. The comparison between the top and bottom panels reinforces a pattern documented in Section~\ref{subsec:baseline} using cognitive psychology questions, namely that larger-scale models tend to generate more rational responses to belief-based questions.

We also examine LLM forecasts from Experiments 2 and 3.  Fig.~\ref{fig_afrouzi_variant} plots $\hat{\rho}$ against $\rho$, with the top panel examining longer-term forecasts $F_{it} x_{t+5}$ from Experiment 2 and the bottom panel examining short-term forecasts $F_{it} x_{t+1}$ from Experiment 3, in which LLMs are given more detailed knowledge about the evolution of $x_t$. 

\begin{center}
    [Place Fig.~\ref{fig_afrouzi_variant} about here]
\end{center}

The comparison between Fig.~\ref{fig_afrouzi} and Fig.~\ref{fig_afrouzi_variant} yields two observations. First, for the three advanced large-scale models of GPT-4, Claude 3 Opus, and Gemini 1.5 Pro, LLMs' longer-term forecasts produce human-like biases that are absent in short-term forecasts: the top panel of Fig.~\ref{fig_afrouzi_variant} shows that, the persistence $\hat{\rho}$ implied by longer-term forecasts is significantly higher than the true persistence $\rho$, and the difference between $\hat{\rho}$ and $\rho$ is larger for lower values of $\rho$. In comparison, as discussed before, the top panel of Fig.~\ref{fig_afrouzi} shows that short-term forecasts are largely rational. Notice from~\cite{Afrouzi2023} that human participants' longer-term forecasts are also more biased than their short-term forecasts. 

Second, the comparison between the bottom panel of Fig.~\ref{fig_afrouzi_variant} and the top panel of Fig.~\ref{fig_afrouzi} shows that providing detailed information about the data generating process can be counterproductive: it gives rise to more human-like biases in LLM responses. Interestingly, this effect is unique to LLMs;~\cite{Afrouzi2023} find that human forecasts are unaffected by the provision of additional information about the evolution of $x_t$. Section~\ref{sec:corrections} provides more discussion of this finding.

\subsection{LLM Responses to Questions from the~\cite{bose2022} Experiment}
\label{subsec:results_exp_econ1}

We next turn to \cite{bose2022}, who design an experiment in which human participants observe stock price trajectories and then make investment decisions. Specifically, the experiment generates price charts for 1,000 stocks randomly selected from the Center for Research in Security Prices (CRSP) database, using the stocks' daily returns from 2017. These price charts vary along two dimensions: the stock's cumulative annual return and its convexity score, which measures the curvature of the stock price trajectory. Each price chart is viewed by four human participants. Each participant is given an endowment of 1,000 monetary units and, after seeing a price trajectory, is asked: ``How much of a 1,000 monetary unit endowment are you willing to invest in the stock for the following 12 months? The remainder will be put in a safe bank account."

\cite{bose2022} construct two novel variables. The first is the ``visual salience" of a stock price trajectory, denoted by ``VS" and defined as a weighted average of the stock's past daily returns; the weights are computed using a machine learning algorithm that captures the visual attention paid to each return. The second is a related measure, $\mathbb{C}\text{orr}$(Returns, VS weights), defined as the correlation between past daily returns and their associated visual salience weights. The authors interpret both variables as reflecting non-traditional preferences---preferences that violate the Expected Utility framework---and show that these variables strongly predict participants' investment amounts, even after controlling for variables derived from alternative frameworks, including cumulative prospect theory (CPT;~\citealp{tversky1992prospect}) and salience theory (\citealp{bordalo2012salience,bordalo2013salience}).

We replicate this setup with LLMs by presenting them with stock price trajectories and eliciting their investment amounts. Specifically, we follow the same procedure to select 1,000 stocks from CRSP and use their daily returns from 2017 to construct price charts; we apply the same machine learning algorithm to compute the two key variables, ``VS" and $\mathbb{C}\text{orr}$(Returns, VS weights); and we construct the same control variables including those related to CPT (\citealp{tversky1992prospect}) and salience theory (\citealp{bordalo2012salience,bordalo2013salience}). Finally, as in the~\cite{Afrouzi2023} experiments, we examine six of the twelve LLMs that support graphical input. Each LLM observes each of the 1,000 stock price charts four times, and in each instance, provides an investment amount, measured as the fraction of the endowed 1,000 monetary units that it is willing to invest in the stock.

\begin{center}
    [Place Table~\ref{tab:bose_investment_amount} about here]
\end{center}

Table~\ref{tab:bose_investment_amount} reports linear regression results that relate elicited investment amounts to a broad set of explantory variables for the advanced smaller-scale and advanced large-scale models of GPT, Claude, and Gemini. The key variables of interest are $\text{VS}\times \mathbbm{1}_{large}$ in Columns (1), (3), and (5) and $\mathbb{C}\text{orr}$(Returns, VS weights)$\times\mathbbm{1}_{large}$ in Columns (2), (4), and (6). The coefficients on these interaction terms are positive and mostly statistically significant.\footnote{For GPT, the coefficient on $\text{VS}\times \mathbbm{1}_{large}$ is positive but not statistically significant.} These results indicate that, as an LLM increases in scale, its investment decisions depend more strongly on non-traditional preferences documented in~\cite{bose2022} for human behavior, reinforcing the pattern identified in Section~\ref{subsec:baseline} using cognitive psychology questions.\footnote{The coefficients on VS and $\mathbb{C}\text{orr}$(Returns, VS weights) are negative. This pattern is consistent with smaller-scale models behaving rationally: in an equilibrium in which behavioral investors overweight stocks with high VS and $\mathbb{C}\text{orr}$(Returns, VS weights), these stocks become overvalued; as such, rational investors' allocations should load negatively on these variables.}

\section{Correcting LLM Biases} \label{sec:corrections}

Section~\ref{sec:results} documents that, for many preference-based and belief-based questions, LLM responses exhibit behavioral biases. In this section, we explore role-priming methods---instructing an LLM to view itself as a particular type of individual---as well as debiasing techniques that aim to correct these biases. 

We begin by examining how role priming affects LLM responses. We consider two versions. The first instructs the LLMs to view themselves as rational investors; the second instructs them to view themselves as real-world retail investors. We implement each version by adding a single sentence at the beginning of the prompt. For the first version, the sentence is ``When answering questions below, please think of yourself as a rational investor who makes decisions using the `expected utility' framework." For the second version, the sentence is ``When answering questions below, please think of yourself as a real-world retail investor who makes economic and financial decisions." 

\begin{center}
    [Place Table~\ref{table:role_priming} about here]
\end{center}

Table~\ref{table:role_priming} presents the effects of role priming on LLM behavior. Panel A reports the treatment effects of priming the LLMs to be rational investors. Averaged across the twelve LLMs, this role priming increases rational responses by 4.3\% for preference-based questions (significant at the 5\% level) and by 3.3\% for belief-based questions (significant at the 10\% level).\footnote{We report treatment effects from regression specifications that include model fixed effects to account for differences across the twelve LLMs. The results are broadly similar without these controls.} Panel B reports the treatment effects of priming the LLMs to be real-world retail investors. Averaged across the twelve LLMs, this role priming reduces rational responses by 3.9\% for preference-based questions (significant at the 5\% level) and has no statistically significant effect on responses to belief-based questions. Overall, Table~\ref{table:role_priming} suggests that instructing LLMs to act as rational investors can reduce biases. 

To better understand the mechanisms through which role priming affects LLM responses, we conduct a mediation analysis that examines whether role priming operates through intervening variables, referred to as mediators.\footnote{A classic example of mediation analysis is provided by~\cite{cohn2012fear}.} Specifically, we address three questions. First, we examine whether role priming affects the confidence levels that LLMs assign to their choices and their self-reported reasoning type, where type A corresponds to intuitive thinking and type B corresponds to analytical thinking and calculations. Second, we assess whether LLM confidence and reasoning type influence their responses. Third, we test whether including LLM confidence and reasoning type as control variables in  regressions of LLM responses on role priming attenuates the estimated effect of role priming. 

\begin{center}
    [Place Table~\ref{table:mediation_role_priming} about here]
\end{center}

Table~\ref{table:mediation_role_priming} presents the results. LLM confidence is measured by ``high confidence," an indicator that equals one if an LLM assigns a confidence level greater than 0.9 to its choice, and reasoning type is measured by ``system 2 thinking," an indicator that equals one if the LLM selects reasoning type B.\footnote{Our analysis includes all twelve LLMs but focuses on preference-based questions only, as Table~\ref{table:role_priming} shows that the effects of role priming on LLMs' responses to belief-based questions are often insignificant.} Panel A restricts the sample to LLM responses generated using either the baseline prompt or a treatment prompt that primes LLMs to be rational investors. Panel B restricts the sample to LLM responses generated using either the baseline prompt or a treatment prompt that primes LLMs to be real-world retail investors.

Columns (1) and (2) show that role priming significantly affects LLM confidence and reasoning type. The rational-investor role-priming prompt increases ``high confidence" by 6.6\% and ``system 2 thinking" by 11.4\%; both effects are significant at the 1\% level. In contrast, the retail-investor role-priming prompt deceases ``high confidence" by 6.0\% and ``system 2 thinking" by 8.0\%, with both effects significant at the 5\% level. Columns (4) and (7) show that ``high confidence" and ``system 2 thinking" significantly influence LLM responses. For example, ``system 2 thinking" increases rational responses by 40\% in Panel A and by 38\% in Panel B, while ``high confidence" decreases human-like responses by 35\% in Panel A and by 58\% in Panel B. Finally, comparisons between Columns (3) and (5)---and between Columns (6) and (8)---show that including LLM confidence and reasoning type as control variables in regressions of LLM responses on role priming significantly attenuates the estimated effect of role priming. Taken together, Table~\ref{table:mediation_role_priming} provides evidence that role priming affects LLM responses through confidence and reasoning type.

Table~\ref{table:debiasing} explores two additional debiasing techniques. The first technique combines the sentence that primes LLMs to be rational investors with a detailed four-step procedure that guides the LLMs to choose a course of action rationally under the Expected Utility framework:

\begin{quote}
``Please be reminded of the procedure of choosing a course of action under the `expected utility' framework. For each course of action: \\
(1) You list all possible wealth outcomes it could result; here, a wealth outcome accounts for existing wealth and any potential changes in wealth. \\
(2) You compute the utility of each wealth outcome, using a globally concave utility function; note that the utility function focuses on total wealth outcomes rather than gains or losses alone. \\
(3) You weigh the utility of each outcome by the probability of the outcome. \\
(4) You sum up across outcomes to obtain the expected utility of the course of action.

\noindent{You repeat the four-step procedure above for each possible course of action and choose the course of action with the highest expected utility. When answering questions below, please provide the concrete steps you take for computing the expected utility of each course of action.}" 
\end{quote}

The second technique combines the sentence that primes LLMs to be rational investors with a summary of the key findings from~\cite{kahneman1979prospect} that describe systematic human biases. The summary is generated by uploading the .pdf form of the original~\cite{kahneman1979prospect} paper to an interactive GPT-4o chat box and then requesting a summary of the paper's key insights. The specific summary is given by:

\begin{quote}
``Please be reminded of prospect theory, a framework that describes human decision-making. The main takeaway from Prospect Theory: An Analysis of Decision under Risk by Daniel Kahneman and Amos Tversky (1979) is that human decision-making under risk systematically deviates from the predictions of traditional expected utility theory. Instead of evaluating choices purely in terms of final wealth states, individuals evaluate gains and losses relative to a reference point.

\noindent{Key Insights:}\\
\noindent{Certainty Effect -- People overweight certain outcomes relative to probable ones, leading to risk aversion in gains and risk-seeking behavior in losses.}\\
\noindent{Loss Aversion -- Losses loom larger than equivalent gains, meaning the psychological impact of losing \$100 is greater than the pleasure of gaining \$100.}\\
\noindent{Diminishing Sensitivity -- The value function is concave for gains and convex for losses, meaning the impact of an additional dollar diminishes as amounts increase.}\\
\noindent{Decision Weights vs. Probabilities -- People do not evaluate probabilities linearly; they tend to overweight small probabilities (making lotteries attractive) and underweight moderate to high probabilities (explaining why they buy insurance).}\\
\noindent{Isolation Effect -- Decision-making is influenced by how choices are framed, leading to inconsistent preferences when identical problems are presented in different ways.
This theory revolutionized behavioral economics by demonstrating that individuals do not always make rational choices based on maximizing expected utility but rather follow heuristics and biases shaped by psychological perceptions of risk and reward."} 
\end{quote}

Importantly, as a debiasing technique, the goal of providing the key findings from~\cite{kahneman1979prospect} is to help LLMs avoid making the same mistakes. Accordingly, we add the following sentence to the end of the above summary: ``As a rational investor, you should avoid making the mistakes described in prospect theory."

\begin{center}
    [Place Table~\ref{table:debiasing} about here]
\end{center}

Table~\ref{table:debiasing} compares the baseline debiasing technique of simply priming LLMs to be rational investors with the two detailed debiasing techniques described above. The analysis in this table focuses only on the first three experimental questions listed in Table~\ref{tab:psych_questions}: these are prospect theory-related questions, one on diminishing sensitivity, one on loss aversion, and one on probability weighting.\footnote{We focus on these prospect theory-related questions for two reasons. First, LLM responses to these questions are often irrational, leaving room for debiasing. Second, one debiasing technique described above provides the key findings from prospect theory, which will mostly likely affect LLM responses to prospect theory-related questions.} Table~\ref{table:debiasing} shows that the provision of the four-step procedure that guides LLMs to behave rationally is ineffective in reducing biases. Moreover, the provision of the key findings from~\cite{kahneman1979prospect} \textit{reduces} rational responses by about 29\% and increases human-like responses by about 20\%. Taken together, these results suggest that the provision of additional information---even if genuinely useful for decision making---does not always improve LLM performance: information overload can hinder an LLM's ability to produce rational responses. This is consistent with our finding in Section~\ref{subsec:results_exp_econ} that providing more information about the data generating process in the~\cite{Afrouzi2023} experiments increases human-like biases in LLM responses.

In summary, simple role priming that instructs LLMs to act as rational investors is effective, although the magnitude of improvement is economically modest.\footnote{Table~\ref{tab:summary_advanced} shows that for preference-based questions, the baseline share of irrational (human-like) responses is often exceeds 50\%, so a reduction of 3-4\% reported in Table~\ref{table:role_priming} is small relative to the baseline.} The two detailed debiasing techniques are either ineffective or counterproductive. Overall, our analysis in Section~\ref{sec:corrections} suggests that debiasing LLMs remains a challenging task. 

\section{Conclusion} \label{sec:conclusion}

Artificial intelligence, especially generative AI epitomized by LLMs, has become increasingly important in social and economic activities. This paper calls for a structured study of the behavioral economics of AI. As a first step, we examine the behavior of four prominent families of LLMs---ChatGPT, Anthropic Claude, Google Gemini, and Meta Llama---by leveraging experimental designs from the cognitive psychology and experimental economics literatures.

We document systematic patterns in the behavioral biases that LLMs exhibit. For cognitive psychology and experimental economics questions that study human preferences, LLM responses become increasingly irrational and human-like as models become more advanced or larger in scale. In contrast, for cognitive psychology and experimental economics questions that study human beliefs, the advanced large-scale LLMs generate responses that are largely rational. Moreover, we observe substantial heterogeneity in responses across the four LLM families. 

We further explore role-priming and debiasing methods that may affect LLM behavior. A role-priming prompt that instructs LLMs to behave as rational investors who make decisions according to the Expected Utility framework is effective in reducing biases, while a prompt that instructs LLMs to behave as real-world retail investors leads to less rational responses. In both cases, role priming affects responses through changes in LLM confidence and reasoning type, although its overall impact is economically modest. Finally, providing genuinely useful information---either a detailed procedure that guides LLMs to choose a course of action rationally under the Expected Utility framework or a summary of key findings from~\cite{kahneman1979prospect} that describe human biases---does not reduce LLM biases and can even be counterproductive. \\

\vspace{0.2in}

\clearpage

\begin{doublespacing}   

\bibliographystyle{jfe}

\bibliography{ref_ai_bias}

@article{korinek2023generative,
  title={{Generative AI for Economic Research: Use Cases and Implications for Economists}},
  author={Korinek, Anton},
  journal={Journal of Economic Literature},
  volume={61},
  number={4},
  pages={1281--1317},
  year={2023},
  publisher={American Economic Association 2014 Broadway, Suite 305, Nashville, TN 37203-2425}
}

@techreport{vidal2023,
  title={How {AI} and {LLMs} are {S}treamlining {F}inancial {S}ervices},
  author={Vidal, Nicolas},
  year={2023},
  type={Forbes}
}

@BOOK{tegmark2018life,
  author      = {Tegmark, Max},
  title       = {{Life 3.0: Being Human in the Age of Artificial Intelligence}},
  publisher   = {Random House Audio Publishing Group},
  year        = 2017,
}

@techreport{tomlinsonetal2024,
  title={Changing the {G}ame: {H}ow {AI} is {P}oised to {T}ransform {B}anking, {C}apital {M}arkets},
  author={Tomlinson, Neil and Laughridge, Kevin and Dockar, Barry},
  year={2024},
  type={Wall {S}treet {J}ournal}
}

@article{kahneman1973prediction,
  title={On the {P}sychology of {P}rediction},
  author={Kahneman, Daniel and Tversky, Amos},
  journal={Psychological Review},
  volume={80},
  number={4},
  pages={237-251},
  year={1973}
}

@article{kahneman1979prospect,
  title={Prospect {T}heory: {A}n {A}nalysis of {D}ecision under {R}isk},
  author={Kahneman, Daniel and Tversky, Amos},
  journal={Econometrica},
  volume={47},
  number={2},
  pages={263-292},
  year={1979}
}

@article{tversky1992prospect,
  title={{Advances in Prospect Theory: Cumulative Representation of Yncertainty}},
  author={Tversky, Amos and Kahneman, Daniel},
  journal={Journal of Risk and Uncertainty},
  volume={5},
  pages={297-323},
  year={1992}
}

@article{bordalo2012salience,
  title={{Salience Theory of Choice Under Risk}},
  author={Pedro Bordalo and Nicola Gennaioli and Andrei Shleifer},
  journal={Quarterly Journal of Economics},
  volume={127},
  number = {3},
  pages={1243-1285},
  year={2012}
}

@article{bordalo2013salience,
  title={{Salience and Asset Prices}},
  author={Pedro Bordalo and Nicola Gennaioli and Andrei Shleifer},
  journal={American Economic Review},
  volume={103},
  number = {3},
  pages={623-628},
  year={2013}
}

@article{cohn2012fear,
  title={{Evidence for Countercyclical Risk Aversion: An Experiment with Financial Professionals}},
  author={Alain Cohn and Jan Engelmann and Ernst Fehr and Michel André Maréchal},
  journal={American Economic Review},
  volume={105},
  number = {2},
  pages={860-885},
  year={2012}
}

@techreport{ma_etal_2023,
  title={{Is ChatGPT Humanly Irrational}?},
  author={Ma, Ding and Zhang, Tongda and Saunders, Michael},
  year={2023},
  type={{Working Paper}}
}

@techreport{bybee_2025,
  title={{The Ghost in the Machine: Generating Beliefs with Large Language Models}},
  author={Leland Bybee},
  year={2025},
  type={{Working Paper}}
}

@techreport{manning_horton_2025,
  title={{General Social Agents}},
  author={Benjamine S. Manning and John J. Horton},
  year={2025},
  type={{Working Paper}}
}

@techreport{hansen_et_al_2025,
  title={{Simulating the Survey of Professional Forecasters}},
  author={Anne Lundgaard Hansen and John J. Horton and Sophia Kazinnik and Daniela Puzzello and Ali Zarifhonarvar},
  year={2025},
  type={{Working Paper}}
}

@techreport{cook_et_al_2025,
  title={{What Do LLMs Want?}},
  author={Thomas R. Cook and Zach Modig and Sophia Kazinnik and Nathan Palmer},
  year={2025},
  type={{Working Paper}}
}

@article{chen_etal_2023_emergence,
  title={The {E}mergence of {E}conomic {R}ationality of {GPT}},
  author={Chen, Yiting and Liu, Tracy Xiao and Shan, You and Zhong, Songfa},
  journal={Proceedings of the National Academy of Sciences},
  volume={120},
  number={51},
  pages={e2316205120},
  year={2023}
}

@article{stiennon2020learning,
  title     = {{Learning to Summarize from Human Feedback}},
  author    = {Nisan Stiennon and Long Ouyang and Jeff Wu and Daniel M. Ziegler and Ryan Lowe and et al.},
  journal={Conference on Neural Information Processing Systems},
  volume={34},
  pages={3008--3021},
  year = {2020}
}

@article{fan_etal_2024_can,
  title={Can {L}arge {L}anguage {M}odels {S}erve as {R}ational {P}layers in {G}ame {T}heory? {A} {S}ystematic {A}nalysis},
  author={Fan, Caoyun and Chen, Jindou and Jin, Yaohui and He, Hao},
  journal={AAAI Conference on Artificial Intelligence},
  volume={38},
  number={16},
  pages={17960--17967},
  year={2024}
}

@techreport{bai2022,
      title={{Constitutional AI: Harmlessness from AI Feedback}}, 
      author={Yuntao Bai and Saurav Kadavath and Sandipan Kundu and Amanda Askell and Jackson Kernion and et al},
      year={2022},
      type={{Working Paper}}
}

@techreport{chen_etal_2024_manager,
  title={A {M}anager and an {AI} {W}alk into a {B}ar: {D}oes {C}hat{GPT} {M}ake {B}iased {D}ecisions {L}ike {W}e {D}o?},
  author={Chen, Yang and Kirshner, Samuel and Ovchinnikov, Anton and Andiappan, Meena and Jenkin, Tracy},
  year={2024},
  type={{Working Paper}}
}

@article{mei_etal_2024_turing,
  title={A {T}uring {T}est of {W}hether {AI} {C}hatbots are {B}ehaviorally {S}imilar to {H}umans},
  author={Mei, Qiaozhu and Xie, Yutong and Yuan, Walter and Jackson, Matthew O},
  journal={Proceedings of the National Academy of Sciences},
  volume={121},
  number={9},
  pages={e2313925121},
  year={2024}, 
}

@techreport{bauer_etal_2023_decoding,
  title={{D}ecoding {GPT}'s {H}idden `{R}ationality' of {C}ooperation},
  author={Bauer, Kevin and Liebich, Lena and Hinz, Oliver and Kosfeld, Michael},
  year={2023},
  type={{Working Paper}}
}

@article{brookins_2024_playing,
  title={Playing {G}ames {w}ith {GPT}: {W}hat {C}an {W}e {L}earn {A}bout a {L}arge {L}anguage {M}odel {f}rom {C}anonical {S}trategic {G}ames?},
  author={Brookins, Philip and DeBacker, Jason},
  journal={Economics Bulletin},
  volume={44},
  number={1},
  pages={25--37},
  year={2024}
}

@article{frederick2002,
  title={{Time Discounting and Time Preference: A Critical Review}},
  author={Shane Frederick and George Loewenstein and Ted O'Donoghue},
  journal={Journal of Economic Literature},
  volume={40},
  number={2},
  pages={351--401},
  year={2002},
  publisher={American Economic Association}
}

@article{rapoport1997randomization,
  title={Randomization in {I}ndividual {C}hoice {B}ehavior.},
  author={Rapoport, Amnon and Budescu, David V},
  journal={Psychological Review},
  volume={104},
  number={3},
  pages={603-617},
  year={1997},
  publisher={American Psychological Association}
}

@article{rapoport1992generation,
  title={Generation of {R}andom {S}eries in {T}wo-{P}erson {S}trictly {C}ompetitive {G}ames.},
  author={Rapoport, Amnon and Budescu, David V},
  journal={Journal of Experimental Psychology: General},
  volume={121},
  number={3},
  pages={352-363},
  year={1992},
  publisher={American Psychological Association}
}

@article{tversky1981framing,
  title={The {F}raming of {D}ecisions and the {P}sychology of {C}hoice},
  author={Tversky, Amos and Kahneman, Daniel},
  journal={Science},
  volume={211},
  pages={453--458},
  year={1981}
}

@article{ellsberg1961,
	Author = {Daniel Ellsberg},
	Journal = {Quarterly Journal of Economics},
	Number = {4},
	Pages = {643--669},
	Publisher = {Oxford University Press},
	Title = {{Risk, Ambiguity, and the Savage Axioms}},
	Volume = {75},
	Year = {1961}}

@article{Afrouzi2023,
  title = {Overreaction in {E}xpectations: {E}vidence and {T}heory},
  author = {Afrouzi, Hassan and Kwon, Spencer Y and Landier,  Augustin and Ma, Yueran and Thesmar, David},
  journal = {Quarterly Journal of Economics},
  number = {3},
  pages = {1713–1764},
  publisher = {Oxford University Press},
  volume = {138},
  year = {2023}}

@article{lian2018,
  title = {{Low Interest Rates and Risk-Taking: Evidence from Individual Investment Decisions}},
  author = {Lian, Chen and Ma, Yueran and Wang, Carmen},
  journal = {Review of Financial Studies},
  number = {6},
  pages = {2107–2148},
  publisher = {Oxford University Press},
  volume = {32},
  year = {2018}
}

@article{bose2022,
  title = {{Decision Weights for Experimental Asset Prices Based on Visual Salience}},
  author = {Bose, Devdeepta and Cordes, Henning and Nolte, Sven and Schneider, Judith and Camerer, Colin},
  journal = {Review of Financial Studies},
  number = {11},
  pages = {5904–5126},
  publisher = {Oxford University Press},
  volume = {35},
  year = {2022}}

@article{rabin2002inference,
	Author = {Rabin, Matthew},
	Journal = {Quarterly Journal of Economics},
	Number = {3},
	Pages = {775--816},
	Title = {{Inference by Believers in the Law of Small Numbers}},
	Volume = {117},
	Year = {2002}}

@BOOK{thaler:2008,
  author      = {Richard H. Thaler and Cass R. Sunstein},
  title       = {{Nudge: Improving Decisions About Health, Wealth, and Happiness}},
  publisher   = {Yale University Press},
  year        = 2008,
}

@BOOK{vonneumann:1944,
  author      = {John {Von Neumann} and Oskar Morgenstern},
  title       = {{Theory of Games and Economic Behavior}},
  publisher   = {Princeton University Press},
  year        = 1944,
}

@article{dellavigna2022,
	Author = {Stefano DellaVigna and Elizabeth Linos},
	Journal = {Econometrica},
	Number = {1},
	Pages = {81--116},
	Title = {{RCTs to Scale: Comprehensive Evidence from Two Nudge Units}},
	Volume = {90},
	Year = {2022}}

@incollection{Barberis:2018,
  author      = "Nicholas Barberis",
  title       = "Psychology-{B}ased {M}odels of {A}sset {P}rices and {T}rading {V}olume",
  editor      = "Douglas Bernheim and Stefano DellaVigna and David Laibson",
  pages       = "79--175",
  booktitle   = "Handbook of Behavioral Economics",
  publisher   = "North Holland, Amsterdam",
  year        = 2018,
}

@incollection{barberis2003survey,
  author      = "Nicholas Barberis and Richard Thaler",
  title       = "A {S}urvey of {B}ehavioral {F}inance",
  editor      = "George Constantinides and Milton Harris and Rene M. Stulz",
  pages       = "1053--1128",
  booktitle   = "Handbook of the Economics of Finance",
  publisher   = "North Holland, Amsterdam",
  year        = 2003,
}

@incollection{choi2004,
  author      = "James J Choi and David Laibson and Brigitte C Madrian and Andrew Metrick",
  title       = "{For Better or for Worse: Default Effects and 401(k) Savings Behavior}",
  editor      = "David A. Wise",
  pages       = "81--126",
  booktitle   = "Perspectives on the Economics of Aging",
  publisher   = "University of Chicago Press",
  year        = 2004,
}

@techreport{charness2023generation,
  title={{Generation Next: Experimentation with AI}},
  author={Charness, Gary and Jabarian, Brian and List, John A},
  year={2023},
  type={{Working Paper}}
}

@article{bail2024can,
  title={{Can Generative AI Improve Social Science?}},
  author={Bail, Christopher A},
  journal={Proceedings of the National Academy of Sciences},
  volume={121},
  number={21},
  pages={e2314021121},
  year={2024}
}

@article{shiffrin2023psychology,
  title     = {{Probing the Psychology of AI Models}},
  author    = {Richard Shiffrin and Melanie Mitchell},
  journal={Proceedings of the National Academy of Sciences},
  volume    = {120},
  number    = {10},
  pages     = {e2300963120},
  year      = {2023}
}

@article{binz_schulz_2023,
  title     = {{Using Cognitive Psychology to Understand GPT-3}},
  author    = {Marcel Binz and Eric Schulz},
  journal={Proceedings of the National Academy of Sciences},
  volume    = {120},
  number    = {6},
  pages     = {e2218523120},
  year      = {2023}
}

@techreport{ouyang2024ethical,
  title={{How Ethical Should AI Be? How AI Alignment Shapes Risk Preferences of LLMs}},
  author={Ouyang, Shumiao and Yun, Hayong and Zheng, Xingjian},
  year={2024},
  type={{Working Paper}}
}

@techreport{chen2024does,
  title={{What Does ChatGPT Make of Historical Stock Returns? Extrapolation and Miscalibration in LLM Stock Return Forecasts}},
  author={Chen, Shuaiyu and Green, T Clifton and Gulen, Huseyin and Zhou, Dexin},
  year={2025},
  type={{Working Paper}}
}

@techreport{bowen2024measuring,
  title={{Measuring and Mitigating Racial Disparities in Large Language Model Mortgage Underwriting}},
  author={Bowen, Donald E and Price, S McKay and Stein, Luke C.D. and Yang, Ke},
  year={2025},
  type={{Working Paper}}
}

@techreport{vafa2024large,
  title={{Do Large Language Models Perform the Way People Expect? Measuring the Human Generalization Function}},
  author={Vafa, Keyon and Rambachan, Ashesh and Mullainathan, Sendhil},
  year={2024},
  type={{Working Paper}}
}

@article{chang2024survey,
  title={{A Survey on Evaluation of Large Language Models}},
  author={Chang, Yupeng and Wang, Xu and Wang, Jindong and Wu, Yuan and Yang, Linyi and et al},
  journal={ACM Transactions on Intelligent Systems and Technology},
  volume={15},
  number={3},
  pages={1--45},
  year={2024}
}

@techreport{shazeer2017,
      title={{Outrageously Large Neural Networks: The Sparsely-Gated Mixture-of-Experts Layer}}, 
      author={Noam Shazeer and Azalia Mirhoseini and Krzysztof Maziarz and Andy Davis and Quoc Le and et al},
      year={2017},
      type={{Working Paper}}
}

@article{BarHillel1979,
  title = {{The Role of Sample Size in Sample Evaluation}},
  author = {Bar-Hillel,  Maya},
  journal = {Organizational Behavior and Human Performance},
  volume = {24},
  number = {2},
  pages = {245–257},
  year = {1979}
}

@article{Tversky1974,
  title = {{Judgment under Uncertainty: Heuristics and Biases}},
  author = {Tversky, Amos and Kahneman, Daniel},
  journal = {Science},
  volume = {185},
  number = {4157},
  pages = {1124–1131},
  year = {1974},
}

@article{Well1990,
  title = {{Understanding the Effects of Sample Size on the Variability of the Mean}},
  author = {Well,  Arnold D and Pollatsek,  Alexander and Boyce,  Susan J},
  journal = {Organizational Behavior and Human Decision Processes},
  volume = {47},
  number = {2},
  pages = {289–312},
  year = {1990}
}

@article{Tversky1983,
  title={{Extensional Versus Intuitive Reasoning: The Conjunction Fallacy in Probability Judgment}},
  author={Amos Tversky and Daniel Kahneman},
  journal={Psychological Review},
  year={1983},
  volume={90},
  pages={293-315}
}

@incollection{Wason_book_1972,
  author    = {Wason, Peter C. and Johnson-Laird, Philip N.},
  booktitle = {Psychology of Reasoning: Structure and Content},
  editor =    {Wason, Peter C.},
  pages     = {171--181},
  publisher = {Harvard University Press},
  title     = {{Immediate Inferences with Quantifiers}},
  year      = {1972}
}

@techreport{Brown_gpt3.5,
  doi = {10.48550/ARXIV.2005.14165},
  url = {https://arxiv.org/abs/2005.14165},
  author = {Brown,  Tom B. and Mann,  Benjamin and Ryder,  Nick and Subbiah,  Melanie and Kaplan,  Jared and et al},
  title = {{Language Models are Few-Shot Learners}},
  year = {2020},
  type={{Working Paper}}
}

@article{moore2008trouble,
  title={{The Trouble with Overconfidence}},
  author={Moore, Don A. and Healy, Paul J.},
  journal={Psychological Review},
  volume={115},
  number={2},
  pages={502--517},
  year={2008},
  publisher={American Psychological Association}
}

@article{deaves2009experimental,
  title={{An Experimental Test of the Impact of Overconfidence and Gender on Trading Activity}},
  author={Deaves, Richard and L{\"u}ders, Erik and Luo, Guo Ying},
  journal={Review of Finance},
  volume={13},
  number={3},
  pages={555--575},
  year={2009},
}
\end{doublespacing}




\onehalfspacing
\newpage

\begingroup
\small

\makeatletter
\let\orig@floatboxreset\@floatboxreset
\def\@floatboxreset{%
  \orig@floatboxreset
  \small
}
\makeatother


\small

\section*{Figures and Tables}

\setcounter{figure}{0}
\begin{figure}[!htbp]

\scalebox{0.95}{
    \colorbox{blue!10}{
        \begin{minipage}{\textwidth}
Instructions: \\
Consider the following scenarios and respond according to the template provided. Please treat each scenario as completely separate from the other. The output should be a markdown code snippet formatted in the following schema, including the leading and trailing ``\texttt{```json}" and ``\texttt{```}" and should not include any note or comment: \\
\texttt{```json} \\
\{ \\
\hspace{0.5cm} \texttt{"Scenario A": \{} \\
\hspace{1cm} \texttt{"Choice": string,} \\
\hspace{1cm} \texttt{"Confidence": float,} \\
\hspace{1cm} \texttt{"Explanation": string,} \\
\hspace{1cm} \texttt{"Reasoning": string} \\
\hspace{0.5cm} \texttt{\},} \\
\hspace{0.5cm} \texttt{"Scenario B": \{} \\
\hspace{1cm} \texttt{"Choice": string,} \\
\hspace{1cm} \texttt{"Confidence": float,} \\
\hspace{1cm} \texttt{"Explanation": string,} \\
\hspace{1cm} \texttt{"Reasoning": string} \\
\hspace{0.5cm} \texttt{\}} \\
\} \\
\texttt{```} \\

Scenario A: \\
In addition to whatever you own, you have been given \$1,000. You now need to choose between the following two options: option A (\$1,000, 0.5), meaning winning \$1,000 with 0.5 probability and winning zero with 0.5 probability, versus option B (\$500), meaning winning \$500 with certainty. Please answer as shown above. Indicate the choice you prefer (``A" or ``B"), your confidence level (a number between 0 and 1), a brief explanation for your choice (in less than 50 words), and your reasoning type (``A" if your reasoning is based more on intuitive thinking, and ``B" if your reasoning is based more on analytical thinking and calculations).\\

\vspace{0.05in}

Scenario B: \\
Next, please consider a different scenario; please treat it as a completely separate scenario from the one you were just asked about. Specifically, please consider the following scenario. In addition to whatever you own, you have been given \$2,000. You now need to choose between the following two options: option A (--\$1,000, 0.5), meaning losing \$1,000 with 0.5 probability and losing zero with 0.5 probability, versus option B: (--\$500), meaning losing \$500 with certainty. Please answer as shown above. Indicate the choice you prefer (``A" or ``B"), your confidence level (a number between 0 and 1), a brief explanation for your choice (in less than 50 words), and your reasoning type (``A" if your reasoning is based more on intuitive thinking, and ``B" if your reasoning is based more on analytical thinking and calculations).
\end{minipage}
}}
\caption{\textbf{Example of prompt: Diminishing sensitivity of prospect theory.}}

\vspace{0.2in}
This figure presents an example of a prompt that elicits LLM responses to a question designed by~\cite{kahneman1979prospect} to document diminishing sensitivity, a key element of prospect theory.
    \label{prompt: ds_pt}
\end{figure}

\clearpage

\begin{figure}[h!]

\centering
	\subfloat[\hspace{6em} Panel A: Preference-based questions]{\includegraphics[scale=0.18]{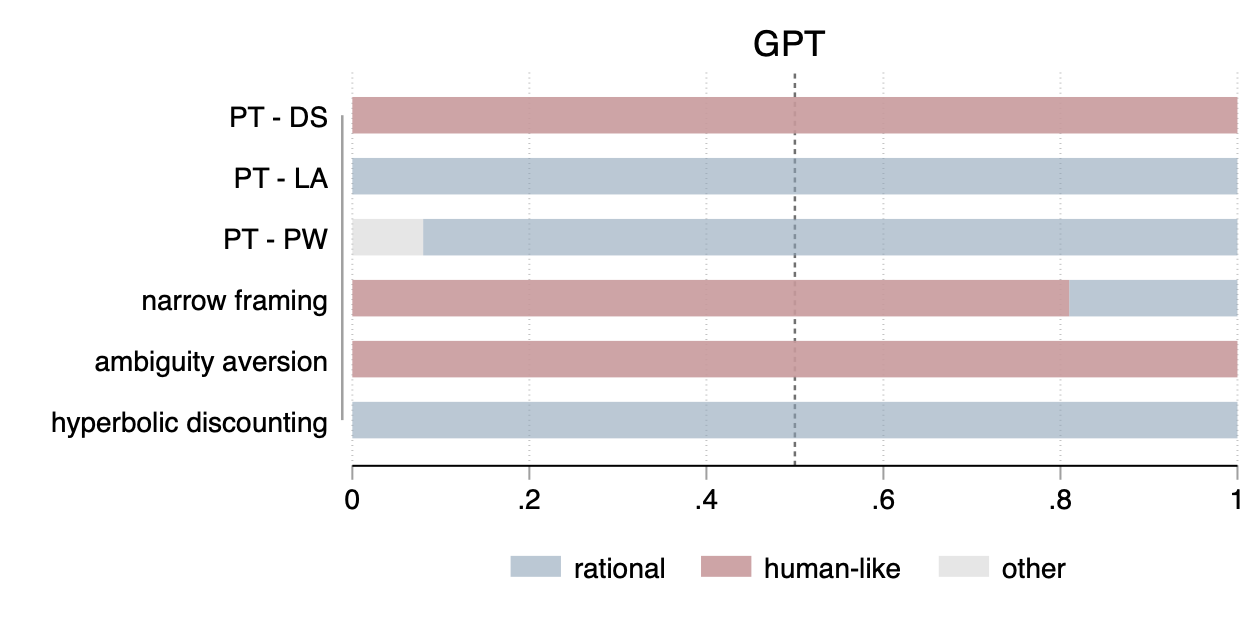}}
    \subfloat[\hspace{6em} Panel B: Belief-based questions]{\includegraphics[scale=0.18]{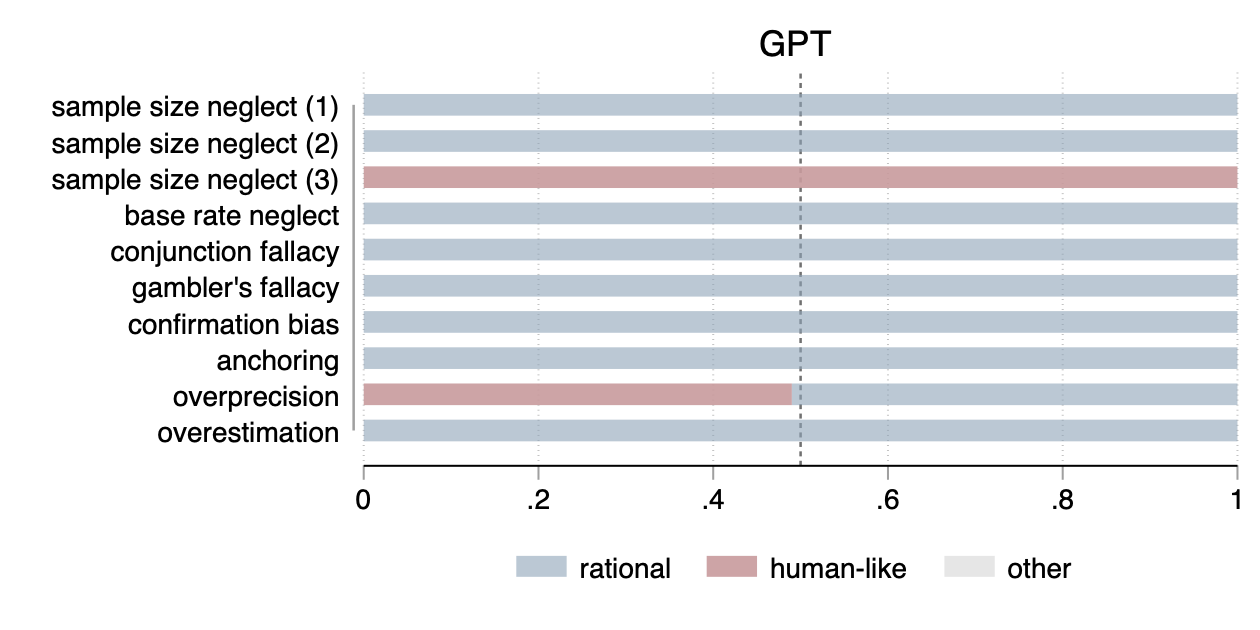}}\\
    \subfloat{\includegraphics[scale=0.18]{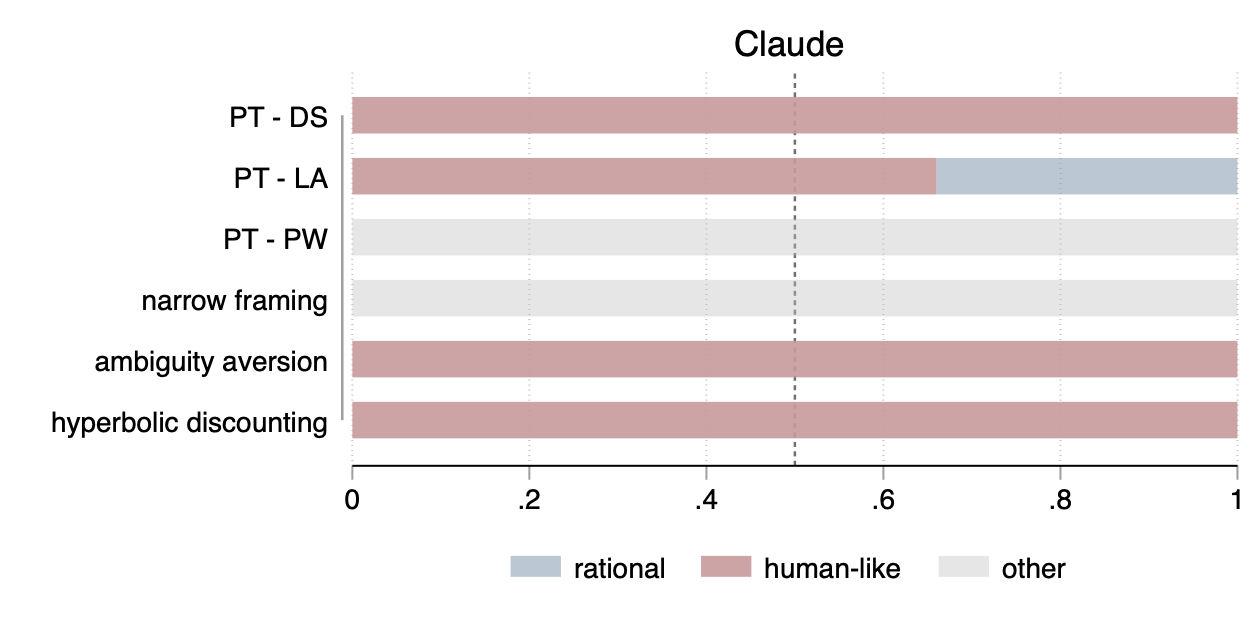}}
	\subfloat{\includegraphics[scale=0.18]{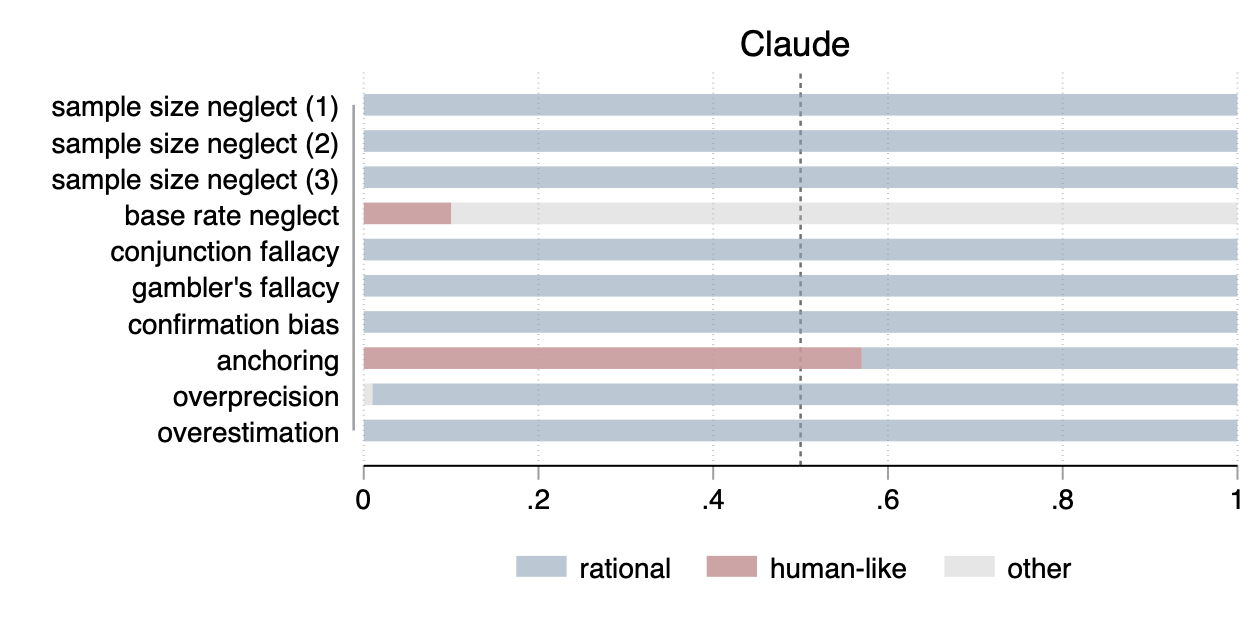}}\\
    \subfloat{\includegraphics[scale=0.18]{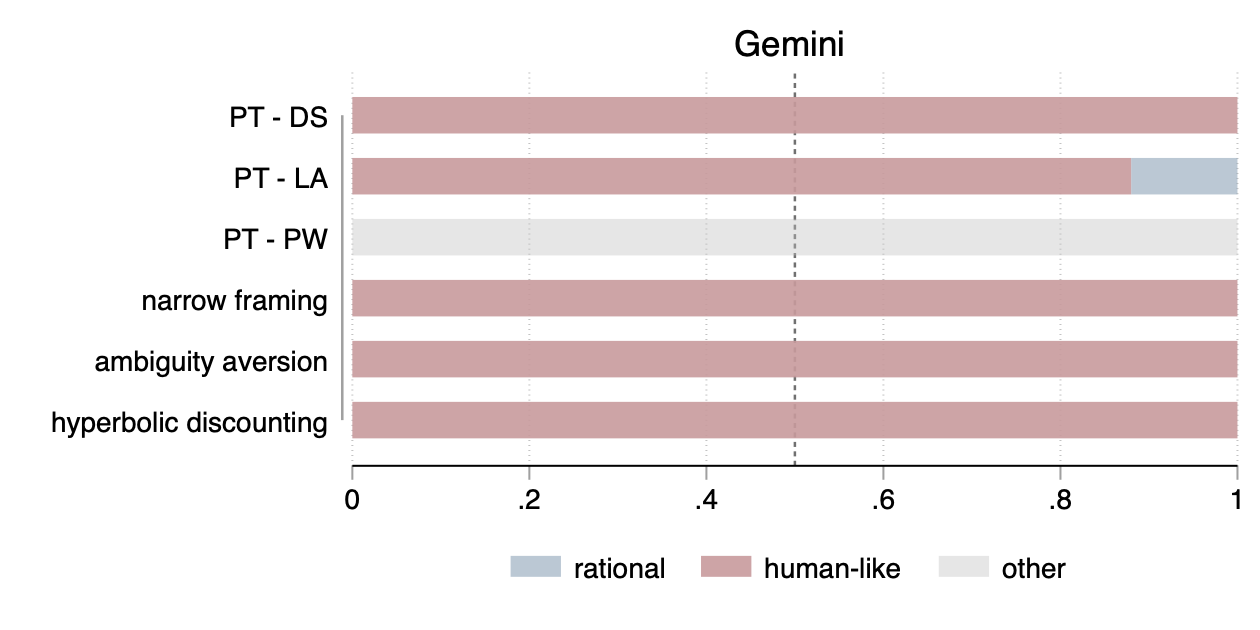}}
    \subfloat{\includegraphics[scale=0.18]{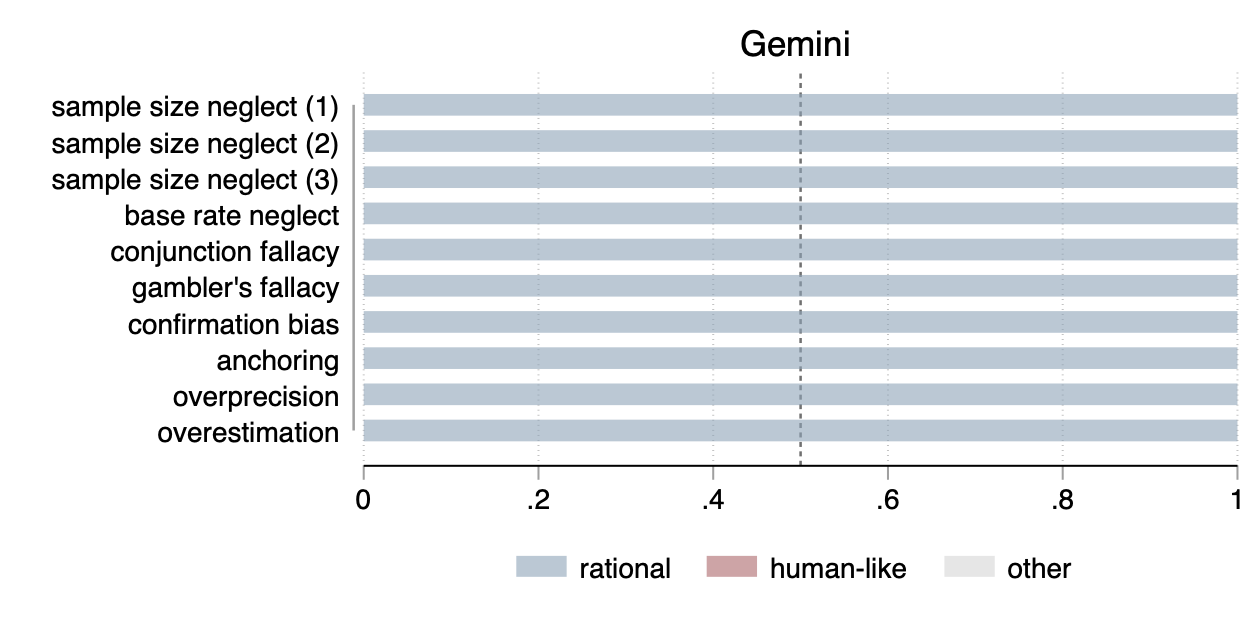}}\\
    \subfloat{\includegraphics[scale=0.18]{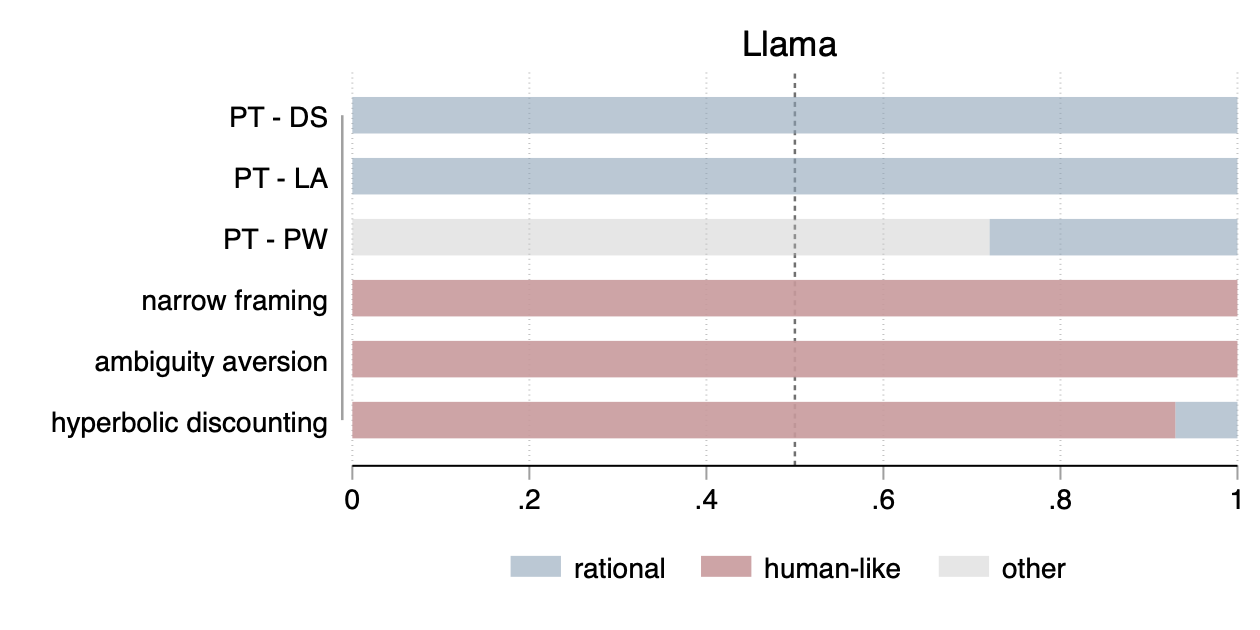}}
	\subfloat{\includegraphics[scale=0.18]{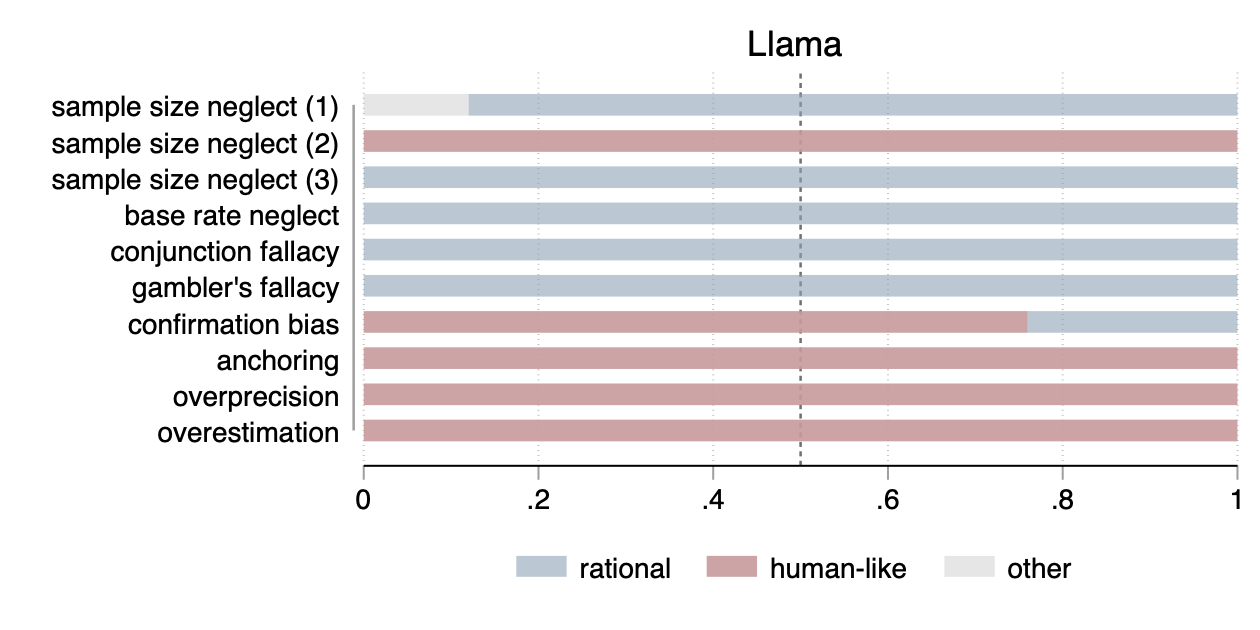}}

\caption{\textbf{Proportion of LLM responses: Advanced large-scale models.}}
	\label{fig_large_advanced}

\vspace{0.1in}
\caption*{
This figure plots the proportions of LLM responses categorized as rational (blue), human-like (red), or other (gray) for the four advanced large-scale LLMs: GPT-4, Claude 3 Opus, Gemini 1.5 Pro, and Llama 3 70B. The left panel presents results for the six preference-based questions, while the right panel presents results for the ten belief-based questions.}
	
\end{figure}

\clearpage 
\begin{landscape}
\begin{figure}[h!]

	\begin{center}
	\subfloat[\hspace{10em}Panel A: Preference-based questions]{\includegraphics[scale=0.25]{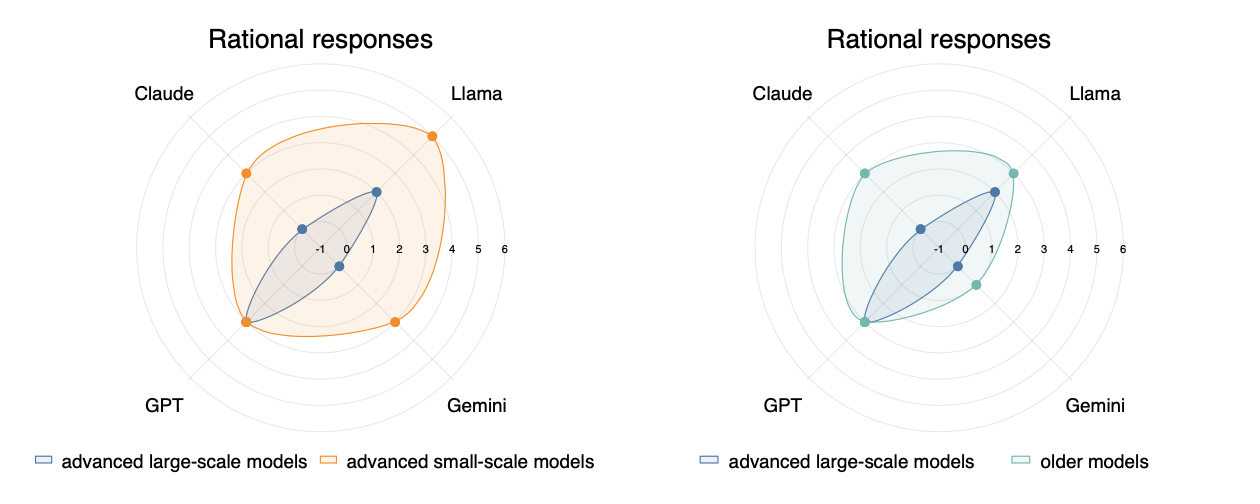}}
    \subfloat[\hspace{10em}Panel B: Belief-based questions]{\includegraphics[scale=0.25]{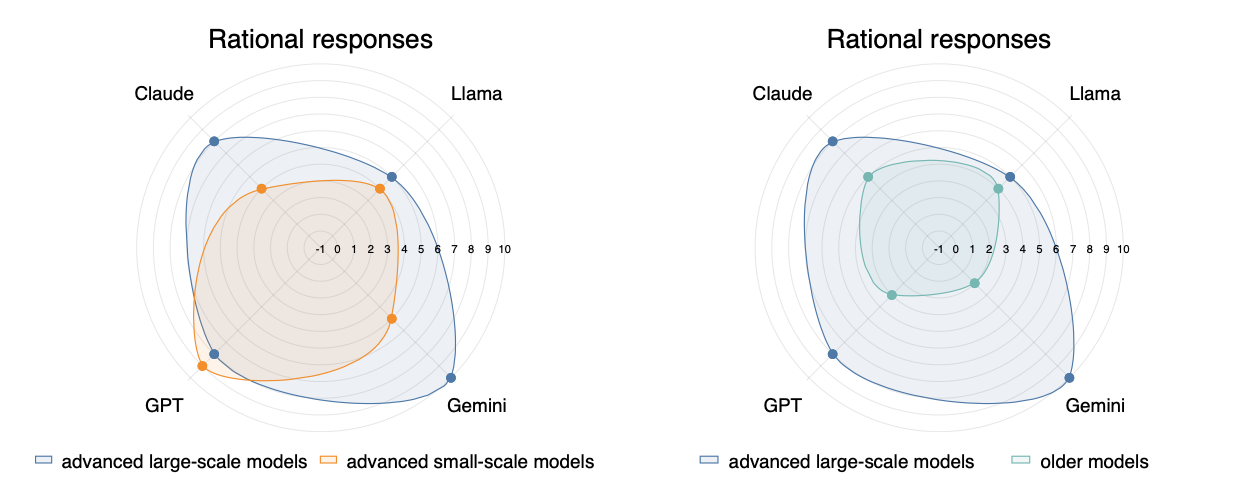}}
    \\~\\
    \vspace{2em}
    \subfloat{\includegraphics[scale=0.25]{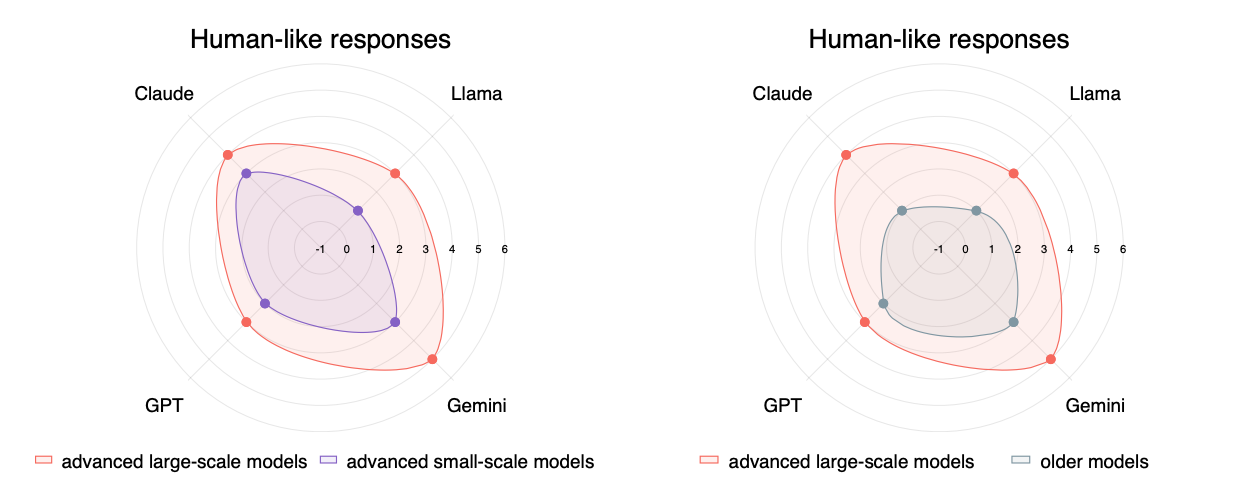}}
	\subfloat{\includegraphics[scale=0.25]{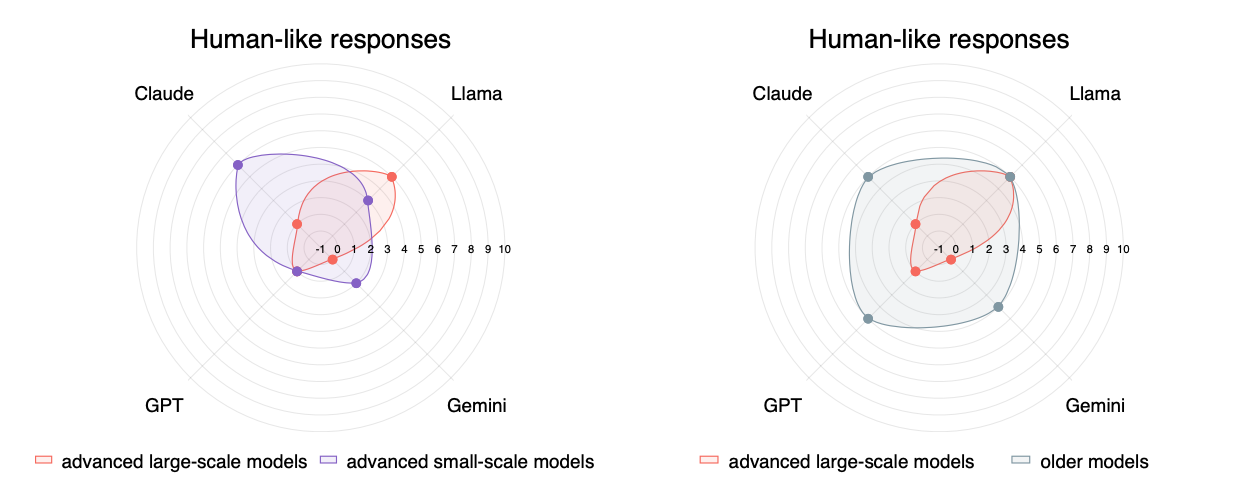}}
	\end{center}
\caption{\textbf{Heterogeneity in LLM responses across model generations and model scales.}}

\vspace{0.2in}
This figure presents radar charts that compare the number of questions for which each LLM produces predominantly rational responses (top row) or human-like responses (bottom row), separately for preference-based questions (left panel) and belief-based questions (right panel). Comparisons are made between advanced large-scale models and advanced smaller-scale models, and between advanced large-scale models and older models.

	\label{fig_heterogeneity} 	

\end{figure}
\end{landscape}

\clearpage

\begin{figure}[h!]
\vspace{-.2in}
    \begin{center}
        \includegraphics[scale=.7]{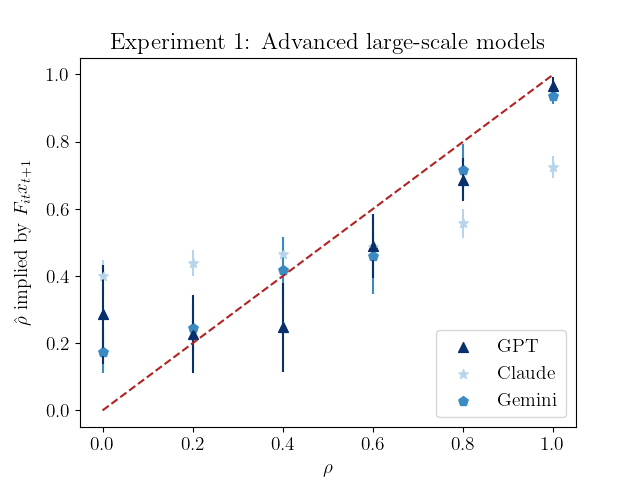} 
        \\
        \vspace{0.2in}
        \includegraphics[scale=.7]{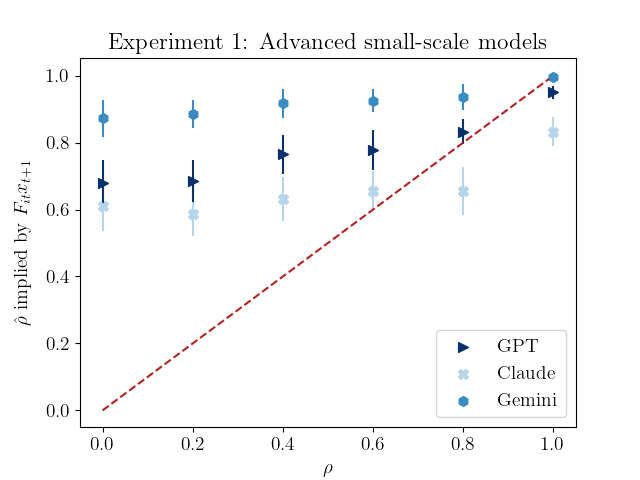}
    \end{center}

\caption{\textbf{LLM forecasts: Experiment 1 in~\cite{Afrouzi2023}.}}

\vspace{0.2in}
The figure plots the perceived persistence $\hat{\rho}$ against the true $\rho$. Here, $\hat{\rho}$ is estimated using LLM forecasts collected from the setting of Experiment 1 in~\cite{Afrouzi2023}: the top panel reports results for the three advanced large-scale models of GPT-4, Claude 3 Opus, and Gemini 1.5 Pro, while the bottom panel reports results for the three advanced small-scale models of GPT-4o, Claude 3 Haiku, and Gemini 1.5 Flash. For each estimated $\hat{\rho}$, the vertical bar shows the 95\% confidence interval. The procedure for estimating $\hat{\rho}$ is described in Section~\ref{subsec:results_exp_econ} of the main text. The red dashed line is a 45-degree line, which represents the persistence implied by full information rational expectations (FIRE).

	\label{fig_afrouzi} 	
\end{figure}

\clearpage

\begin{figure}[h!]
\vspace{-.2in}
    \begin{center}
        \includegraphics[scale=.7]{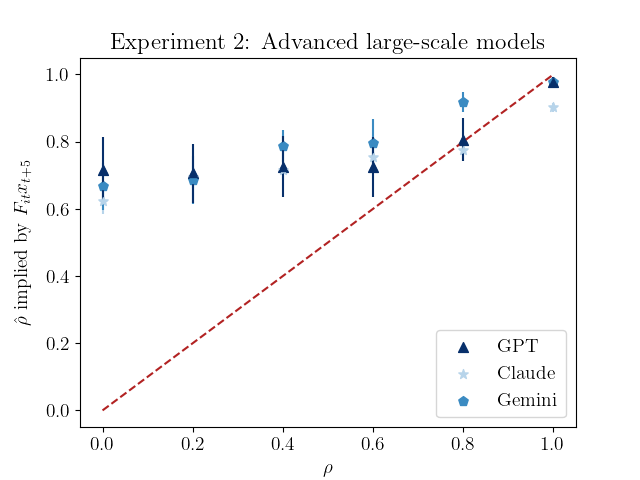} 
        \\
        \vspace{0.2in}
        \includegraphics[scale=.7]{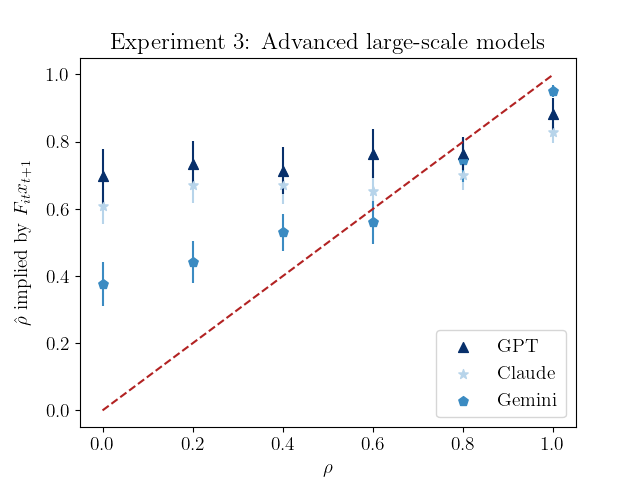}
    \end{center}

\caption{\textbf{LLM forecasts: Experiments 2 and 3 in~\cite{Afrouzi2023}.}}

\vspace{0.2in}

The figure plots the perceived persistence $\hat{\rho}$ against the true $\rho$. Here, $\hat{\rho}$ is estimated using the LLM forecasts collected from the settings of Experiments 2 and 3 in~\cite{Afrouzi2023}: the top panel examines LLMs forecasts collected from the setting of Experiment 2, and the bottom panel examines LLM forecasts collected from the setting of Experiment 3. Both panels report results for the three advanced large-scale models of GPT-4, Claude 3 Opus, and Gemini 1.5 Pro. For each estimated $\hat{\rho}$, the vertical bar shows the 95\% confidence interval. The procedure for estimating $\hat{\rho}$ is described in Section~\ref{subsec:results_exp_econ} of the main text. The red dashed line is a 45-degree line, which represents the persistence implied by full information rational expectations (FIRE).

	\label{fig_afrouzi_variant} 	
\end{figure}

\small
\setcounter{table}{0}
\makeatletter 
\renewcommand{\thetable}{\arabic{table}}

\clearpage
\begin{table}[htbp!]
    \caption{{\bf Summary of experimental questions from cognitive psychology.}}
    This table summarizes the sixteen experimental questions drawn from the cognitive psychology literature. Specifically, Question 1 is based on Problems 11 and 12 in~\cite{kahneman1979prospect} (page 273). Question 2 is based on an example of loss aversion discussed in~\cite{barberis2003survey} (page 1069). Question 3 is based on Problems 14 and $14^{\prime}$ in~\cite{kahneman1979prospect} (page 281). Question 4 is a modified version of Problem 10 in~\cite{tversky1981framing} (page 457). Question 5 is based on an example of hyperbolic discounting discussed in~\cite*{frederick2002} (page 361). Question 6 is based on Questions 3 and 4 of the~\cite{ellsberg1961} experiment (pages 650 to 651). Question 7 is based on an experiment discussed in~\cite{Tversky1974} that documents sample size neglect as a form of representativeness heuristic (page 1125). Question 8 is based on Experiment 2 in~\cite*{Well1990} (page 297). Question 9 is based on Problem 10 in~\cite{BarHillel1979} (page 255). Question 10 is based on an experiment designed in~\cite{kahneman1973prediction} (page 241). Question 11 is based on an experiment designed in~\cite{Tversky1983} (pages 297 and 299). Question 12 is based on an experiment discussed in~\cite{rabin2002inference} (page 781). Question 13 is based on a selection task discussed in~\cite{Wason_book_1972} (page 173). Question 14 is based on an experiment discussed in~\cite{Tversky1974} that documents anchoring (page 1128). Question 15 is based on a set of general knowledge questions adapted from Appendix C of~\cite*{deaves2009experimental} (page 2). Finally, Question 16 follows the procedure discussed on pages 508 to 509 of~\cite{moore2008trouble} to document overestimation. 
    \\
    \label{tab:psych_questions}

    \vspace{0.0in} 
    \centering
    \caption*{\centering Panel A: Questions that study the psychology of preferences}
    \footnotesize
    \begin{tabular}{ccc}
        \hline
        \hline
        Question number & Documented bias & Note \\
        \hline
        1 & prospect theory - diminishing sensitivity & risk preferences \\
        2 & prospect theory - loss aversion & risk preferences \\
        3 & prospect theory - probability weighting & risk preferences \\
        4 & narrow framing & risk preferences \\
        5 & ambiguity aversion & risk preferences \\
        6 & hyperbolic discounting & time preferences \\
        \hline
        \hline
    \end{tabular}

    \vspace{0.1in} 
    \centering
    \caption*{\centering Panel B: Questions that study the psychology of beliefs}
    \footnotesize
    \begin{tabular}{cc}
        \hline
        \hline
        Question number & Documented bias \\
        \hline
        7 & sample size neglect (1)\\
        8 & sample size neglect (2)\\
        9 & sample size neglect (3)\\
        10 & base rate neglect \\
        11 & conjunction fallacy \\
        12 & gambler's fallacy \\
        13 & confirmation bias \\
        14 & anchoring \\
        15 & overconfidence - overprecision \\
        16 & overconfidence - overestimation \\
        \hline
        \hline
    \end{tabular}
    
\end{table}

\clearpage

\begin{landscape}
\begin{table}[htbp!]
\caption{\bf Description of large language models.}
This table describes the twelve LLMs examined in our analysis. We group these models into four LLM families: ChatGPT, Anthropic Claude, Google Gemini, and Meta Llama. For each family, we treat the advanced large-scale models as baselines: GPT-4, Claude 3 Opus, Gemini 1.5 Pro, and Llama 3 70B. We also analyze their smaller-scale versions---GPT-4o, Claude 3 Haiku, Gemini 1.5 Flash, and Llama 3 8B---as well as their predecessors---GPT-3.5 Turbo, Claude 2, Gemini 1.0 Pro, and Llama 2 70B. RLHF and RLAI denote ``Reinforcement Learning from Human Feedback" and ``Reinforcement Learning from AI," respectively. MMLU denotes ``Massive Multitask Language Understanding" and it provides a benchmark score for evaluating LLM capabilities. Vision indicates whether a model supports graphical inputs.\\


\footnotesize
\begin{centering}
    \begin{tabular}{lccccccc}
    \tabularnewline
        \hline 
        \hline 
         Model & Release year & Number of parameters & Number of tokens & Instruction &  Context window & MMLU & Vision  \\
         \hline
         GPT-3.5 Turbo & 2022 & 175 B & 300 B & RLHF & 16,385 & 70 & No \\
         GPT-4 & 2023 & 1T$^{*}$ & 13T$^{*}$ & RLHF & 128,000 & 86.5 & Yes \\
         GPT-4o & 2024 & - & 13T$^{*}$ & RLHF & 128,000 & 88.7 & Yes \\
         \hline
         Claude 2  & 2023 & 200 B$^{*}$ & - & RLAI + RLHF & 100,000 & 78.5 & No \\
         Claude 3 Opus & 2024 & 1T$^{*}$ & - & RLAI + RLHF & 200,000 & 86.8 &  Yes \\
         Claude 3 Haiku  & 2024 & 20B$^{*}$ & - & RLAI + RLHF & 200,000 & 75.2 &  Yes \\
         \hline 
         Gemini 1.0 Pro  & 2024 & 100 B$^{*}$ & - & RLHF & 32,000 & - & Yes \\
         Gemini 1.5 Pro  & 2024 & 1T$^{*}$ & - & RLHF &  128,000 & 81.9 & Yes \\
         Gemini 1.5 Flash  & 2024 & 30 B$^{*}$ & - & RLHF &  128,000 & 81.0 & Yes \\
         \hline 
         Llama 2 70B  & 2023 & 70 B & 2 T & RLHF & 4,096 & 68.9 & No \\
         Llama 3 70B  & 2024 & 70 B & 15 T & RLHF& 8,200 & 80.2 & No \\
         Llama 3 8B  & 2024 & 8 B & 15 T & RLHF & 8,200 & 68.4 & No \\
         
    \hline 
    \hline 
    \end{tabular}
    \par\end{centering}
    
    \vspace{0.1in}
    $^{*}$These numbers are unofficial and estimated. 
    \label{tab:LLM_description}
\end{table}
\end{landscape}

\clearpage
\begin{landscape}

\begin{table}[htbp!]
\caption{{\bf Rational responses versus human-like responses: Advanced large-scale models.}}
This table reports the proportion of responses classified as rational or human-like for the four advanced large-scale LLMs: GPT-4, Claude 3 Opus, Gemini 1.5 Pro, and Llama 3 70B. Panel A presents results for the six preference-based questions, and Panel B presents results for the ten belief-based questions. The abbreviations ``PT - DS," ``PT - LA," and ``PT - PW" denote ``prospect theory - diminishing sensitivity," ``prospect theory - loss aversion," and ``prospect theory - probability weighting," respectively. Numbers in parentheses are $p$-values from a binomial test with the null hypothesis that the proportion of rational or human-like responses is less than or equal to 50\%. $^{***}p<0.01$, $^{**}p<0.05$ and $^{*}p<0.1$.\\

\vspace{0.1in}

\label{tab:summary_advanced}
\footnotesize
\scalebox{0.8}{
\begin{tabularx}{1.7\textwidth}{lllllHHlllllHHlllllHHlllllHH}

\hline
\hline
\multicolumn{28}{l}{Panel A: Preference-based questions} \\
\midrule
& \multicolumn{6}{c}{GPT}  & &\multicolumn{6}{c}{Claude} &  & \multicolumn{6}{c}{Gemini} &  & \multicolumn{6}{c}{Llama} \\
\cmidrule(r){2-7} \cmidrule(r){9-14} \cmidrule(r){16-21} \cmidrule(r){23-28}
& \multicolumn{2}{l}{\%rational}  & \multicolumn{2}{l}{\%human-like}  & & &  & \multicolumn{2}{l}{\%rational}  & \multicolumn{2}{l}{\%human-like}   & &  &  & \multicolumn{2}{l}{\%rational}  & \multicolumn{2}{l}{\%human-like} & & &  & \multicolumn{2}{l}{\%rational}  & \multicolumn{2}{l}{\%human-like} & &  \\
\midrule
PT - DS & 0.00 & (1.000) & 1.00 & (0.000)*** & 0.00 & (1.000) &  & 0.00 & (1.000) & 1.00 & (0.000)*** & 0.00 & (1.000) &  & 0.00 & (1.000) & 1.00 & (0.000)*** & 0.00 & (1.000) &  & 1.00 & (0.000)*** & 0.00 & (1.000) & 0.00 & (1.000) \\
PT - LA & 1.00 & (0.000)*** & 0.00 & (1.000) & 0.00 & (1.000) &  & 0.34 & (1.000) & 0.66 & (0.001)*** & 0.00 & (1.000) &  & 0.12 & (1.000) & 0.88 & (0.000)*** & 0.00 & (1.000) &  & 1.00 & (0.000)*** & 0.00 & (1.000) & 0.00 & (1.000) \\
PT - PW & 0.92 & (0.000)*** & 0.00 & (1.000) & 0.08 & (1.000) &  & 0.00 & (1.000) & 0.00 & (1.000) & 1.00 & (0.000)*** &  & 0.00 & (1.000) & 0.00 & (1.000) & 1.00 & (0.000)*** &  & 0.28 & (1.000) & 0.00 & (1.000) & 0.72 & (0.000)*** \\
narrow framing & 0.19 & (1.000) & 0.81 & (0.000)*** & 0.00 & (1.000) &  & 0.00 & (1.000) & 0.00 & (1.000) & 1.00 & (0.000)*** &  & 0.00 & (1.000) & 1.00 & (0.000)*** & 0.00 & (1.000) &  & 0.00 & (1.000) & 1.00 & (0.000)*** & 0.00 & (1.000) \\
ambiguity aversion & 0.00 & (1.000) & 1.00 & (0.000)*** & 0.00 & (1.000) &  & 0.00 & (1.000) & 1.00 & (0.000)*** & 0.00 & (1.000) &  & 0.00 & (1.000) & 1.00 & (0.000)*** & 0.00 & (1.000) &  & 0.00 & (1.000) & 1.00 & (0.000)*** & 0.00 & (1.000) \\
hyperbolic discounting & 1.00 & (0.000)*** & 0.00 & (1.000) & 0.00 & (1.000) &  & 0.00 & (1.000) & 1.00 & (0.000)*** & 0.00 & (1.000) &  & 0.00 & (1.000) & 1.00 & (0.000)*** & 0.00 & (1.000) &  & 0.07 & (1.000) & 0.93 & (0.000)*** & 0.00 & (1.000) \\
\midrule
\\
\multicolumn{28}{l}{Panel B: Belief-based questions} \\

\midrule
 & \multicolumn{6}{c}{GPT}  & &\multicolumn{6}{c}{Claude} &  & \multicolumn{6}{c}{Gemini} &  & \multicolumn{6}{c}{Llama} \\
\cmidrule(r){2-7} \cmidrule(r){9-14} \cmidrule(r){16-21} \cmidrule(r){23-28}
& \multicolumn{2}{l}{\%rational}  & \multicolumn{2}{l}{\%human-like}  & & &  & \multicolumn{2}{l}{\%rational}  & \multicolumn{2}{l}{\%human-like}   & &  &  & \multicolumn{2}{l}{\%rational}  & \multicolumn{2}{l}{\%human-like} & & &  & \multicolumn{2}{l}{\%rational}  & \multicolumn{2}{l}{\%human-like} & & \\
\midrule
sample size neglect (1) & 1.00 & (0.000)*** & 0.00 & (1.000) & 0.00 & (1.000) &  & 1.00 & (0.000)*** & 0.00 & (1.000) & 0.00 & (1.000) &  & 1.00 & (0.000)*** & 0.00 & (1.000) & 0.00 & (1.000) &  & 0.88 & (0.000)*** & 0.00 & (1.000) & 0.12 & (1.000) \\
sample size neglect (2) & 1.00 & (0.000)*** & 0.00 & (1.000) & 0.00 & (1.000) &  & 1.00 & (0.000)*** & 0.00 & (1.000) & 0.00 & (1.000) &  & 1.00 & (0.000)*** & 0.00 & (1.000) & 0.00 & (1.000) &  & 0.00 & (1.000) & 1.00 & (0.000)*** & 0.00 & (1.000) \\
sample size neglect (3) & 0.00 & (1.000) & 1.00 & (0.000)*** & 0.00 & (1.000) &  & 1.00 & (0.000)*** & 0.00 & (1.000) & 0.00 & (1.000) &  & 1.00 & (0.000)*** & 0.00 & (1.000) & 0.00 & (1.000) &  & 1.00 & (0.000)*** & 0.00 & (1.000) & 0.00 & (1.000) \\
base rate neglect & 1.00 & (0.000)*** & 0.00 & (1.000) & 0.00 & (1.000) &  & 0.00 & (1.000) & 0.10 & (1.000) & 0.90 & (0.000)*** &  & 1.00 & (0.000)*** & 0.00 & (1.000) & 0.00 & (1.000) &  & 1.00 & (0.000)*** & 0.00 & (1.000) & 0.00 & (1.000) \\
conjunction fallacy & 1.00 & (0.000)*** & 0.00 & (1.000) & 0.00 & (1.000) &  & 1.00 & (0.000)*** & 0.00 & (1.000) & 0.00 & (1.000) &  & 1.00 & (0.000)*** & 0.00 & (1.000) & 0.00 & (1.000) &  & 1.00 & (0.000)*** & 0.00 & (1.000) & 0.00 & (1.000) \\
gambler's fallacy & 1.00 & (0.000)*** & 0.00 & (1.000) & 0.00 & (1.000) &  & 1.00 & (0.000)*** & 0.00 & (1.000) & 0.00 & (1.000) &  & 1.00 & (0.000)*** & 0.00 & (1.000) & 0.00 & (1.000) &  & 1.00 & (0.000)*** & 0.00 & (1.000) & 0.00 & (1.000) \\
confirmation bias & 1.00 & (0.000)*** & 0.00 & (1.000) & 0.00 & (1.000) &  & 1.00 & (0.000)*** & 0.00 & (1.000) & 0.00 & (1.000) &  & 1.00 & (0.000)*** & 0.00 & (1.000) & 0.00 & (1.000) &  & 0.24 & (1.000) & 0.76 & (0.000)*** & 0.00 & (1.000) \\
anchoring & 1.00 & (0.000)*** & 0.00 & (1.000) & 0.00 & (1.000) &  & 0.43 & (0.933) & 0.57 & (0.097)* & 0.00 & (1.000) &  & 1.00 & (0.000)*** & 0.00 & (1.000) & 0.00 & (1.000) &  & 0.00 & (1.000) & 1.00 & (0.000)*** & 0.00 & (1.000) \\
overprecision & 0.51 & (0.460) & 0.49 & (0.618) & 0.00 & (1.000) &  & 0.99 & (0.000)*** & 0.00 & (1.000) & 0.01 & (1.000) &  & 1.00 & (0.000)*** & 0.00 & (1.000) & 0.00 & (1.000) &  & 0.00 & (1.000) & 1.00 & (0.000)*** & 0.00 & (1.000) \\
overestimation & 1.00 & (0.000)*** & 0.00 & (1.000) & 0.00 & (1.000) &  & 1.00 & (0.000)*** & 0.00 & (1.000) & 0.00 & (1.000) &  & 1.00 & (0.000)*** & 0.00 & (1.000) & 0.00 & (1.000) &  & 0.00 & (1.000) & 1.00 & (0.000)*** & 0.00 & (1.000) \\

\hline
\hline
\end{tabularx}
}
\end{table}
\end{landscape}

\clearpage
\begin{table}[htbp!]
\caption{{\bf Heterogeneity in responses across LLM families.}\\
}
This table reports marginal effects from the probit regressions:
\begin{equation*}
\Pr(Y_{iqk}=1) = \Phi(\alpha + \beta_1 \cdot Claude_i + \beta_2 \cdot Gemini_i + \beta_3 \cdot Llama_i + \epsilon_{iqk})
\end{equation*}
for model $i$, question $q$, and iteration $k$, where $\Phi(\cdot)$ denotes the cumulative distribution function of a standard Normal random variable. For Columns (1) and (3), $Y_{iqk}$ is a binary variable that takes the value of one if model $i$'s response to question $q$ in iteration $k$ is classified as rational, and zero otherwise. For Columns (2) and (4), $Y_{iqk}$ is a binary variable that takes the value of one if model $i$'s response to question $q$ in iteration $k$ is classified as human-like, and zero otherwise. For both cases, the independent variables---$Claude_i$, $Gemini_i$, and $Llama_i$---are indicators for the respective LLM families, with the LLM family of GPT serving as the omitted baseline category. The reported coefficients represent the change in the predicted probability of observing $Y_{iqk} = 1$ that is associated with changing the LLM from GPT to Claude, Gemini, or Llama. Standard errors, clustered at the question level, are reported in parentheses. $^{***}p<0.01$, $^{**}p<0.05$ and $^{*}p<0.1$.\\

\vspace{0.1in}

\label{tab:across_family}

\footnotesize
\begin{tabularx}{\linewidth}{lYYYY}

\hline
\hline
 & (1) & (2) & (3) & (4) \\
\midrule
Dep. var:  & \multicolumn{4}{c}{LLM response is characterized as} \\
& Rational & Human-like & Rational & Human-like \\
\cmidrule(r){2-2}\cmidrule(lr){3-3} \cmidrule(lr){4-4}
\cmidrule(l){5-5}
Sample: & \multicolumn{2}{c}{Preference-based questions}  & \multicolumn{2}{c}{Belief-based questions} \\
\cmidrule(r){1-1} \cmidrule(r){2-3} \cmidrule(l){4-5}
Claude & --0.126 & --0.0483 & --0.0997 & 0.126 \\
 & (0.083) & (0.118) & (0.084) & (0.102) \\
Gemini & --0.229*** & 0.167** & --0.0800 & 0.0107 \\
 & (0.065) & (0.077) & (0.049) & (0.051) \\
Llama & 0.0816 & --0.141 & --0.250** & 0.210** \\
 & (0.150) & (0.127) & (0.098) & (0.088) \\
 \hline
Baseline LLM family: & \multicolumn{4}{c}{GPT} \\
\hline
Observations & 7,150 & 7,150 & 12,000 & 12,000 \\
Pseudo $R$-squared & 0.043 & 0.037 & 0.025 & 0.026 \\

\hline
\hline
\end{tabularx}
\end{table}

\clearpage
\begin{landscape}
\begin{table}[htbp!]
\caption{\bf Heterogeneity in responses across model generations and model scales.}
This table reports marginal effects from the probit regressions specified in equations~\eqref{eq:probit_across_generation} and~\eqref{eq:probit_across_scale} in the main text. For Columns (1), (2), (5), and (6), the dependent variable is a binary variable that takes the value of one if an LLM response is classified as rational, and zero otherwise. For Columns (3), (4), (7), and (8), the dependent variable is a binary variable that takes the value of one if an LLM response is classified as human-like, and zero other wise. Columns (1) to (4) report results for preference-based questions and Columns (5) to (8) report results for belief-based questions. Panel A compares advanced large-scale  models with older models. In this case, the sample is restricted to LLM responses from either the advanced large-scale models or the older models, and the key independent variable is an indicator for the advanced models, with older models serving as the baseline. Panel B compares large-scale models with smaller ones. In this case, the sample is restricted to LLM responses from either the advanced large-scale models or the advanced smaller-scale models, and the key independent variable is an indicator for the large-scale models. Standard errors, clustered at the question level, are reported in parentheses. $^{***}p<0.01$, $^{**}p<0.05$ and $^{*}p<0.1$.\\

\vspace{0.1in}

\label{tab:heterogneity}

\footnotesize
\scalebox{0.97}{
\begin{tabularx}{1.4\textwidth}{lYYYYcYYYY}
\hline
\hline
\multicolumn{10}{l}{Panel A: Advanced models versus older models} \\
\midrule
 & (1) & (2) & (3) & (4) &  & (5) & (6) & (7) & (8) \\
\midrule
Dep. var:  & \multicolumn{9}{c}{LLM response is characterized as} \\
 & Rational & Rational & Human-like & Human-like &  & Rational & Rational & Human-like & Human-like \\
\cmidrule(r){2-2} \cmidrule(lr){3-3} \cmidrule(lr){4-4} \cmidrule(l){5-5}
\cmidrule(r){7-7} \cmidrule(lr){8-8} \cmidrule(lr){9-9} \cmidrule(l){10-10}
Sample: & \multicolumn{4}{c}{Preference-based questions} & &\multicolumn{4}{c}{Belief-based questions} \\
\cmidrule(r){1-1} \cmidrule{2-5} \cmidrule{7-10}
Advanced & --0.223* & --0.231** & 0.272** & 0.273** &  & 0.407*** & 0.409*** & --0.327*** & --0.333*** \\
 & (0.121) & (0.116) & (0.127) & (0.126) &  & (0.127) & (0.125) & (0.104) & (0.102) \\
LLM family FE & No & Yes & No & Yes &  & No & Yes & No & Yes \\
Observations & 4,800 & 4,800 & 4,800 & 4,800 &  & 8,000 & 8,000 & 8,000 & 8,000 \\
Pseudo $R$-squared & 0.042 & 0.120 & 0.055 & 0.107 &  & 0.133 & 0.162 & 0.097 & 0.134 \\

\midrule
\\
\multicolumn{10}{l}{Panel B: Large-scale models versus smaller-scale models} \\
\midrule
& (1) & (2) & (3) & (4) &  & (5) & (6) & (7) & (8) \\
\midrule
Dep. var:  & \multicolumn{9}{c}{LLM response is characterized as} \\
 & Rational & Rational & Human-like & Human-like &  & Rational & Rational & Human-like & Human-like \\
\cmidrule(r){2-2} \cmidrule(lr){3-3} \cmidrule(lr){4-4} \cmidrule(l){5-5}
\cmidrule(r){7-7} \cmidrule(lr){8-8} \cmidrule(lr){9-9} \cmidrule(l){10-10}
Sample: & \multicolumn{4}{c}{Preference-based questions} & &\multicolumn{4}{c}{Belief-based questions} \\
\cmidrule(r){1-1} \cmidrule{2-5} \cmidrule{7-10}
Large & --0.321*** & --0.331*** & 0.212 & 0.216 &  & 0.240*** & 0.239*** & --0.155** & --0.157** \\
 & (0.093) & (0.091) & (0.130) & (0.132) &  & (0.092) & (0.090) & (0.074) & (0.073) \\
LLM family FE & No & Yes & No & Yes &  & No & Yes & No & Yes \\
Observations & 4,750 & 4,750 & 4,750 & 4,750 &  & 8,000 & 8,000 & 8,000 & 8,000 \\
Pseudo $R$-squared & 0.081 & 0.153 & 0.033 & 0.066 &  & 0.054 & 0.144 & 0.029 & 0.117 \\

\hline
\hline
\end{tabularx}}
\end{table}
\end{landscape}

\begin{table}[!htbp] 
\caption{\bf LLM investment decisions: Experiment in~\cite{bose2022}.}
This table reports linear regression results that relate investment amounts elicited from LLMs---measured as the fraction of the endowed 1,000 monetary units that the LLM is willing to invest in a stock---to a broad set of explanatory variables. The key variables are VS, $\mathbb{C}\text{orr}$(Returns, VS weights), and their interaction with $\mathbbm{1}_{large}$, where VS is a weighted average of the stock's past daily returns, with  weights computed using the SAM algorithm described in~\cite{bose2022} that captures the visual attention paid to each return; $\mathbb{C}\text{orr}$(Returns, VS weights) is the correlation between past daily returns and their associated visual salience weights; and $\mathbbm{1}_{large}$ is an indicator for the large-scale models. All control variables are constructed following the procedure in~\cite{bose2022}. Robust standard errors are reported in parentheses. $^{***} p<0.01$, $^{**} p<0.05$ and $^{*} p<0.1$.
\label{tab:bose_investment_amount}
\begin{center}
\footnotesize
\vspace{-0.1cm}
\scalebox{0.92}{
\begin{tabular}{lcccccc}
   \tabularnewline \midrule \midrule
   Dep. var: & \multicolumn{6}{c}{Investment amount (\%)}\\
                                                  & \multicolumn{2}{c}{GPT} & \multicolumn{2}{c}{Claude} & \multicolumn{2}{c}{Gemini} \\ 
\cmidrule(r){2-3} \cmidrule(r){4-5} \cmidrule(r){6-7}                               & (1)                      & (2)                            & (3)                      & (4)                            & (5)                   & (6)\\  
   \midrule   
  
   VS                                             & --0.0105               &                             & --0.0002                  &                                & --0.0301$^{***}$          &   \\   
                                                  & (0.0065)              &                             & (0.0048)                 &                                & (0.0052)                 &   \\   
   CPT                                            & 0.0841$^{***}$        &                             & 0.0218$^{***}$           &                                & 0.0780$^{***}$           &   \\   
                                                  & (0.0081)              &                             & (0.0060)                 &                                & (0.0065)                 &   \\   
   Salience                                       & 0.0017                &                             & --0.0037                  &                                & --0.0249$^{***}$          &   \\   
                                                  & (0.0053)              &                             & (0.0043)                 &                                & (0.0048)                 &   \\     
   VS $\times$ $\mathbbm{1}_{large}$              & 0.0089                &                             & 0.0173$^{**}$            &                                & 0.0225$^{**}$            &   \\   
                                                  & (0.0073)              &                             & (0.0081)                 &                                & (0.0099)                 &   \\   
   CPT $\times$ $\mathbbm{1}_{large}$             & --0.0567$^{***}$       &                             & --0.0184$^{*}$            &                                & --0.0039                  &   \\   
                                                  & (0.0092)              &                             & (0.0103)                 &                                & (0.0123)                 &   \\   
   Salience $\times$ $\mathbbm{1}_{large}$        & --0.0098               &                             & 0.0131$^{*}$             &                                & 0.0166$^{*}$             &   \\   
                                                  & (0.0061)              &                             & (0.0074)                 &                                & (0.0087)                 &   \\   
   Corr(Returns, VS weights)                          &                       & --0.0315$^{***}$             &                          & --0.0072$^{***}$                &                          & --0.0280$^{***}$\\   
                                                  &                       & (0.0023)                    &                          & (0.0016)                       &                          & (0.0021)\\   
   Corr(Returns, CPT weights)                         &                       & 0.0255$^{***}$              &                          & 0.0091$^{***}$                 &                          & 0.0182$^{***}$\\   
                                                  &                       & (0.0044)                    &                          & (0.0033)                       &                          & (0.0039)\\   
   Corr(Returns, Salience weights)                    &                       & 0.0206$^{***}$              &                          & --0.0007                        &                          & --0.0008\\   
                                                  &                       & (0.0039)                    &                          & (0.0031)                       &                          & (0.0037)\\   
   Corr(Returns, VS weights) $\times$ $\mathbbm{1}_{large}$  & & 0.0260$^{***}$              &                          & 0.0082$^{***}$                 &                          & 0.0099$^{***}$\\   
                                                  &                       & (0.0025)                    &                          & (0.0031)                       &                          & (0.0030)\\   
   Corr(Returns, CPT weights) $\times$ $\mathbbm{1}_{large}$  & & --0.0085                     &                          & --0.0029                        &                          & 0.0224$^{***}$\\   
                                                  &                       & (0.0052)                    &                          & (0.0066)                       &                          & (0.0061)\\   
   Corr(Returns, Salience weights) $\times$ $\mathbbm{1}_{large}$    &                       & --0.0273$^{***}$             &                          & 0.0069                         &                          & --0.0002\\   
                                                  &                       & (0.0046)                    &                          & (0.0060)                       &                          & (0.0059)\\   
   Constant                                       & 0.5865$^{***}$        & 0.5865$^{***}$              & 0.4806$^{***}$           & 0.4806$^{***}$                 & 0.5990$^{***}$           & 0.5990$^{***}$\\
                                                  & (0.0015)              & (0.0015)                    & (0.0013)                 & (0.0013)                       & (0.0014)                 & (0.0014)\\   
   $\mathbbm{1}_{large}$                          & --0.0908$^{***}$       & --0.0908$^{***}$             & --0.1568$^{***}$          & --0.1568$^{***}$                & 0.0710$^{***}$           & 0.0710$^{***}$\\   
                                                  & (0.0018)              & (0.0017)                    & (0.0024)                 & (0.0024)                       & (0.0022)                 & (0.0022)\\ 
   \midrule
   Controls                                       & Yes                   & Yes                         & Yes                      & Yes                            & Yes                      & Yes\\ 
   Observations                                   & 8,000                 & 8,000                       & 8,000                    & 8,000                          & 8,000                    & 8,000\\
    Adjusted $R$-squared                                 & 0.717               & 0.722                     & 0.458                  & 0.457                        & 0.692                  & 0.695\\    
   \midrule \midrule
\end{tabular}}
\end{center}
\end{table}

\clearpage
\begin{landscape}
\begin{table}[htbp]
\caption{\bf Treatment effects of role-priming prompts.}
\label{table:role_priming}

This table reports marginal effects from a series of probit regressions. For Columns (1), (2), (5), and (6), the dependent variable is a binary variable that takes the value of one if an LLM response is classified as rational, and zero otherwise; for Columns (3), (4), (7), and (8), the dependent variable is a binary variable that takes the value of one if an LLM response is classified as human-like, and zero otherwise. Regressions in Columns (1) to (4) are for preference-based questions, and regressions in Columns (5) to (8) are for belief-based questions; each regression uses responses from all twelve LLMs. Panel A restricts the sample to LLM responses generated using either the baseline prompt or a treatment prompt that primes LLMs to be rational investors. Panel B restricts the sample to LLM responses generated using either the baseline prompt or a treatment prompt that primes LLMs to be real-world retail investors. For both panels, the key independent variable is an indicator for the treatment prompt, with the baseline prompt serving as the omitted baseline category. Standard errors, clustered at the question level, are reported in parentheses. $^{***}p<0.01$, $^{**}p<0.05$ and $^{*}p<0.1$.\\

\vspace{0.1in}

\footnotesize
\begin{tabularx}{1.4\textwidth}{lYYYYcYYYY}
\hline
\hline
\multicolumn{10}{l}{Panel A: Role-priming prompt (rational investor)}\\
\midrule
& (1) & (2) & (3) & (4) &  & (5) & (6) & (7) & (8) \\
\midrule
Dep. var:  & \multicolumn{9}{c}{LLM response is characterized as} \\
 & Rational & Rational & Human-like & Human-like &  & Rational & Rational & Human-like & Human-like \\
\cmidrule(r){2-2} \cmidrule(lr){3-3} \cmidrule(lr){4-4} \cmidrule(l){5-5}
\cmidrule(r){7-7} \cmidrule(lr){8-8} \cmidrule(lr){9-9} \cmidrule(l){10-10}
Sample: & \multicolumn{4}{c}{Preference-based questions} & &\multicolumn{4}{c}{Belief-based questions} \\
\cmidrule(r){1-1} \cmidrule{2-5} \cmidrule{7-10}
Role-priming prompt & 0.0439*** & 0.0430** & --0.0418* & --0.0405* &  & 0.0331* & 0.0325* & --0.0087 & --0.0067 \\
 & (0.017) & (0.017) & (0.021) & (0.021) &  & (0.019) & (0.019) & (0.025) & (0.024) \\
Model FE & No & Yes & No & Yes &  & No & Yes & No & Yes \\
Observations & 14,308 & 14,308 & 14,308 & 14,308 &  & 23,993 & 23,993 & 23,993 & 23,993 \\
Pseudo $R$-squared & 0.001 & 0.155 & 0.001 & 0.098 &  & 0.001 & 0.184 & 0.000 & 0.153 \\

\midrule
\\
\multicolumn{10}{l}{Panel B: Role-priming prompt (retail investor)}\\
\midrule
& (1) & (2) & (3) & (4) &  & (5) & (6) & (7) & (8) \\
\midrule
Dep. var:  & \multicolumn{9}{c}{LLM response is characterized as} \\
 & Rational & Rational & Human-like & Human-like &  & Rational & Rational & Human-like & Human-like \\
\cmidrule(r){2-2} \cmidrule(lr){3-3} \cmidrule(lr){4-4} \cmidrule(l){5-5}
\cmidrule(r){7-7} \cmidrule(lr){8-8} \cmidrule(lr){9-9} \cmidrule(l){10-10}
Sample: & \multicolumn{4}{c}{Preference-based questions} & &\multicolumn{4}{c}{Belief-based questions} \\
\cmidrule(r){1-1} \cmidrule{2-5} \cmidrule{7-10}
Role-priming prompt & --0.0361* & --0.0388** & 0.0150 & 0.0152 &  & 0.0010 & --0.0021 & 0.0052 & 0.0084 \\
 & (0.019) & (0.019) & (0.024) & (0.025) &  & (0.018) & (0.018) & (0.020) & (0.020) \\
Model FE & No & Yes & No & Yes &  & No & Yes & No & Yes \\
Observations & 14,310 & 14,310 & 14,310 & 14,310 &  & 23,999 & 23,999 & 23,999 & 23,999 \\
Pseudo $R$-squared & 0.001 & 0.165 & 0.000 & 0.101 &  & 0.000 & 0.163 & 0.000 & 0.143 \\
\hline
\hline
\end{tabularx}
\end{table}
\end{landscape}

\clearpage
\begin{landscape}
\begin{table}[htbp]
\caption{\bf Treatment effects of role-priming prompts: Mediation analysis.}
\label{table:mediation_role_priming}

This table reports marginal effects from a series of probit regressions that analyze the mechanisms through which role priming affects LLM responses. For Column (1), the dependent variable is an indicator for ``high confidence," which equals one if an LLM assigns a confidence level greater than 0.9 to its choice (the median confidence level in our sample). For Column (2), the dependent variable is an indicator for ``system 2 thinking," which equals one if the LLM selects reasoning type B. For Columns (3) to (5), the dependent variable is a binary variable that takes the value of one if an LLM response is classified as rational, and zero otherwise; for Columns (6) to (8), the dependent variable is a binary variable that takes the value of one if an LLM response is classified as human-like, and zero otherwise. All regressions include responses from all twelve LLMs but focus on preference-based questions only. Panel A restricts the sample to LLM responses generated using either the baseline prompt or a treatment prompt that primes LLMs to be rational investors. Panel B restricts the sample to LLM responses generated using either the baseline prompt or a treatment prompt that primes LLMs to be real-world retail investors. For both panels, key independent variables include an indicator for the treatment prompt (with the baseline prompt serving as the omitted category), a ``high confidence" indicator, and a ``system 2 thinking" indicator. Standard errors, clustered at the question level, are reported in parentheses. $^{***}p<0.01$, $^{**}p<0.05$ and $^{*}p<0.1$.\\

\vspace{0.05in}

\footnotesize
\scalebox{0.7}{
\begin{tabularx}{1.96\textwidth}{lYYYYYYYY}
\hline
\hline
\multicolumn{9}{l}{Panel A: Role-priming prompt (rational investor)}\\
\midrule
& (1) & (2) & (3) & (4) & (5) & (6) & (7) & (8) \\
\midrule
Dep. var:  & High confidence & System 2 thinking & \multicolumn{6}{c}{LLM response is characterized as} \\
& & & Rational & Rational & Rational & Human-like & Human-like & Human-like \\
\cmidrule(lr){4-4} \cmidrule(l){5-5} \cmidrule(lr){6-6}
\cmidrule(lr){7-7} \cmidrule(lr){8-8} \cmidrule(l){9-9}
Sample: & \multicolumn{8}{c}{Preference-based questions} \\
\cmidrule{1-9}
Role-priming prompt & 0.0656*** & 0.114*** & 0.0430** &  & 0.000718 & --0.0405* &  & 0.00417 \\
& (0.020) & (0.041) & (0.017) &  & (0.010) & (0.021) &  & (0.015) \\
High confidence & &  &  & 0.143 & 0.143 &  & --0.350* & --0.351* \\
&  &  &  & (0.142) & (0.141) &  & (0.187) & (0.187) \\
System $2$ thinking & &  &  & 0.400*** & 0.400*** &  & --0.201* & --0.202* \\
&  &  &  & (0.124) & (0.124) &  & (0.105) & (0.104) \\
Model FE & Yes & Yes & Yes & Yes & Yes & Yes & Yes & Yes \\
Observations & 11,911 & 13,108 & 14,308 & 14,307 & 14,307 & 14,308 & 14,307 & 14,307 \\
Pseudo $R$-squared & 0.166 & 0.278 & 0.155 & 0.217 & 0.217 & 0.098 & 0.165 & 0.165 \\

\midrule
\\
\multicolumn{9}{l}{Panel B: Role-priming prompt (retail investor)}\\
\midrule
& (1) & (2) & (3) & (4) & (5) & (6) & (7) & (8) \\
\midrule
Dep. var:  & High confidence & System 2 thinking & \multicolumn{6}{c}{LLM response is characterized as} \\
& & & Rational & Rational & Rational & Human-like & Human-like & Human-like \\
\cmidrule(lr){4-4} \cmidrule(l){5-5} \cmidrule(lr){6-6}
\cmidrule(lr){7-7} \cmidrule(lr){8-8} \cmidrule(l){9-9}
Sample: & \multicolumn{8}{c}{Preference-based questions} \\
\cmidrule{1-9}
Role-priming prompt & --0.0601** & --0.0797** & --0.0388** &  & --0.00529 & 0.0152 &  & --0.0182 \\
 & (0.031) & (0.033) & (0.019) &  & (0.025) & (0.025) &  & (0.034) \\
High confidence &  &  &  & 0.322*** & 0.321** &  & --0.581*** & --0.583*** \\
 &  &  &  & (0.123) & (0.125) &  & (0.180) & (0.182) \\
System 2 thinking &  &  &  & 0.377** & 0.376** &  & --0.199 & --0.202 \\
 &  &  &  & (0.156) & (0.156) &  & (0.129) & (0.129) \\
Model FE & Yes & Yes & Yes & Yes & Yes & Yes & Yes & Yes \\
Observations & 10,728 & 14,310 & 14,310 & 14,310 & 14,310 & 14,310 & 14,310 & 14,310 \\
Pseudo $R$-squared & 0.135 & 0.301 & 0.165 & 0.256 & 0.256 & 0.101 & 0.199 & 0.199 \\
\hline
\hline
\end{tabularx}
}
\end{table}
\end{landscape}

\clearpage
\begin{table}[htbp!]
\caption{\bf Comparison of debiasing techniques: Prospect theory-related questions.}
\label{table:debiasing}

This table reports marginal effects from a series of probit regressions. For Columns (1) and (2), the dependent variable is a binary variable that takes the value of one if an LLM response is classified as rational, and zero otherwise; for Columns (3) and (4), the dependent variable is a binary variable that takes the value of one if an LLM response is classified as human-like, and zero otherwise. Regressions are estimated using LLM responses to prospect theory-related questions only; each regression uses responses from all twelve LLMs. Panel A restricts the sample to LLM responses generated using either the baseline prompt (omitted category) or a treatment prompt that primes LLMs to be rational investors. Panel B restricts the sample to LLM responses generated using either the rational-investor role-priming prompt (omitted category) or an instruction-based prompt that combines the sentence that primes LLMs to be rational investors with a detailed four-step procedure that guides LLMs to choose actions rationally. Panel C restricts the sample to the LLM responses generated using either the rational-investor role-priming prompt (omitted category) or a knowledge-enrichment prompt that combines the sentence that primes LLMs to be rational investors with a summary of the key findings from~\cite{kahneman1979prospect} that describes systematic human biases. Standard errors, clustered at the question level, are reported in parentheses. $^{***}p<0.01$, $^{**}p<0.05$ and $^{*}p<0.1$.\\

\vspace{0.1in}

\footnotesize

\scalebox{0.72}{
\begin{tabularx}{1.37\linewidth}{lYYYY}
\hline
\hline
\multicolumn{5}{l}{Panel A: Role-priming prompt (rational investor)}\\
\midrule
& (1) & (2) & (3) & (4) \\
\midrule
Dep. var:  & \multicolumn{4}{c}{LLM response is characterized as} \\
& Rational & Rational & Human-like & Human-like \\
\cmidrule(r){2-2}\cmidrule(lr){3-3} \cmidrule(lr){4-4}
\cmidrule(l){5-5}
Sample: & \multicolumn{4}{c}{Prospect theory-related questions} \\
\midrule
Role-priming prompt & 0.0375*** & 0.0401*** & --0.0225** & --0.0267** \\
 & (0.007) & (0.007) & (0.011) & (0.012) \\
Model FE & No & Yes & No & Yes \\
Observations & 7,195 & 7,195 & 7,195 & 6,595 \\
Pseudo $R$-squared & 0.001 & 0.231 & 0.001 & 0.150 \\
\midrule
\\
\multicolumn{5}{l}{Panel B: Instruction-based prompt}\\
\midrule
 & (1) & (2) & (3) & (4) \\
\midrule
Dep. var:  & \multicolumn{4}{c}{LLM response is characterized as} \\
& Rational & Rational & Human-like & Human-like \\
\cmidrule(r){2-2}\cmidrule(lr){3-3} \cmidrule(lr){4-4}
\cmidrule(l){5-5}
Sample: & \multicolumn{4}{c}{Prospect theory-related questions} \\
\midrule
Instruction-based prompt & --0.0987 & --0.0985 & 0.0836 & 0.0874 \\
 & (0.078) & (0.074) & (0.085) & (0.087) \\
Model FE & No & Yes & No & Yes \\
Observations & 7,195 & 7,195 & 7,195 & 6,595 \\
Pseudo $R$-squared & 0.007 & 0.241 & 0.008 & 0.211 \\
\midrule
\\
\multicolumn{5}{l}{Panel C: Knowledge-enrichment prompt}\\
\midrule
 & (1) & (2) & (3) & (4) \\
\midrule
Dep. var:  & \multicolumn{4}{c}{LLM response is characterized as} \\
& Rational & Rational & Human-like & Human-like \\
\cmidrule(r){2-2}\cmidrule(lr){3-3} \cmidrule(lr){4-4}
\cmidrule(l){5-5}
Sample: & \multicolumn{4}{c}{Prospect theory-related questions} \\
\midrule
Knowledge-enrichment prompt &  --0.291*** & --0.287*** & 0.203* & 0.204* \\
 & (0.055) & (0.052) & (0.112) & (0.106) \\
Model FE & No & Yes & No & Yes \\
Observations & 7,191 & 7,191 & 7,191 & 7,191 \\
Pseudo $R$-squared & 0.070 & 0.262 & 0.038 & 0.149 \\
\hline
\hline
\end{tabularx}
}

\end{table}

\clearpage

\appendix
\setcounter{page}{1}

\section*{\centering Internet Appendix for \\
    ``Behavioral Economics of AI: LLM Biases and Corrections"}
\begin{center}
    PIETRO BINI, LIN WILLIAM CONG, XING HUANG, and LAWRENCE J. JIN
\end{center}

\vspace{0.1in}
This Internet Appendix contains the following two sections: 
\vspace{0.1in}
\begin{itemize}
    \item Section I: Additional Figures
    \item Section II: Prompt Design
\end{itemize}


\renewcommand{\thetable}{IA.\Roman{table}}
\renewcommand{\thefigure}{IA.\arabic{figure}}
\setcounter{table}{0}
\setcounter{figure}{0}
\begin{center}
    \subsection*{\normalfont \textbf{I. Additional Figures}}
\end{center}
\begin{figure}[H]

	\begin{center}
    \subfloat[\hspace{6em} Panel A: Preference-based questions]{\includegraphics[scale=0.18]{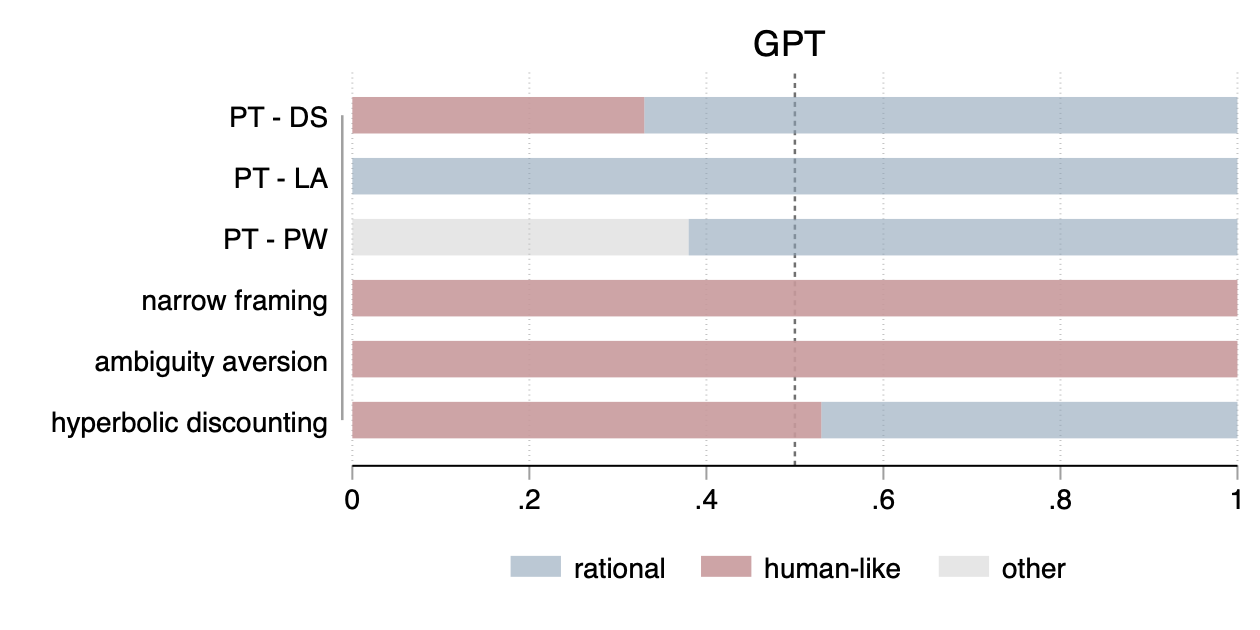}}
    \subfloat[\hspace{6em} Panel B: Belief-based questions]{\includegraphics[scale=0.18]{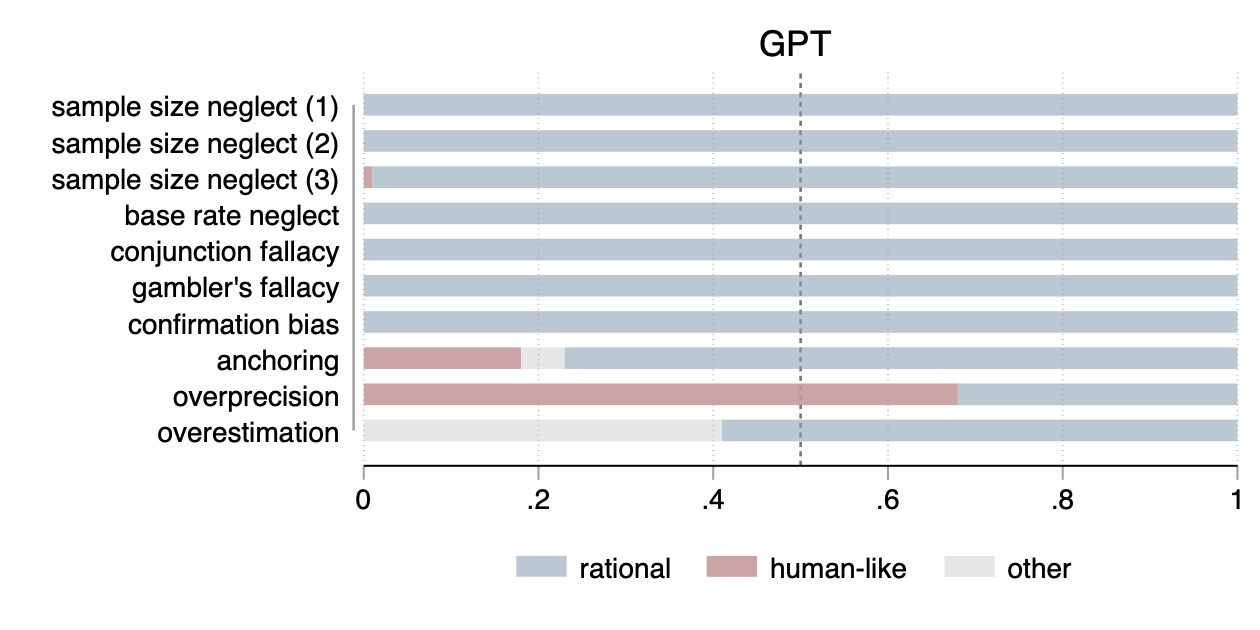}}\\
    \subfloat{\includegraphics[scale=0.18]{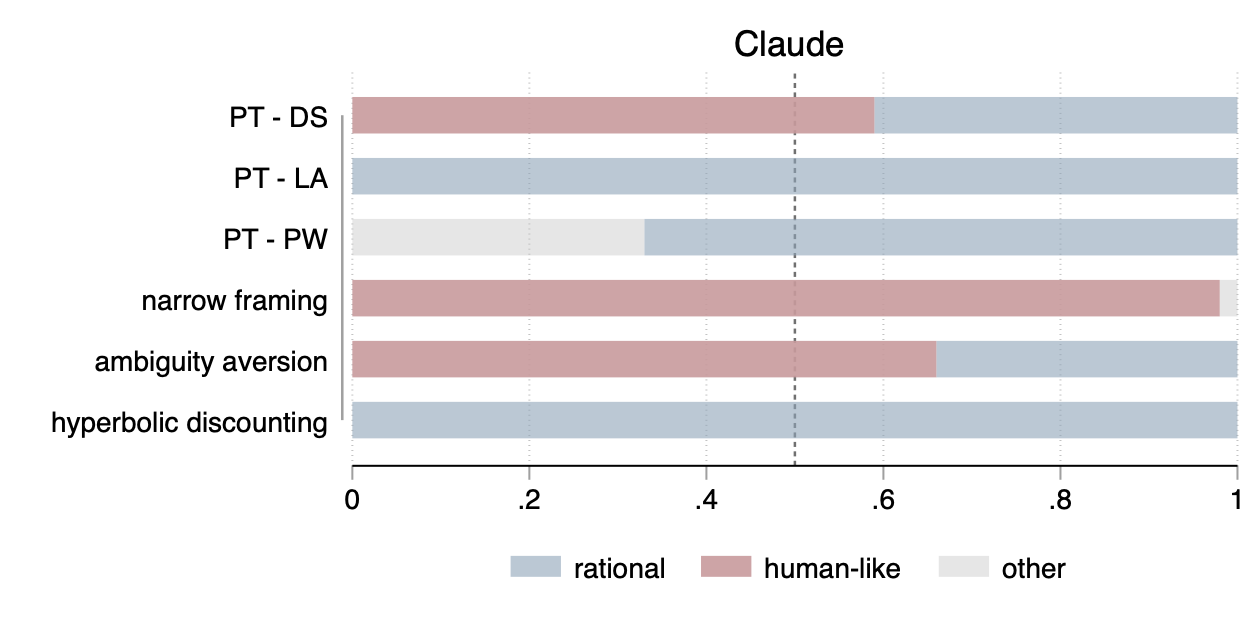}}
	\subfloat{\includegraphics[scale=0.18]{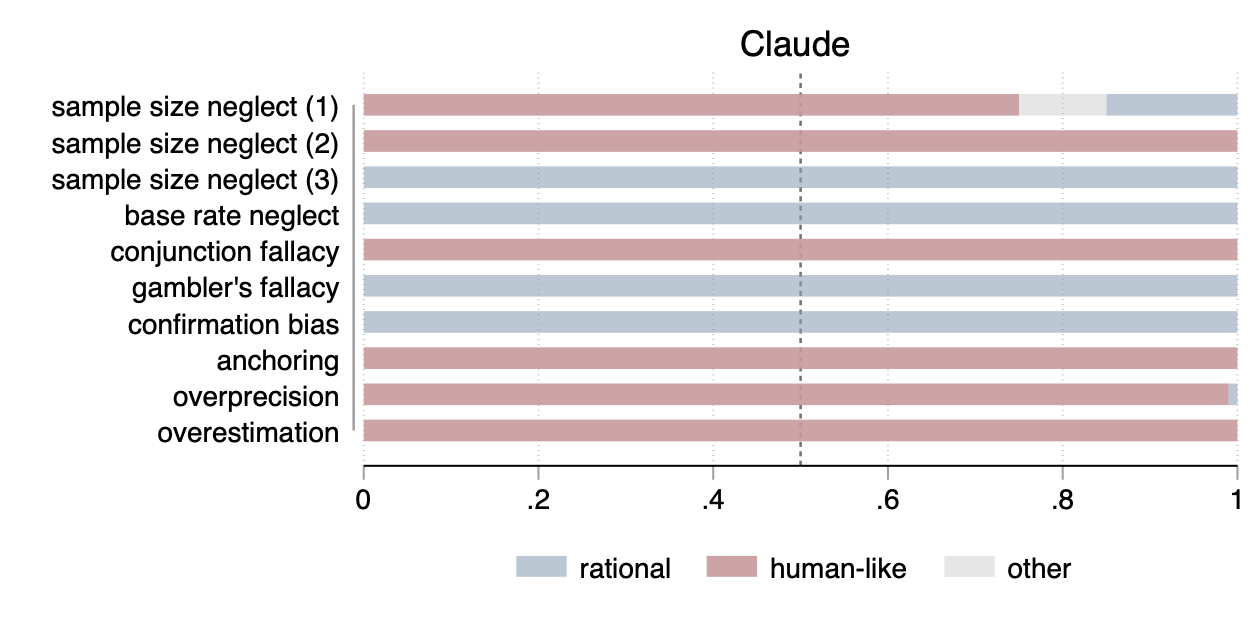}}\\
    \subfloat{\includegraphics[scale=0.18]{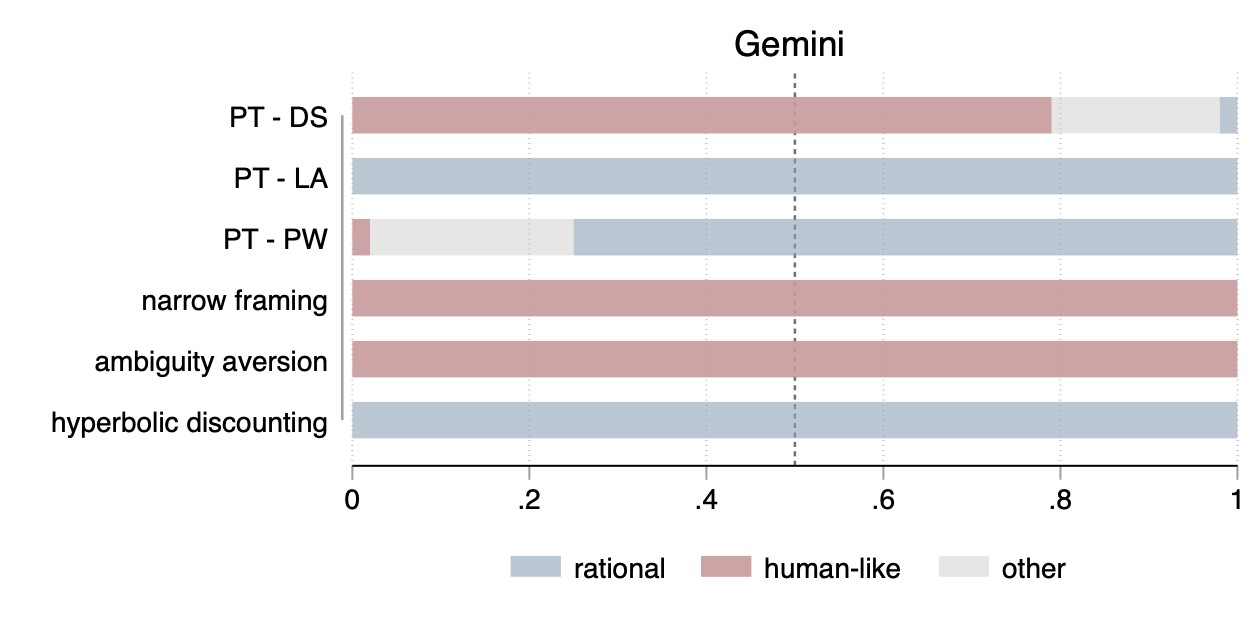}}
    \subfloat{\includegraphics[scale=0.18]{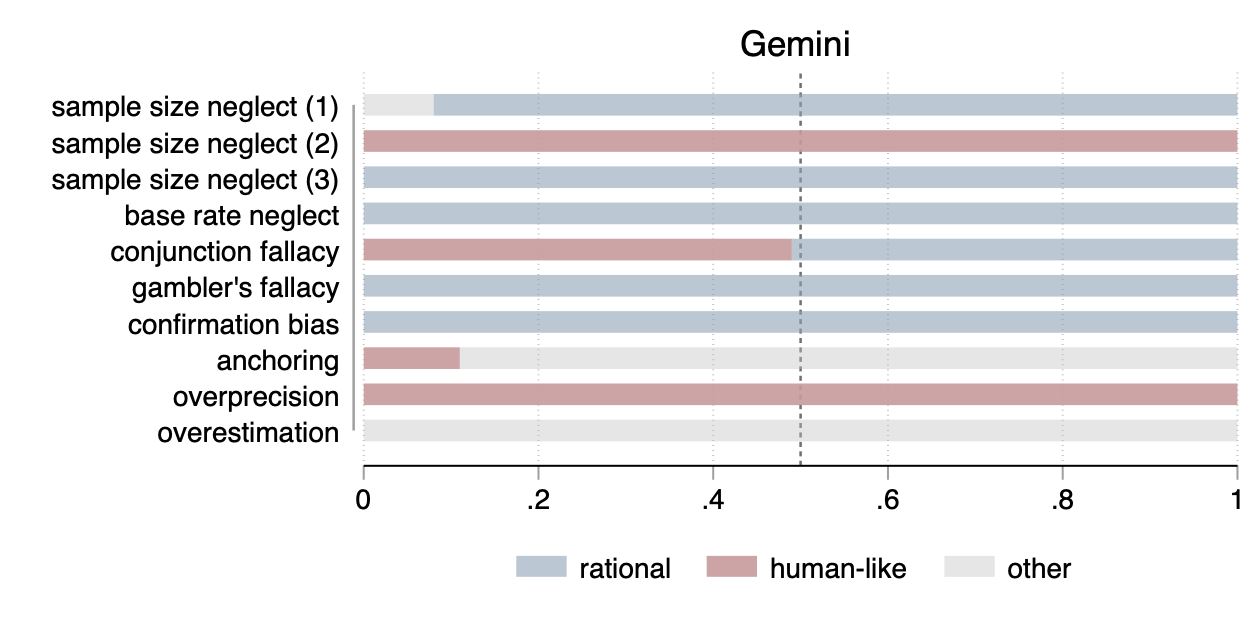}}\\
    \subfloat{\includegraphics[scale=0.18]{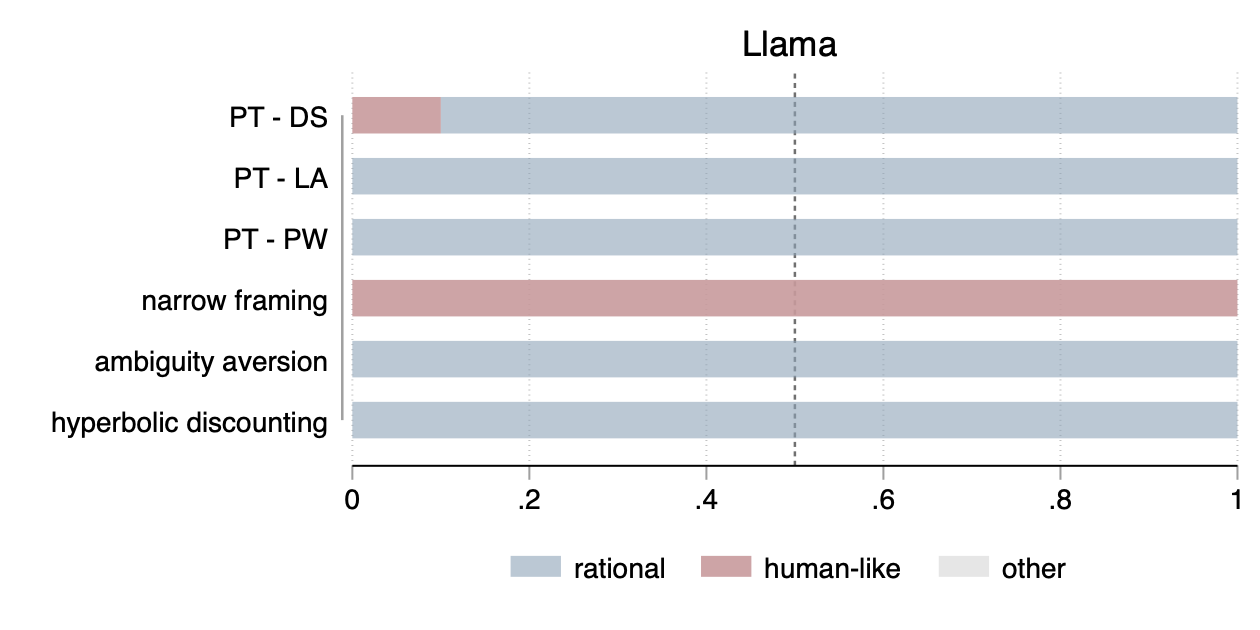}}
	\subfloat{\includegraphics[scale=0.18]{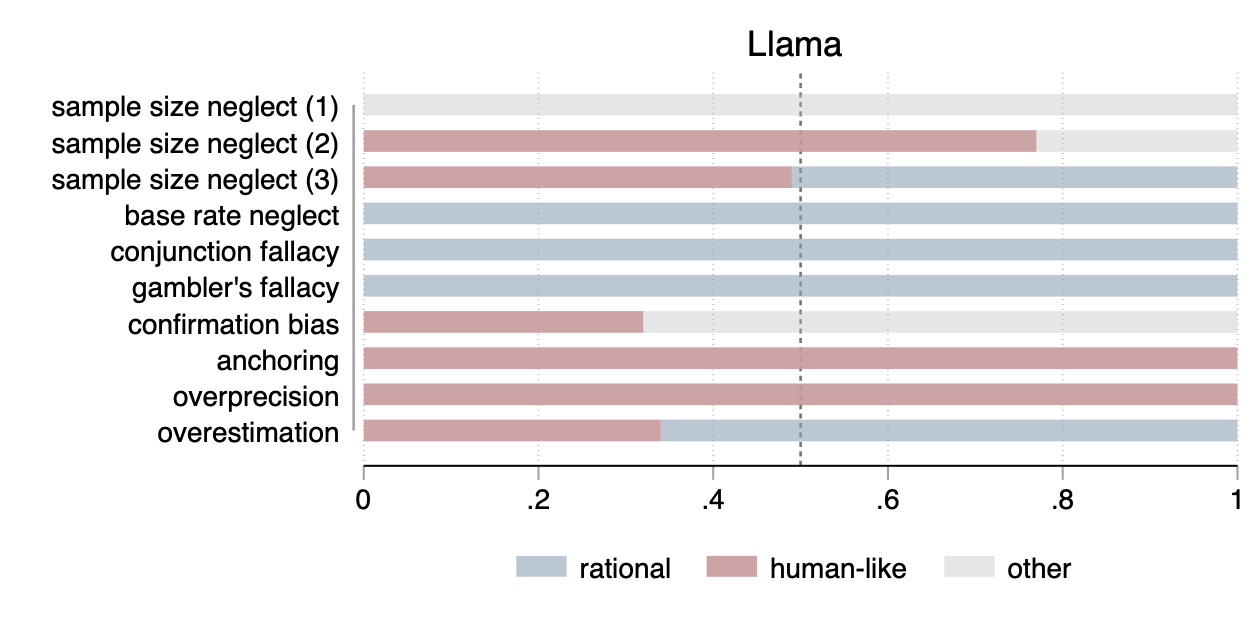}}
	\end{center}
    \caption{\textbf{Proportion of LLM responses: Advanced small-scale models.}}

    \vspace{0.2in}
This figure plots the proportions of LLM responses categorized as rational (blue), human-like (red), or other (gray), for the four advanced small-scale LLMs: GPT-4o, Claude 3 Haiku, Gemini 1.5 Flash, and Llama 3 8B. The left panel presents results for the six preference-based questions, while the right panel presents results for the ten belief-based questions.
    \label{fig_small_advanced} 
\end{figure}

\clearpage 
\begin{figure}[H]

	\begin{center}
    \subfloat[\hspace{6em} Panel A: Preference-based questions]{\includegraphics[scale=0.18]{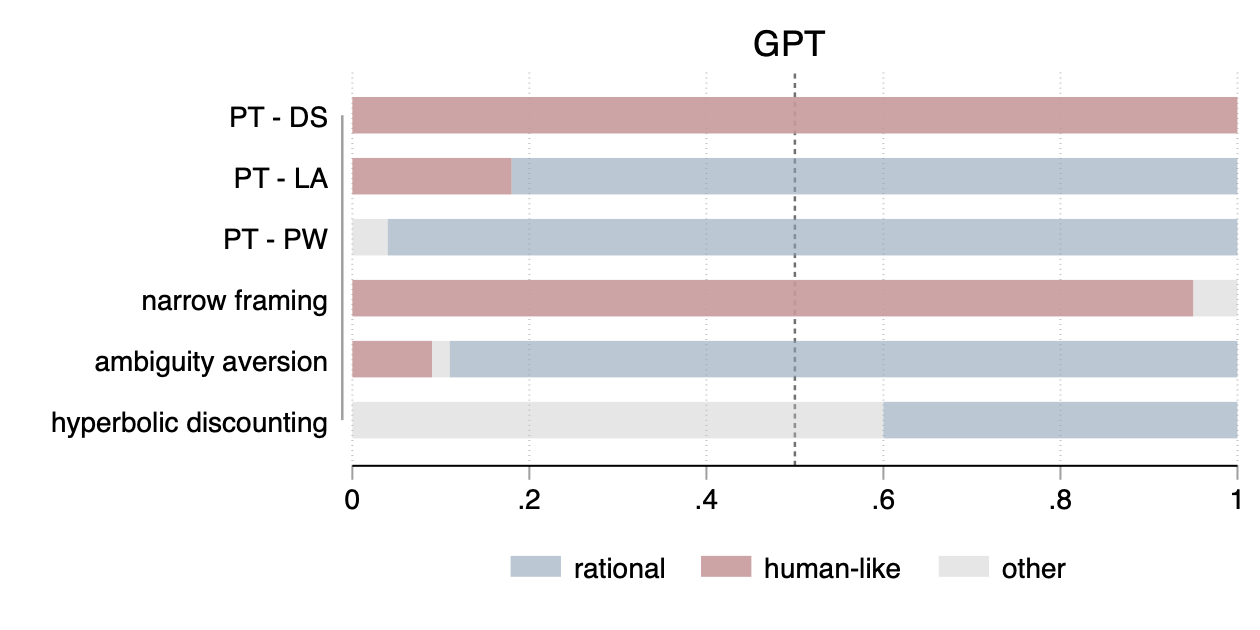}}
    \subfloat[\hspace{6em} Panel B: Belief-based questions]{\includegraphics[scale=0.18]{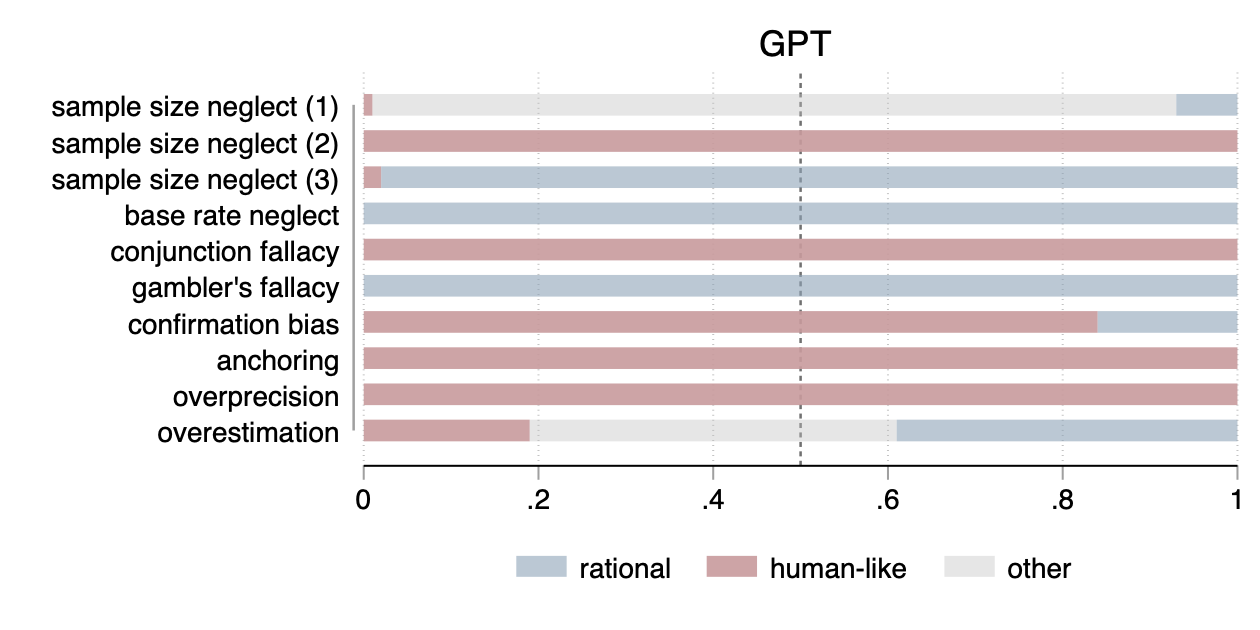}}\\
    \subfloat{\includegraphics[scale=0.18]{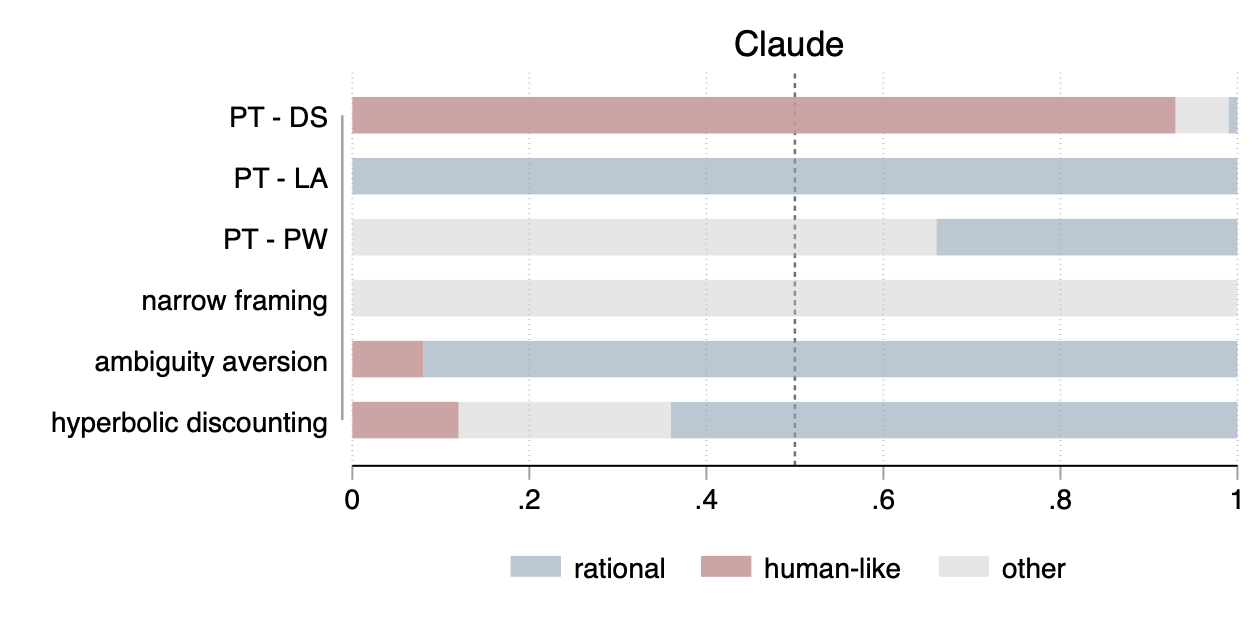}}
	\subfloat{\includegraphics[scale=0.18]{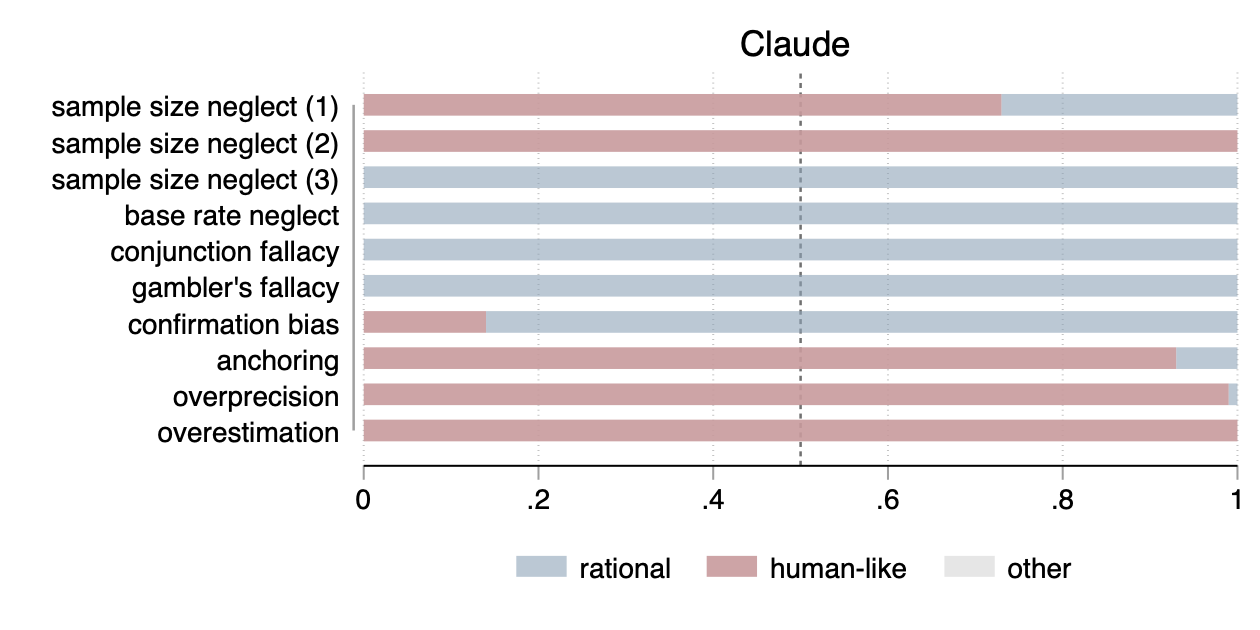}}\\
    \subfloat{\includegraphics[scale=0.18]{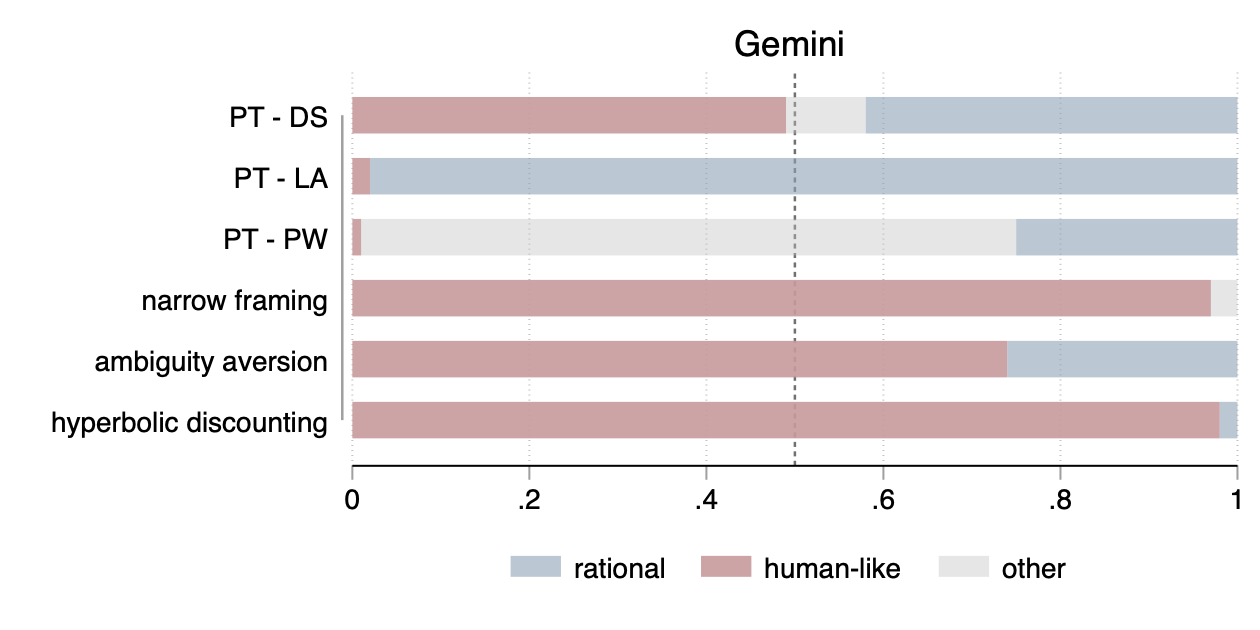}}
    \subfloat{\includegraphics[scale=0.18]{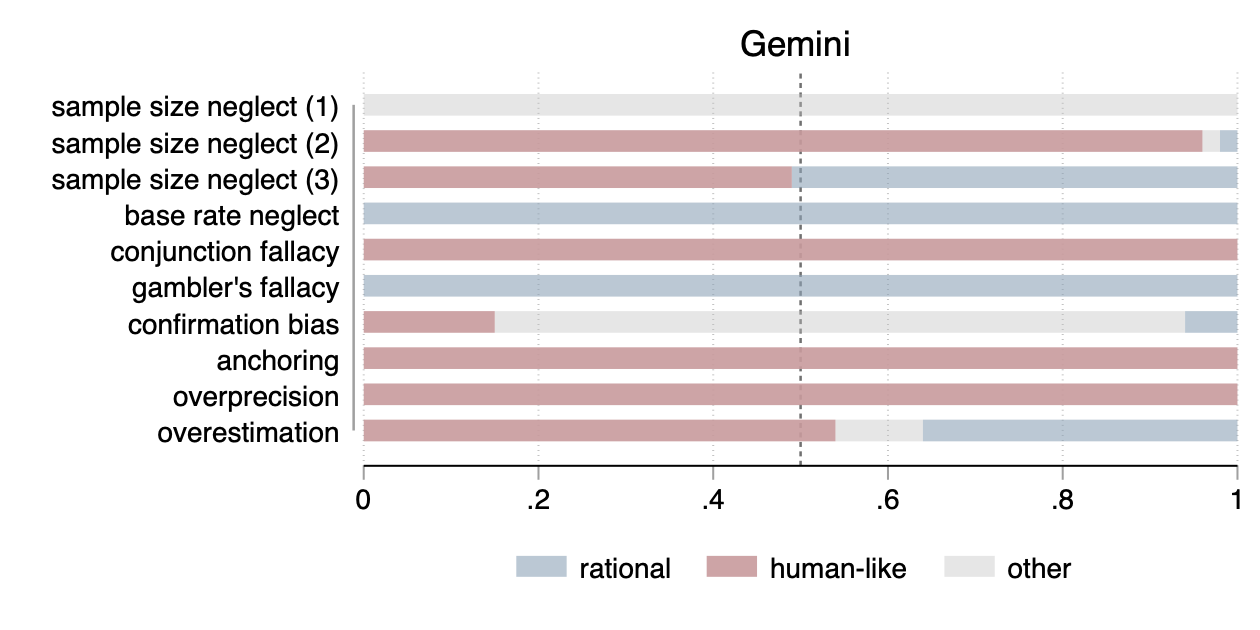}}\\
    \subfloat{\includegraphics[scale=0.18]{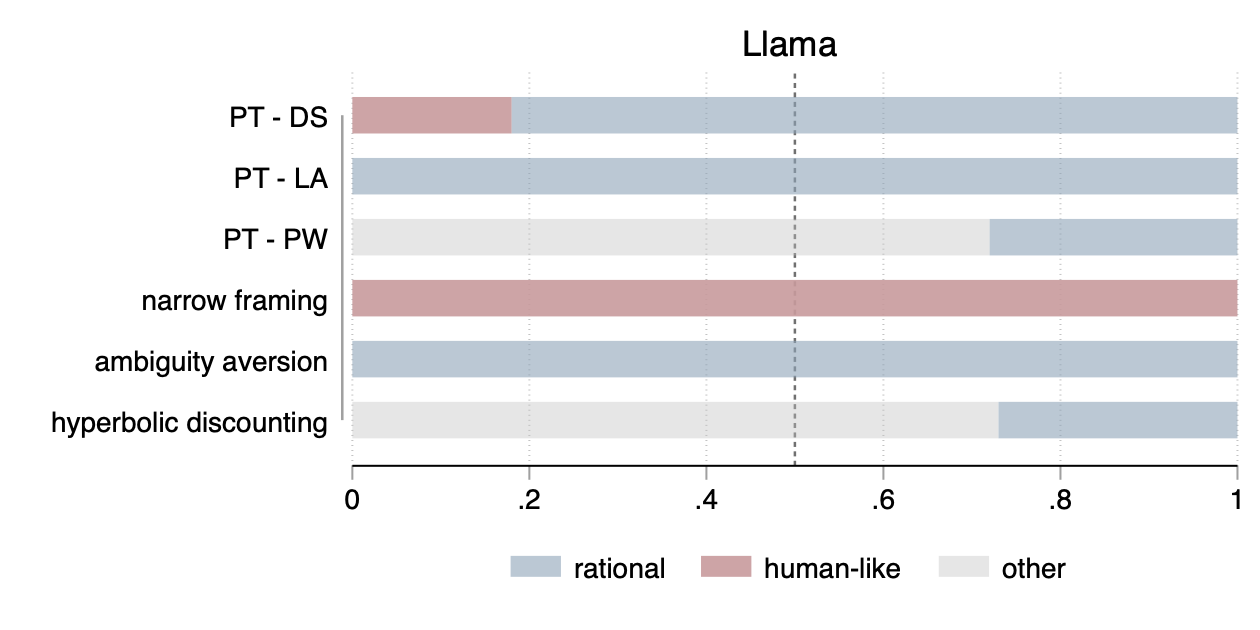}}
	\subfloat{\includegraphics[scale=0.18]{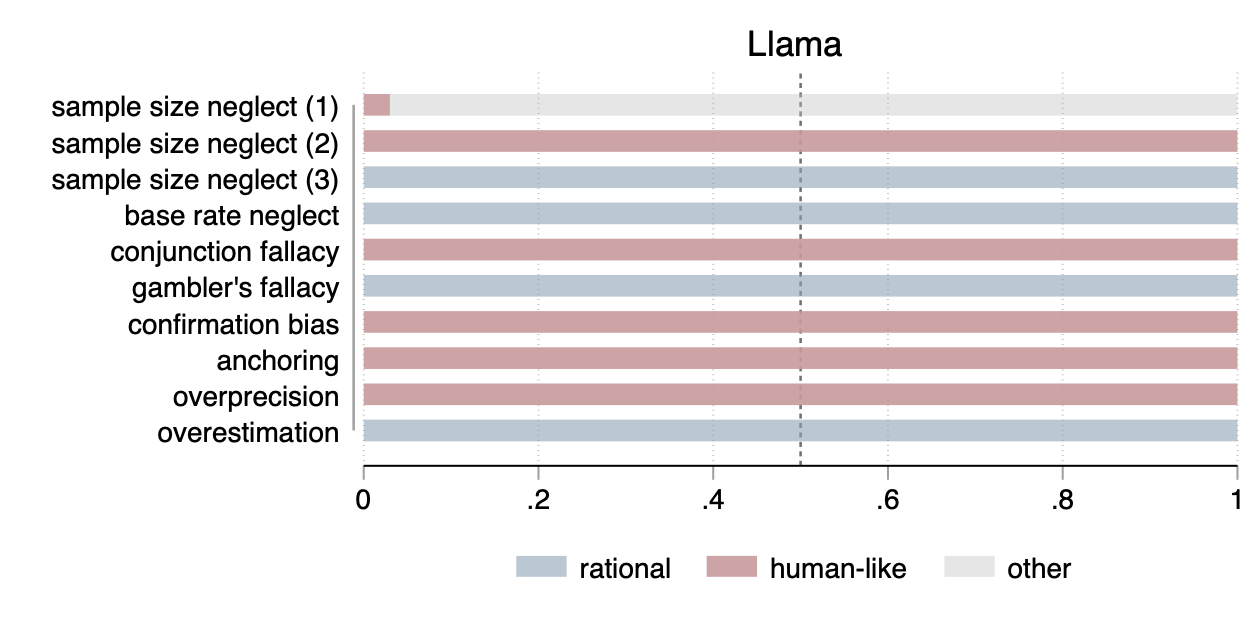}}
	\end{center}
\caption{\textbf{Proportion of LLM responses: Older models.}}

\vspace{0.2in}
This figure plots the proportions of LLM responses categorized as rational (blue), human-like (red), or other (gray), for the four older versions of LLMs: GPT-3.5 Turbo, Claude 2, Gemini 1.0 Pro, and Llama 2 70B. The left panel presents results for the six preference-based questions, while the right panel presents results for the ten belief-based questions.
	\label{fig_older} 	
\end{figure}

\clearpage
\begin{center}
    \subsection*{\normalfont \textbf{II. Prompt Design}}
\end{center}
\renewcommand{\thetable}{IB.\Roman{table}}
\renewcommand{\thefigure}{IB.\arabic{figure}}
\setcounter{table}{0}
\setcounter{figure}{0}

\begin{figure}[!htbp]
\scalebox{0.95}{
    \colorbox{blue!10}{
        \begin{minipage}{\textwidth}
Instructions: \\
Consider the following scenarios and respond according to the template provided. Please treat each scenario as completely separate from the other. The output should be a markdown code snippet formatted in the following schema, including the leading and trailing ``\texttt{```json}" and ``\texttt{```}" and should not include any note or comment: \\
\texttt{```json} \\
\{ \\
\hspace{0.5cm} \texttt{"Scenario A": \{} \\
\hspace{1cm} \texttt{"Choice": string,} \\
\hspace{1cm} \texttt{"Confidence": float,} \\
\hspace{1cm} \texttt{"Explanation": string,} \\
\hspace{1cm} \texttt{"Reasoning": string} \\
\hspace{0.5cm} \texttt{\},} \\
\hspace{0.5cm} \texttt{"Scenario B": \{} \\
\hspace{1cm} \texttt{"Choice": string,} \\
\hspace{1cm} \texttt{"Confidence": float,} \\
\hspace{1cm} \texttt{"Explanation": string,} \\
\hspace{1cm} \texttt{"Reasoning": string} \\
\hspace{0.5cm} \texttt{\}} \\
\} \\
\texttt{```} \\

Scenario A: \\
In addition to whatever you own, you have been given \$1,000. You now need to choose between the following two options: option A (\$1,000, 0.5), meaning winning \$1,000 with 0.5 probability and winning zero with 0.5 probability, versus option B (\$500), meaning winning \$500 with certainty. Please answer as shown above. Indicate the choice you prefer (``A" or ``B"), your confidence level (a number between 0 and 1), a brief explanation for your choice (in less than 50 words), and your reasoning type (``A" if your reasoning is based more on intuitive thinking, and ``B" if your reasoning is based more on analytical thinking and calculations).\\

\vspace{0.05in}

Scenario B: \\
Next, please consider a different scenario; please treat it as a completely separate scenario from the one you were just asked about. Specifically, please consider the following scenario. In addition to whatever you own, you have been given \$2,000. You now need to choose between the following two options: option A (--\$1,000, 0.5), meaning losing \$1,000 with 0.5 probability and losing zero with 0.5 probability, versus option B: (--\$500), meaning losing \$500 with certainty. Please answer as shown above. Indicate the choice you prefer (``A" or ``B"), your confidence level (a number between 0 and 1), a brief explanation for your choice (in less than 50 words), and your reasoning type (``A" if your reasoning is based more on intuitive thinking, and ``B" if your reasoning is based more on analytical thinking and calculations).
\end{minipage}
}}
\caption{\textbf{Prompt for Question 1: Diminishing sensitivity of prospect theory.}}

\vspace{0.2in}
This figure presents a prompt that elicits LLM responses to a question designed by~\cite{kahneman1979prospect} to document diminishing sensitivity, a key element of prospect theory.
    \label{prompt: q1}
\end{figure}
\begin{figure}[!htbp]
\scalebox{0.95}{
    \colorbox{blue!10}{
        \begin{minipage}{\textwidth}

Instructions: \\
Consider the following question and respond according to the template provided. The output should be a markdown code snippet formatted in the following schema, including the leading and trailing ``\texttt{```json}" and ``\texttt{```}" and should not include any note or comment: \\
\texttt{```json} \\
\{ \\
\hspace{1cm} \texttt{"Choice": string,} \\
\hspace{1cm} \texttt{"Confidence": float,} \\
\hspace{1cm} \texttt{"Explanation": string,} \\
\hspace{1cm} \texttt{"Reasoning": string} \\
\} \\
\texttt{```} \\

Question: \\
Would you accept or turn down a 50:50 bet to win \$110 or lose \$100? \\

\vspace{0.05in}

Response Format: \\
Please answer as shown above. Indicate the choice you prefer (``Accept" or ``Turn down"), your confidence level (a number between 0 and 1), a brief explanation for your choice (in less than 50 words), and your reasoning type (``A" if your reasoning is based more on intuitive thinking, and ``B" if your reasoning is based more on analytical thinking and calculations).

\end{minipage}
}}
\caption{\textbf{Prompt for Question 2: Loss aversion of prospect theory.}}

\vspace{0.2in}
This figure presents a prompt that elicits LLM responses to a question discussed in~\cite{barberis2003survey} that documents loss aversion, a key element of prospect theory.
    \label{prompt: q2}
\end{figure}
\begin{figure}[!htbp]
\scalebox{0.95}{
    \colorbox{blue!10}{
        \begin{minipage}{\textwidth}
Instructions: \\
Consider the following scenarios and respond according to the template provided. Please treat each scenario as completely separate from the other. The output should be a markdown code snippet formatted in the following schema, including the leading and trailing ``\texttt{```json}" and ``\texttt{```}" and should not include any note or comment: \\
\texttt{```json} \\
\{ \\
\hspace{0.5cm} \texttt{"Scenario A": \{} \\
\hspace{1cm} \texttt{"Choice": string,} \\
\hspace{1cm} \texttt{"Confidence": float,} \\
\hspace{1cm} \texttt{"Explanation": string,} \\
\hspace{1cm} \texttt{"Reasoning": string} \\
\hspace{0.5cm} \texttt{\},} \\
\hspace{0.5cm} \texttt{"Scenario B": \{} \\
\hspace{1cm} \texttt{"Choice": string,} \\
\hspace{1cm} \texttt{"Confidence": float,} \\
\hspace{1cm} \texttt{"Explanation": string,} \\
\hspace{1cm} \texttt{"Reasoning": string} \\
\hspace{0.5cm} \texttt{\}} \\
\} \\
\texttt{```} \\

Scenario A: \\
Please consider the following scenario. Choose between the following two options: option A (\$5,000, 0.001), meaning receiving \$5,000 with 0.001 probability and receiving zero with 0.999 probability, versus option B (\$5), meaning receiving \$5 with certainty. Please answer as shown above. Indicate the choice you prefer (``A" or ``B"), your confidence level (a number between 0 and 1), a brief explanation for your choice (in less than 50 words), and your reasoning type (``A" if your reasoning is based more on intuitive thinking, and ``B" if your reasoning is based more on analytical thinking and calculations). \\

\vspace{0.05in}

Scenario B: \\
Next, please consider a different scenario; please treat it as a completely separate scenario from the one you were just asked about. Specifically, choose between the following two options: option A (--\$5,000, 0.001), meaning losing \$5,000 with 0.001 probability and losing zero with 0.999 probability, versus option B (--\$5), meaning losing \$5 with certainty. Please answer as shown above. Indicate the choice you prefer (``A" or ``B"), your confidence level (a number between 0 and 1), a brief explanation for your choice (in less than 50 words), and your reasoning type (``A" if your reasoning is based more on intuitive thinking, and ``B" if your reasoning is based more on analytical thinking and calculations).
\end{minipage}
}}
\caption{\textbf{Prompt for Question 3: Probability weighting of prospect theory.}}

\vspace{0.2in}
This figure presents a prompt that elicits LLM responses to a question designed by~\cite{kahneman1979prospect} to document probability weighting, a key element of prospect theory.
    \label{prompt: q3}
\end{figure}
\begin{figure}[!htbp]
\scalebox{0.95}{
    \colorbox{blue!10}{
        \begin{minipage}{\textwidth}
Instructions: \\
Consider the following scenarios and respond according to the template provided. Please treat each scenario as completely separate from the other. The output should be a markdown code snippet formatted in the following schema, including the leading and trailing ``\texttt{```json}" and ``\texttt{```}" and should not include any note or comment: \\
\texttt{```json} \\
\{ \\
\hspace{0.5cm} \texttt{"Scenario A": \{} \\
\hspace{1cm} \texttt{"Choice": string,} \\
\hspace{1cm} \texttt{"Confidence": float,} \\
\hspace{1cm} \texttt{"Explanation": string,} \\
\hspace{1cm} \texttt{"Reasoning": string} \\
\hspace{0.5cm} \texttt{\},} \\
\hspace{0.5cm} \texttt{"Scenario B": \{} \\
\hspace{1cm} \texttt{"Choice": string,} \\
\hspace{1cm} \texttt{"Confidence": float,} \\
\hspace{1cm} \texttt{"Explanation": string,} \\
\hspace{1cm} \texttt{"Reasoning": string} \\
\hspace{0.5cm} \texttt{\}} \\
\} \\
\texttt{```} \\

Scenario A: \\
Please consider the following scenario. Imagine that you are about to purchase a jacket for \$125 and a calculator for \$15. The calculator salesman informs you that the calculator you wish to buy is on sale for \$10 at the other branch of the store, located 5 minutes drive away. Would you make the trip to the other store? Please answer as shown above. Indicate the choice you prefer  (``Yes" making the trip or "No" not making the trip), your confidence level (a number between 0 and 1), a brief explanation for your choice (in less than 50 words), and your reasoning type (``A" if your reasoning is based more on intuitive thinking, and ``B" if your reasoning is based more on analytical thinking and calculations). \\

\vspace{0.05in}

Scenario B: \\
Next, please consider a different scenario; please treat it as a completely separate scenario from the one you were just asked about. Specifically, imagine that you are about to purchase a jacket for \$15 and a calculator for \$125. The calculator salesman informs you that the calculator you wish to buy is on sale for \$120 at the other branch of the store, located 5 minutes drive away. Would you make the trip to the other store? Please answer as shown above. Indicate the choice you prefer  (``Yes" making the trip or ``No" not making the trip), your confidence level (a number between 0 and 1), a brief explanation for your choice (in less than 50 words), and your reasoning type (``A" if your reasoning is based more on intuitive thinking, and ``B" if your reasoning is based more on analytical thinking and calculations).
\end{minipage}
}}
\caption{\textbf{Prompt for Question 4: Narrow framing.}}

\vspace{0.2in}
This figure presents a prompt that elicits LLM responses to a question designed by~\cite{tversky1981framing} to document narrow framing. The original question from~\cite{tversky1981framing} writes ``20 minutes" for both Scenario A and Scenario B; our paper adjusts it to ``5 minutes" to account for inflation between 1981 and 2024 (when we collected our data).
    \label{prompt: q4}
\end{figure}
\begin{figure}[!htbp]
\scalebox{0.95}{
    \colorbox{blue!10}{
        \begin{minipage}{\textwidth}
Instructions: \\
Consider the following scenarios and respond according to the template provided. Please treat each scenario as completely separate from the other. The output should be a markdown code snippet formatted in the following schema, including the leading and trailing ``\texttt{```json}" and ``\texttt{```}" and should not include any note or comment: \\
\texttt{```json} \\
\{ \\
\hspace{0.5cm} \texttt{"Scenario A": \{} \\
\hspace{1cm} \texttt{"Choice": string,} \\
\hspace{1cm} \texttt{"Confidence": float,} \\
\hspace{1cm} \texttt{"Explanation": string,} \\
\hspace{1cm} \texttt{"Reasoning": string} \\
\hspace{0.5cm} \texttt{\},} \\
\hspace{0.5cm} \texttt{"Scenario B": \{} \\
\hspace{1cm} \texttt{"Choice": string,} \\
\hspace{1cm} \texttt{"Confidence": float,} \\
\hspace{1cm} \texttt{"Explanation": string,} \\
\hspace{1cm} \texttt{"Reasoning": string} \\
\hspace{0.5cm} \texttt{\}} \\
\} \\
\texttt{```} \\

Scenario A: \\
Please consider the following scenario. Choose between the following two options: option A, receiving \$100 today, versus option B, receiving \$110 tomorrow. Please answer as shown above. Indicate the choice you prefer (``A" or ``B"), your confidence level (a number between 0 and 1), a brief explanation for your choice (in less than 50 words), and your reasoning type (``A" if your reasoning is based more on intuitive thinking, and ``B" if your reasoning is based more on analytical thinking and calculations). \\

\vspace{0.05in}

Scenario B: \\
Next, please consider a different scenario; please treat it as a completely separate scenario from the one you were just asked about. Specifically, choose between the following two options: option A, receiving \$100 in 30 days, versus option B, receiving \$110 in 31 days. Please answer as shown above. Indicate the choice you prefer (``A" or ``B"), your confidence level (a number between 0 and 1), a brief explanation for your choice (in less than 50 words), and your reasoning type (``A" if your reasoning is based more on intuitive thinking, and ``B" if your reasoning is based more on analytical thinking and calculations).
\end{minipage}
}}
\caption{\textbf{Prompt for Question 5: Hyperbolic discounting.}}

\vspace{0.2in}
This figure presents a prompt that elicits LLM responses to a question discussed in~\cite*{frederick2002} that documents hyperbolic discounting.
    \label{prompt: q5}
\end{figure}
\begin{figure}[!htbp]
\scalebox{0.95}{
    \colorbox{blue!10}{
        \begin{minipage}{\textwidth}
Instructions: \\
Consider the following scenarios and respond according to the template provided. Please treat each scenario as completely separate from the other. The output should be a markdown code snippet formatted in the following schema, including the leading and trailing ``\texttt{```json}" and ``\texttt{```}" and should not include any note or comment: \\
\texttt{```json} \\
\{ \\
\hspace{0.5cm} \texttt{"Scenario A": \{} \\
\hspace{1cm} \texttt{"Choice": string,} \\
\hspace{1cm} \texttt{"Confidence": float,} \\
\hspace{1cm} \texttt{"Explanation": string,} \\
\hspace{1cm} \texttt{"Reasoning": string} \\
\hspace{0.5cm} \texttt{\},} \\
\hspace{0.5cm} \texttt{"Scenario B": \{} \\
\hspace{1cm} \texttt{"Choice": string,} \\
\hspace{1cm} \texttt{"Confidence": float,} \\
\hspace{1cm} \texttt{"Explanation": string,} \\
\hspace{1cm} \texttt{"Reasoning": string} \\
\hspace{0.5cm} \texttt{\}} \\
\} \\
\texttt{```} \\

Scenario A: \\
Please consider the following scenario. There are two urns. Urn C contains 50 red balls and 50 black balls. Urn U contains 100 balls, each either red or black, but with unknown proportion of each color. Choose between the following two bets: R1: draw a ball from Urn C, get \$20 if red, and R2: draw a ball from Urn U, get \$20 if red. Please answer as shown above. Indicate the choice you prefer (``R1" or ``R2"), your confidence level (a number between 0 and 1), a brief explanation for your choice (in less than 50 words), and your reasoning type (``A" if your reasoning is based more on intuitive thinking, and ``B" if your reasoning is based more on analytical thinking and calculations). \\

\vspace{0.05in}

Scenario B: \\
Next, please consider an alternative scenario. Specifically, now choose between the following bets: B1: draw a ball from Urn C, get \$20 if black, and B2: draw a ball from Urn U, get \$20 if black. Please answer as shown above. Indicate the choice you prefer (``B1" or ``B2"), your confidence level (a number between 0 and 1), a brief explanation for your choice (in less than 50 words), and your reasoning type (``A" if your reasoning is based more on intuitive thinking, and ``B" if your reasoning is based more on analytical thinking and calculations).
\end{minipage}
}}
\caption{\textbf{Prompt for Question 6: Ambiguity aversion.}}

\vspace{0.2in}
This figure presents a prompt that elicits LLM responses to a question designed by~\cite{ellsberg1961} to document ambiguity aversion.
    \label{prompt: q6}
\end{figure}
\begin{figure}[!htbp]
\scalebox{0.95}{
    \colorbox{blue!10}{
        \begin{minipage}{\textwidth}

Instructions: \\
Consider the following question and respond according to the template provided. The output should be a markdown code snippet formatted in the following schema, including the leading and trailing ``\texttt{```json}" and ``\texttt{```}" and should not include any note or comment: \\
\texttt{```json} \\
\{ \\
\hspace{1cm} \texttt{"Choice": string,} \\
\hspace{1cm} \texttt{"Confidence": float,} \\
\hspace{1cm} \texttt{"Explanation": string,} \\
\hspace{1cm} \texttt{"Reasoning": string} \\
\} \\
\texttt{```} \\

Question: \\
A certain town is served by two hospitals. In the larger hospital about 45 babies are born each day, and in the smaller hospital about 15 babies are born each day. As you know, about 50 percent of all babies are boys. However, the exact percentage varies from day to day. Sometimes it may be higher than 50 percent, sometimes lower. For a period of 1 year, each hospital recorded the days on which more than 60 percent of the babies born were boys. \\
Which hospital do you think recorded more such days? \\
A: The larger hospital \\
B. The smaller hospital \\
C: About the same (that is, within 5 percent of each other)\\

\vspace{0.05in}

Response Format: \\
Please answer as shown above. Indicate the choice you prefer (``A", ``B" or ``C"), your confidence level (a number between 0 and 1), a brief explanation for your choice (in less than 50 words), and your reasoning type (``A" if your reasoning is based more on intuitive thinking, and ``B" if your reasoning is based more on analytical thinking and calculations).

\end{minipage}
}}
\caption{\textbf{Prompt for Question 7: Sample size neglect.}}

\vspace{0.2in}
This figure presents a prompt that elicits LLM responses to a question designed by~\cite{Tversky1974} to document sample size neglect.
    \label{prompt: q7}
\end{figure}
\begin{figure}[!htbp]
\scalebox{0.95}{
    \colorbox{blue!10}{
        \begin{minipage}{\textwidth}

Instructions: \\
Consider the following question and respond according to the template provided. The output should be a markdown code snippet formatted in the following schema, including the leading and trailing ``\texttt{```json}" and ``\texttt{```}" and should not include any note or comment: \\
\texttt{```json} \\
\{ \\
\hspace{1cm} \texttt{"Choice": string,} \\
\hspace{1cm} \texttt{"Confidence": float,} \\
\hspace{1cm} \texttt{"Explanation": string,} \\
\hspace{1cm} \texttt{"Reasoning": string} \\
\} \\
\texttt{```} \\

Question: \\
When they turn 18, American males must register for the draft at a local post office. In addition to other information, the height of each male is obtained. The national average height of 18-year-old males is 5 feet, 9 inches. \\
Every day for one year, 25 men registered at post office A and 100 men registered at post office B. At the end of each day, a clerk at each post office computed and recorded the average height of the men who had registered there that day. \\
Which would you expect to be true (choose one)? \\
A: The number of days on which the average height was 6 feet or more was greater for post office A than for post office B. \\
B: The number of days on which the average height was 6 feet or more was greater for post office B than for post office A. \\
C: There is no reason to expect that the number of days on which the average height was 6 feet or more was greater for one post office than for the other. \\

\vspace{0.05in}

Response Format: \\
Please answer as shown above. Indicate the choice you prefer (``A", ``B" or ``C"), your confidence level (a number between 0 and 1), a brief explanation for your choice (in less than 50 words), and your reasoning type (``A" if your reasoning is based more on intuitive thinking, and ``B" if your reasoning is based more on analytical thinking and calculations).

\end{minipage}
}}
\caption{\textbf{Prompt for Question 8: Sample size neglect.}}

\vspace{0.2in}
This figure presents a prompt that elicits LLM responses to a question designed by~\cite*{Well1990} to document sample size neglect.
    \label{prompt: q8}
\end{figure}
\begin{figure}[!htbp]
\scalebox{0.95}{
    \colorbox{blue!10}{
        \begin{minipage}{\textwidth}

Instructions: \\
Consider the following question and respond according to the template provided. The output should be a markdown code snippet formatted in the following schema, including the leading and trailing ``\texttt{```json}" and ``\texttt{```}" and should not include any note or comment: \\
\texttt{```json} \\
\{ \\
\hspace{1cm} \texttt{"Choice": string,} \\
\hspace{1cm} \texttt{"Confidence": float,} \\
\hspace{1cm} \texttt{"Explanation": string,} \\
\hspace{1cm} \texttt{"Reasoning": string} \\
\} \\
\texttt{```} \\

Question: \\
You are presented with two covered urns. Both of them contain a mixture of red and green beads. The number of beads is different in the two urns: the small one contains 10 beads altogether, and the large one contains 100 beads altogether. However, the percentage of red and green beads is the same in both urns. The sampling will proceed as follows: You draw a bead blindly from the urn, note its color, and replace it. You mix, draw blindly again, and note down the color again. This goes on to a total of 9 draws from the small urn, or 15 draws from the large urn. \\
In which case do you think your chances for guessing the majority color are better (choose one)? \\
A: The small urn that contains 10 beads. \\
B: The large urn that contains 100 beads. \\

\vspace{0.05in}

Response Format: \\
Please answer as shown above. Indicate the choice you prefer (``A", or ``B"), your confidence level (a number between 0 and 1), a brief explanation for your choice (in less than 50 words), and your reasoning type (``A" if your reasoning is based more on intuitive thinking, and ``B" if your reasoning is based more on analytical thinking and calculations).

\end{minipage}
}}
\caption{\textbf{Prompt for Question 9: Sample size neglect.}}

\vspace{0.2in}
This figure presents a prompt that elicits LLM responses to a question designed by~\cite{BarHillel1979} to document sample size neglect.
    \label{prompt: q9}
\end{figure}
\begin{figure}[!htbp]
\vspace{-.4in}
\scalebox{0.92}{
    \colorbox{blue!10}{
        \begin{minipage}{\textwidth}
Instructions: \\
Consider the following scenarios and respond according to the template provided. Please treat each scenario as completely separate from the other. The output should be a markdown code snippet formatted in the following schema, including the leading and trailing ``\texttt{```json}" and ``\texttt{```}" and should not include any note or comment: \\
\texttt{```json} \\
\{ \\
\hspace{0.5cm} \texttt{"Scenario A": \{} \\
\hspace{1cm} \texttt{"Probability": float,} \\
\hspace{1cm} \texttt{"Confidence": float,} \\
\hspace{1cm} \texttt{"Explanation": string,} \\
\hspace{1cm} \texttt{"Reasoning": string} \\
\hspace{0.5cm} \texttt{\},} \\
\hspace{0.5cm} \texttt{"Scenario B": \{} \\
\hspace{1cm} \texttt{"Probability": float,} \\
\hspace{1cm} \texttt{"Confidence": float,} \\
\hspace{1cm} \texttt{"Explanation": string,} \\
\hspace{1cm} \texttt{"Reasoning": string} \\
\hspace{0.5cm} \texttt{\}} \\
\} \\
\texttt{```} \\

Scenario A: \\
Please consider the following scenario. Consider the following description of Jack: \\
``Jack is a 45 year old man. He is married and has four children. He is generally conservative, careful, and ambitious. He shows no interest in political and social issues and spends most of his free time on his many hobbies which include home carpentry, sailing, and mathematical puzzles." \\
Please note that the above description was randomly drawn from a set of 100 descriptions consisting of 70 engineers and 30 lawyers. Given this description, what is the probability that Jack is one of the 70 engineers in the sample of 100? \\
Please answer as shown above. Indicate the probability that Jack is an engineer (a number between 0 and 1), your confidence level (a number between 0 and 1), a brief explanation for your choice (in less than 50 words), and your reasoning type (``A" if your reasoning is based more on intuitive thinking, and ``B" if your reasoning is based more on analytical thinking and calculations).\\

\vspace{0.05in}

Scenario B: \\
Next, please consider a different scenario; please treat it as a completely separate scenario from the one you were just asked about. Specifically, please consider the following description of Jack: \\
``Jack is a 45 year old man. He is married and has four children. He is generally conservative, careful, and ambitious. He shows no interest in political and social issues and spends most of his free time on his many hobbies which include home carpentry, sailing, and mathematical puzzles." \\
Please note that the above description was randomly drawn from a set of 100 descriptions consisting of 30 engineers and 70 lawyers. Given this description, what is the probability that Jack is one of the 30 engineers in the sample of 100? \\
Please answer as shown above. Indicate the probability that Jack is an engineer (a number between 0 and 1), your confidence level (a number between 0 and 1), a brief explanation for your choice (in less than 50 words), and your reasoning type (``A" if your reasoning is based more on intuitive thinking, and ``B" if your reasoning is based more on analytical thinking and calculations). \\
\end{minipage}
}}
\caption{\textbf{Prompt for Question 10: Base rate neglect.}}

\vspace{0.2in}
This figure presents a prompt that elicits LLM responses to a question designed by~\cite{kahneman1973prediction} to document base rate neglect.
    \label{prompt: q10}
\end{figure}
\begin{figure}[!htbp]
\scalebox{0.95}{
    \colorbox{blue!10}{
        \begin{minipage}{\textwidth}

Instructions: \\
Consider the following scenario and respond according to the template provided. The output should be a markdown code snippet formatted in the following schema, including the leading and trailing ``\texttt{```json}" and ``\texttt{```}" and should not include any note or comment: \\
\texttt{```json} \\
\{ \\
\hspace{1cm} \texttt{"Choice": string,} \\
\hspace{1cm} \texttt{"Confidence": float,} \\
\hspace{1cm} \texttt{"Explanation": string,} \\
\hspace{1cm} \texttt{"Reasoning": string} \\
\} \\
\texttt{```} \\

Scenario: \\
Please consider the following scenario. Linda is 31 years old, single, outspoken and very bright. She majored in philosophy. As a student, she was deeply concerned with issues of discrimination and social justice, and also participated in anti-nuclear demonstrations. \\
Which of the two statements is more probable? \\
A: Linda is a bank teller. \\
B. Linda is a bank teller and is active in the feminist movement. \\
Please answer as shown above. Indicate the choice you prefer (``A" or ``B"), your confidence level (a number between 0 and 1), a brief explanation for your choice (in less than 50 words), and your reasoning type (``A" if your reasoning is based more on intuitive thinking, and ``B" if your reasoning is based more on analytical thinking and calculations).

\end{minipage}
}}
\caption{\textbf{Prompt for Question 11: Conjunction fallacy.}}

\vspace{0.2in}
This figure presents a prompt that elicits LLM responses to a question designed by~\cite{Tversky1983} to document conjunction fallacy.
    \label{prompt: q11}
\end{figure}
\begin{figure}[!htbp]
\scalebox{0.95}{
    \colorbox{blue!10}{
        \begin{minipage}{\textwidth}

Instructions: \\
Consider the following scenario and respond according to the template provided. The output should be a markdown code snippet formatted in the following schema, including the leading and trailing ``\texttt{```json}" and ``\texttt{```}" and should not include any note or comment: \\
\texttt{```json} \\
\{ \\
\hspace{1cm} \texttt{"Probability": float,} \\
\hspace{1cm} \texttt{"Confidence": float,} \\
\hspace{1cm} \texttt{"Explanation": string,} \\
\hspace{1cm} \texttt{"Reasoning": string} \\
\} \\
\texttt{```} \\

Scenario: \\
Please consider the following scenario. Imagine you simulate the random outcome of tossing an unbiased coin 150 times in succession. Suppose the last coin toss gave you a head. What is the probability of getting a tail from the next coin toss? \\
Please answer as shown above. Indicate the probability of getting a tail from the next coin toss (a number between 0 and 1), your confidence level (a number between 0 and 1), a brief explanation for your choice (in less than 50 words), and your reasoning type (``A" if your reasoning is based more on intuitive thinking, and ``B" if your reasoning is based more on analytical thinking and calculations).

\end{minipage}
}}
\caption{\textbf{Prompt for Question 12: Gambler's fallacy.}}

\vspace{0.2in}
This figure presents a prompt that elicits LLM responses to a question discussed in~\cite{rabin2002inference} that documents gambler's fallacy.
    \label{prompt: q12}
\end{figure}
\begin{figure}[!htbp]
\scalebox{0.95}{
    \colorbox{blue!10}{
        \begin{minipage}{\textwidth}

Instructions: \\
Consider the following scenario and respond according to the template provided. The output should be a markdown code snippet formatted in the following schema, including the leading and trailing ``\texttt{```json}" and ``\texttt{```}" and should not include any note or comment: \\
\texttt{```json} \\
\{ \\
\hspace{1cm} \texttt{"Choice": string,} \\
\hspace{1cm} \texttt{"Confidence": float,} \\
\hspace{1cm} \texttt{"Explanation": string,} \\
\hspace{1cm} \texttt{"Reasoning": string} \\
\} \\
\texttt{```} \\

Scenario: \\
Please consider the following scenario. You are shown four cards, marked E, K, 4 and 7. Each card has a letter on one side and a number on the other. You are given the following rule: Every card with a vowel on one side has an even number on the other side. Which cards must you turn over to test whether the rule is true or false? \\
Please answer as shown above. Indicate the cards you turn over to test whether the rule is true or false (one or multiple from E, K, 4 and 7), your confidence level (a number between 0 and 1), a brief explanation for your choice (in less than 50 words), and your reasoning type (``A" if your reasoning is based more on intuitive thinking, and ``B" if your reasoning is based more on analytical thinking and calculations).

\end{minipage}
}}
\caption{\textbf{Prompt for Question 13: Confirmation bias.}}

\vspace{0.2in}
This figure presents a prompt that elicits LLM responses to a question discussed in~\cite{Wason_book_1972} that documents confirmation bias.
    \label{prompt: q13}
\end{figure}
\begin{figure}[!htbp]
\scalebox{0.95}{
    \colorbox{blue!10}{
        \begin{minipage}{\textwidth}
Instructions: \\
Consider the following scenarios and respond according to the template provided. Please treat each scenario as completely separate from the other. The output should be a markdown code snippet formatted in the following schema, including the leading and trailing ``\texttt{```json}" and ``\texttt{```}" and should not include any note or comment: \\
\texttt{```json} \\
\{ \\
\hspace{0.5cm} \texttt{"Scenario A": \{} \\
\hspace{1cm} \texttt{"Direction": string,} \\
\hspace{1cm} \texttt{"Estimate": float,} \\
\hspace{1cm} \texttt{"Confidence": float,} \\
\hspace{1cm} \texttt{"Explanation": string,} \\
\hspace{1cm} \texttt{"Reasoning": string} \\
\hspace{0.5cm} \texttt{\},} \\
\hspace{0.5cm} \texttt{"Scenario B": \{} \\
\hspace{1cm} \texttt{"Direction": string,} \\
\hspace{1cm} \texttt{"Estimate": float,} \\
\hspace{1cm} \texttt{"Confidence": float,} \\
\hspace{1cm} \texttt{"Explanation": string,} \\
\hspace{1cm} \texttt{"Reasoning": string} \\
\hspace{0.5cm} \texttt{\}} \\
\} \\
\texttt{```} \\

Scenario A: \\
Please consider the following scenario. Suppose your objective is to estimate the percentage of African countries in the United Nations. For estimating this quantity, consider that a number between 0 and 100 is drawn randomly by spinning a wheel of fortune. Suppose the number drawn is 10. \\
Please first indicate whether this number of 10 is higher or lower than your estimate on the percentage of African countries in the United Nations. Please then provide your estimate by moving upward or downward from this number of 10. \\
Please answer as shown above. Indicate the direction (``higher" if 10 is higher than your estimate, and ``lower" if 10 is lower than your estimate), your estimate by moving upward or downward from the randomly drawn number of 10 (a number between 0 an 100), your confidence level (a number between 0 and 1), a brief explanation for your choice (in less than 50 words), and your reasoning type (``A" if your reasoning is based more on intuitive thinking, and ``B" if your reasoning is based more on analytical thinking and calculations). \\
\end{minipage}
}}
\end{figure}

\begin{figure}[!htbp]
\scalebox{0.95}{
    \colorbox{blue!10}{
        \begin{minipage}{\textwidth}
Scenario B: \\
Next, please consider a different scenario; please treat it as a completely separate scenario from the one you were just asked about. Specifically, suppose your objective is to estimate the percentage of African countries in the United Nations. For estimating this quantity, consider that a number between 0 and 100 is drawn randomly by spinning a wheel of fortune. Suppose the number drawn is 65. \\
Please first indicate whether this number of 65 is higher or lower than your estimate on the percentage of African countries in the United Nations. Please then provide your estimate by moving upward or downward from this number of 65. \\
Please answer as shown above. Indicate the direction (``higher" if 65 is higher than your estimate, and ``lower" if 65 is lower than your estimate), your estimate by moving upward or downward from the randomly drawn number of 65 (a number between 0 an 100), your confidence level (a number between 0 and 1), a brief explanation for your choice (in less than 50 words), and your reasoning type (``A" if your reasoning is based more on intuitive thinking, and ``B" if your reasoning is based more on analytical thinking and calculations).
\end{minipage}
}}
\caption{\textbf{Prompt for Question 14: Anchoring.}}

\vspace{0.2in}
This figure presents a prompt that elicits LLM responses to a question designed by~\cite{Tversky1974} to document anchoring.
    \label{prompt: q14}
\end{figure}
\begin{figure}[!htbp]
\vspace{-0.7in}
\scalebox{0.9}{
    \colorbox{blue!10}{
        \begin{minipage}{\textwidth}
Instructions: \\
Consider the following questions and respond according to the template provided. The output should be a markdown code snippet formatted in the following schema, including the leading and trailing ``\texttt{```json}" and ``\texttt{```}" and should not include any note or comment: \\
\texttt{```json} \\
\{ \\
\hspace{0.5cm} \texttt{\{"Questions": [} \\
\hspace{1cm} \texttt{\{"Question number": 1, "Lower bound": float, "Upper bound": float\},} \\
\hspace{1cm} \texttt{\{"Question number": 2, "Lower bound": float, "Upper bound": float\},} \\
\hspace{1cm} \texttt{...} \\
\hspace{1cm} \texttt{],} \\
\hspace{0.5cm} \texttt{"Reasoning": string} \\
\} \\
\texttt{```} \\

Questions: \\
For the following series of questions with clear-cut numerical answers, please provide 90\% confidence intervals. Such an interval has a lower and an upper bound such that you are 90\% sure that the correct answer lies in this interval. Note that, if your intervals are too wide, the correct answer will fall in your interval more than 90\% of the time while, if your intervals are too narrow, the correct answer will fall in your interval less than 90\% of the time. \\

\vspace{0.05in}

1.\hspace{1cm}World population total growth between 1990 and 2000 (in percentage terms) \\
2.\hspace{1cm}Year in which Newton discovered universal gravitation \\
3.\hspace{1cm}Number of nations in OPEC \\
4.\hspace{1cm}Number of medals that Greece won at the first Olympic Summer Experiments in 1896 \\
5.\hspace{1cm}Year in which Bell patented the telephone \\ 
6.\hspace{1cm}Percentage of total area in world covered by water \\
7.\hspace{1cm}Height of Sears Tower (now known as the Willis Tower) in Chicago (in feet) including the highest antenna on top of the building \\
8.\hspace{1cm}Number of nations in NATO \\
9.\hspace{1cm}Age of sun in billions of years \\
10.\hspace{1cm}Number of joints in human body \\
11.\hspace{1cm}GDP per capita in France (in thousands of \$US) in 2000 \\
12.\hspace{1cm}Current number of member states in the United Nations General Assembly \\
13.\hspace{1cm}Year in which Mozart wrote his first symphony \\
14.\hspace{1cm}Gestation (conception to birth) period of an Asian elephant (in days) \\
15.\hspace{1cm}Elevation (in feet above sea level) of Mt. Everest \\
16.\hspace{1cm}Number of babies born in world in 2001 (per 1000 people) \\
17.\hspace{1cm}World-wide life expectancy at birth in 2001 \\
18.\hspace{1cm}Land area in the world (in millions of sq mile as of 2017) \\
19.\hspace{1cm}Greatest depth (in feet) of the Pacific Ocean \\
20.\hspace{1cm}Number of calories in 8-ounce russet potato (flesh and skin) according to United States Department of Agriculture \\

\vspace{0.05in}

Response Format: \\
Please answer as shown above. For each question, write the answers as question number (number between 1 and 20), lower bound (a precise number), upper bound (a precise number). For the answers you just provided, please also provide your reasoning type (``A" if your reasoning is based more on intuitive thinking, and ``B" if your reasoning is based more on analytical thinking and calculations).
\end{minipage}
}}
\caption{\textbf{Prompt for Question 15: Overconfidence - overprecision.}}

\vspace{0.2in}
This figure presents a prompt that elicits LLM responses to a set of general knowledge questions; the questions are adapted from Appendix C of~\cite*{deaves2009experimental} and used to document overprecision.
    \label{prompt: q15}
\end{figure}
\begin{figure}[!htbp]
\scalebox{0.92}{
    \colorbox{blue!10}{
       \begin{minipage}{\textwidth}
Instructions: \\
Consider the following questions and respond according to the template provided. The output should be a markdown code snippet formatted in the following schema, including the leading and trailing ``\texttt{```json}" and ``\texttt{```}" and should not include any note or comment: \\
\texttt{```json} \\
\{ \\
\hspace{0.5cm} \texttt{\{"Questions": [} \\
\hspace{1cm} \texttt{\{"Question number": 1, "Choice": string\},} \\
\hspace{1cm} \texttt{\{"Question number": 2, "Choice": string\},} \\
\hspace{1cm} \texttt{...]} \\
\hspace{1cm} \texttt{,} \\
\hspace{0.5cm} \texttt{"Reasoning": string,} \\
\hspace{0.5cm} \texttt{"Accuracy": int} \\
\} \\
\texttt{```} \\

Questions: \\
Here are ten questions about investment: \\
1 - If you buy a company's stock \\
A.\hspace{1cm}You own a part of the company \\
B.\hspace{1cm}You have lent money to the company \\
C.\hspace{1cm}You are liable for the company's debts \\
D.\hspace{1cm}The company will return your original investment to you with interest \\

\vspace{0.05in}

2 - If you buy a company's bond \\
A.\hspace{1cm}You own a part of the company \\
B.\hspace{1cm}You have lent money to the company \\
C.\hspace{1cm}You are liable for the company's debts \\
D.\hspace{1cm}You can vote on shareholder resolutions \\

\vspace{0.05in}

3 - If a company files for bankruptcy, which of the following securities is most at risk of becoming virtually worthless? \\
A.\hspace{1cm}The company's preferred stock \\
B.\hspace{1cm}The company's common stock \\
C.\hspace{1cm}The company's bonds \\

\vspace{0.05in}

4 - In general, investments that are riskier tend to provide higher returns over time than investments with less risk. \\
A.\hspace{1cm}True \\
B.\hspace{1cm}False \\

\vspace{0.05in}

5 - Over the 30 years ending in December 2019 in the US, the best average returns have been generated by: \\
A.\hspace{1cm}Stocks \\
B.\hspace{1cm}Bonds \\
C.\hspace{1cm}CDs \\
D.\hspace{1cm}Money market accounts \\
E.\hspace{1cm}Precious metals \\

\end{minipage}
}
}
\end{figure}

\begin{figure}[!htbp]
\scalebox{0.92}{
    \colorbox{blue!10}{
      \begin{minipage}{\textwidth}

6 - What has been the approximate average annual return of the S\&P 500 stock index for the 50 years ending in December 2019 (not adjusted for inflation)? \\
A.\hspace{1cm}--10\% \\ 
B.\hspace{1cm}--5\% \\
C.\hspace{1cm}+5\% \\
D.\hspace{1cm}+10\% \\
E.\hspace{1cm}+15\% \\ 
F.\hspace{1cm}+20\% \\

\vspace{0.05in}

7 - Which of the following best explains the distinction between nominal returns and real returns? \\
A.\hspace{1cm}Nominal returns are pre-tax returns; real returns are after-tax returns \\
B.\hspace{1cm}Nominal returns are what an investment is expected to earn; real returns are what an investment actually earns \\
C.\hspace{1cm}Nominal returns are not adjusted for inflation; real returns are adjusted for inflation \\
D.\hspace{1cm}Nominal returns are not adjusted for fees and expenses; real returns are adjusted for fees and expenses \\

\vspace{0.05in}

8 - Which of the following best explains why many municipal bonds pay lower yields than other government bonds? \\
A.\hspace{1cm}Municipal bonds are lower risk \\
B.\hspace{1cm}There is a greater demand for municipal bonds \\
C.\hspace{1cm}Municipal bonds can be tax-free \\

\vspace{0.05in}

9 - You invest \$500 to buy \$1,000 worth of stock on margin. The value of the stock drops by 50\%. You sell it. Approximately how much of your original \$500 investment are you left with in the end? \\
A.\hspace{1cm}\$500 \\
B.\hspace{1cm}\$250 \\
C.\hspace{1cm}\$0 \\

\vspace{0.05in}

10 - Which is the best definition of selling short? \\
A.\hspace{1cm}Selling shares of a stock shortly after buying it \\
B.\hspace{1cm}Selling shares of a stock before it has reached its peak \\
C.\hspace{1cm}Selling shares of a stock at a loss \\
D.\hspace{1cm}Selling borrowed shares of a stock \\

\vspace{0.05in}

Response Format: \\
Please answer as shown above. For each question, write your answers as question number (number between 1 and 10) and your choice (one of either ``A", ``B", ``C", ``D", ``E", or ``F") . For the answers you just provided, please also provide your reasoning type (``A" if your reasoning is based more on intuitive thinking, and ``B" if your reasoning is based more on analytical thinking and calculations) and provide an estimate of the accuracy of your answers reflecting how many questions you believe you answered correctly (an integer between 0 and 10).
\end{minipage}
}
}
\caption{\textbf{Prompt for Question 16: Overconfidence - overestimation.}}

\vspace{0.2in}
This figure presents a prompt that elicits LLM responses to ten investment-related questions, along with the LLMs' estimates of their responses' accuracy. The prompt follows the procedure discussed in~\cite{moore2008trouble} to document overestimation, and the ten questions are based on the ``investing knowledge quiz'' designed by the Financial Industry Regulatory Authority.
 
    \label{prompt: q16}
\end{figure}

\end{document}